\documentclass[aps,prx,reprint]{revtex4-2}
\usepackage{float}
\usepackage{graphicx,xcolor} 
\usepackage[normalem]{ulem}
\usepackage{amsmath}
\usepackage{booktabs} 
\usepackage{multirow} 
\usepackage[unicode=true,
bookmarks=true,bookmarksnumbered=false,bookmarksopen=false,
breaklinks=false,pdfborder={0 0 1},backref=false,colorlinks=true]
{hyperref}
\hypersetup{
	linkcolor=magenta, urlcolor=blue, citecolor=blue, pdfstartview={FitH}, hyperfootnotes=false, unicode=true} 

\usepackage[normalem]{ulem}

\begin{document}
\title{System-Level Design of Scalable Fluxonium Quantum Processors with Double-Transmon Couplers}

\date{\today}

\author{Guo Xuan Chan}
\affiliation{Quantum Science Center of Guangdong-Hong Kong-Macao Greater Bay Area, Shenzhen, China}

\author{Wangwei Lan}
\affiliation{Quantum Science Center of Guangdong-Hong Kong-Macao Greater Bay Area, Shenzhen, China}

\author{Tenghui Wang}
\affiliation{Quantum Science Center of Guangdong-Hong Kong-Macao Greater Bay Area, Shenzhen, China}

\author{Xizheng Ma}
\affiliation{Quantum Science Center of Guangdong-Hong Kong-Macao Greater Bay Area, Shenzhen, China}

\author{Chunqing Deng}
 \email{dengchunqing@quantumsc.cn}

\affiliation{Quantum Science Center of Guangdong-Hong Kong-Macao Greater Bay Area, Shenzhen, China}

\author{Lijing Jin}%
 \email{jinlijing@quantumsc.cn}
\affiliation{Quantum Science Center of Guangdong-Hong Kong-Macao Greater Bay Area, Shenzhen, China}

\begin{abstract}

Fluxonium qubits combine long coherence times with strong anharmonicity, making them a promising platform for scalable superconducting quantum processors. Recent experiments have demonstrated high-fidelity operations in multi-qubit processors while suppressing stray qubit interactions using fluxonium–transmon–fluxonium (FTF) architectures. However, scaling such systems to larger arrays is constrained by a trade-off between achievable coupling strength, crosstalk suppression and qubit–qubit spacing required for wiring in a two-dimensional architecture.
Multimode couplers, such as the double-transmon coupler (DTC), provide a promising pathway to overcome this limitation by enabling stronger interactions without compromising qubit spacing and isolation. Here, we develop a quantitative design framework for fluxonium-based quantum processors employing DTCs. 
Central to this work is a frequency-partitioned architecture that places qubit transitions, tunable-coupler excitations, and resonator modes in well-separated spectral regions. This structured allocation reduces parameter interdependence and enables the concurrent optimization of gate operations, readout, and qubit reset.
By formulating device design as a multi-objective optimization problem under realistic experimental constraints and fabrication-induced disorder, we develop a tractable sequential workflow and determine a feasible parameter regime that simultaneously supports high-fidelity single- and two-qubit gates, fast qubit reset, and robust dispersive readout.
Within this framework, we further identify key design principles for scalable operation, including strong qubit–coupler capacitive coupling to suppress leakage during microwave-activated gates, and a large frequency detuning between the coupler's on- and off-modes to mitigate spectator-induced crosstalk during parallel operations.
These results establish a system-level architectural methodology that links circuit parameters to processor-level performance, and provide an experimentally actionable pathway toward scalable fluxonium quantum processors.

\end{abstract}

\maketitle

\section{Introduction}

Achieving scalable superconducting quantum computation is essential for fault-tolerant quantum information processing. Over the past decade, this effort has largely converged on architectures based on transmon qubits~\cite{Koch.07}, 
whose relative simplicity, robust design, and ease of integration have enabled landmark demonstrations of quantum advantage~\cite{Arute.19, Wu.21} and the implementation of quantum error-correction protocols~\cite{Acharya.25, He.25, Besedin.26, Wang.26}.
Despite this progress, scalability remains fundamentally limited by the competing demands of strong, controllable interactions and strict qubit isolation in multi-qubit systems.
In transmon-based architectures, weak anharmonicity combined with dense capacitive coupling gives rise to spectral crowding and drive-induced crosstalk that intensify with increasing system size. While tunable couplers~\cite{Yan.18} mitigate static interactions, high-fidelity parallel operations rely on extensive multi-qubit calibration to compensate for spectator-dependent errors~\cite{Klimov.24, Valles.25}. This reliance introduces a calibration overhead that scales unfavorably with system size, reflecting a fundamental constraint on scalable operation.

Alternatively, fluxonium qubits~\cite{Manucharyan.09} operate in a distinct regime characterized by exceptionally large anharmonicity, strongly suppressed low-frequency charge-dipole matrix elements, and intrinsically enhanced resilience to environmental noise. These properties enable long coherence times~\cite{Somoroff.23, Wang.25} and intrinsically suppress unwanted interactions within the computational subspace. 
Recent experiments have demonstrated high-fidelity two-qubit gates between fluxonium qubits mediated by transmon-based couplers~\cite{Ding.23}. However, it remains unclear whether these coupling strategies can be extended beyond few-qubit demonstrations toward scalable quantum processor architectures. Although one-dimensional chains have recently been realized~\cite{Zhan.26}, generalization to two-dimensional grids introduces new architectural constraints, including increased qubit spacing for control wiring and the simultaneous need to preserve strong, tunable interactions among neighboring qubits without incurring additional crosstalk or calibration overhead.

\begin{table*}[ht]
\label{tab:architecture_comparison}
\renewcommand{\arraystretch}{1.5}
\begin{tabular}{p{2.1cm} @{\hspace{0.3cm}} p{4.8cm} @{\hspace{0.3cm}} p{4.8cm} @{\hspace{0.3cm}} p{5cm}}
\hline \hline
\textbf{Architecture} & \textbf{Coupling and Gate} & \textbf{Advantages} & \textbf{Bottlenecks} \tabularnewline
\hline
$\mathrm{F}-\mathrm{F}$ 
& Capacitive; \newline iSWAP, CR \cite{Bao.22, Dogan.23, Huang.26}
& Long coherence; low leakage 
& Always-on $ZZ$ interaction; high wiring density \tabularnewline

$\mathrm{F}-\mathrm{F}$ 
& Inductive or \newline Inductive + Capacitive; \newline CR, CZ \cite{Nguyen.22, Chakraborty.25, Lin.25.1, Lin.25.2, Ma.24} 
& Long coherence; low leakage 
& Flux crosstalk; high wiring density \tabularnewline

$\mathrm{F}-\mathrm{R}-\mathrm{F}$ 
& Resonator mediated; \newline MAP, RIP \cite{Rosenfeld.24, Xiong.25} 
& Long-range coupling; low wiring overhead 
& Always-on interaction \tabularnewline

$\mathrm{F}-\mathrm{T_\mathrm{G}}-\mathrm{F}$ 
& Grounded transmon mediated; \newline MAP \cite{Ding.23, Singh.26, Zwanenburg.26}
& Strong, tunable coupling 
& Microwave crosstalk, high wiring density  \tabularnewline

$\mathrm{F}-\mathrm{T_\mathrm{F}}-\mathrm{F}$ 
& Floating transmon mediated; \newline MAP \cite{Heunisch.25}
& Reduced spectator crosstalk 
& Weak coupling strength \tabularnewline

\hline
$\mathbf{\mathrm{F}-\mathrm{DTC}-\mathrm{F}}$ \newline \textbf{(this work)} 
& Double-transmon mediated; \newline MAP 
& Strong, controllable interaction; \newline intrinsic suppression of microwave and spectator induced crosstalk
& Multi-parameter optimization; design complexity \textbf{(central focus of this work)} \tabularnewline

\hline \hline
\end{tabular}
\caption{Comparison of representative fluxonium-based multi-qubit architectures.
Architecture components are abbreviated as follows: F, fluxonium; R, resonator; $\mathrm{T_G}$, grounded tunable transmon; $\mathrm{T_F}$, floating tunable transmon; and DTC, double-transmon coupler. Gate schemes are abbreviated as follows: CR, cross-resonance; CZ, controlled-$Z$; MAP, microwave-activated controlled-phase; and RIP, resonator-induced phase.}
\end{table*}

These considerations motivate an alternative architectural paradigm beyond single-mode coupler designs. Here we pursue this direction by integrating fluxonium qubits with a double-transmon coupler (DTC). The DTC mediates long-range, tunable interactions while intrinsically suppressing microwave leakage and spectator-induced crosstalk, and it further preserves effective isolation between computational and noncomputational modes~\cite{Goto.22, Kubo.24, Campbell.23, Li.24, Li.25, Zhao.25.1, Cai.25}. When combined with the distinctive spectral structure of fluxonium qubits, this architecture enables structured frequency allocation, reduces parameter interdependence, and improves robustness against fabrication-induced disorder, thereby offering a viable route toward scalable multi-qubit quantum processors.

Despite these advantages, the rich spectral structure of fluxonium qubits, together with their pronounced sensitivity to circuit parameters, renders systematic device-level design inherently nontrivial. In contrast to all-transmon architectures, where frequency allocation for qubits and couplers is carried out in a relatively straightforward manner, fluxonium-based systems incorporating multi-mode couplers require a more globally constrained design strategy in which qubit and coupler spectra are strongly interdependent.
Previous studies have primarily focused on suppressing spectator-induced crosstalk and enabling high-fidelity two-qubit gates~\cite{Zhao.25.1}. However, several practical considerations that become increasingly relevant in large-scale implementations, such as active reset protocols, high-fidelity qubit readout, and realistic fabrication-induced parameter variations, are often not explicitly incorporated.
To address these limitations, we develop a quantitative theoretical framework that directly connects microscopic Hamiltonian parameters to experimentally relevant performance metrics under realistic measurement and fabrication constraints. 
Within this framework, we derive explicit design criteria for leakage suppression, crosstalk mitigation, and disorder robustness, and identify regimes in which these requirements can be simultaneously satisfied at the system-design level, thereby providing concrete guidance toward scalable fluxonium-based quantum architectures.

The remainder of this paper is organized as follows. In Sec.~\ref{sec:whyF-DTC-F}, we critically examine existing fluxonium-based multi-qubit coupling schemes and identify the key limitations that constrain scalability. 
We then propose the double-transmon coupler as an architecture capable of satisfying both the strong coupling needed for fast two-qubit gates and the isolation requirements for scalability. 
In Sec.~\ref{sec:ProblemFormulation}, we introduce the Hamiltonian of the Fluxonium–DTC–Fluxonium system and formulate the parameter design as an inverse problem, linking device parameters to a target set of criteria inspired by DiVincenzo’s standards. In Sec.~\ref{sec:MetrologyandImplementation}, we develop a systematic, metrology-informed design framework that explicitly incorporates practical calibration protocols into the optimization of circuit parameters. We conclude in Sec.~\ref{sec:Conclusion}.

\section{Fluxonium–Double Transmon Coupler–Fluxonium Architecture: A Scalable Building Block}\label{sec:whyF-DTC-F}

To realize scalable fluxonium-based quantum processors, the elementary building block must ensure robust inter-block isolation, thereby suppressing spurious interactions and preserving the integrity and independent controllability of each subsystem. Specifically, such platforms should satisfy three critical criteria: (1) suppressed microwave driving crosstalk, ensuring that high-fidelity parallel gate operations can be executed without extensive calibration overhead or complex pulse-shaping techniques; (2) minimized flux-bias crosstalk, guaranteeing that tuning a specific element into an operational regime remains strictly localized and does not degrade the coherence or performance of adjacent qubits; and (3) negligible spectator-induced crosstalk, ensuring that targeted gate operations remain entirely decoupled from the states or dynamics of neighboring qubits. An additional requirement is that the spacing between neighboring qubits must not be excessively reduced, as overly compact layouts introduce substantial routing complexity. This constraint becomes particularly critical in large-scale quantum processors, where wiring congestion and layout limitations can significantly hinder scalability.
Various multi-qubit architectures have been proposed in the literature, which can be broadly classified into three main categories: direct coupling \cite{Bao.22, Dogan.23, Huang.26, Nguyen.22, Lin.25.1, Lin.25.2, Ma.24, Chakraborty.25}, single-mode couplers \cite{Chakraborty.25, Zwanenburg.26, Ding.23, Singh.26, Lange.25, Rosenfeld.24, Kugut.25, Heunisch.25, Xiong.25}, and double-mode couplers \cite{Weiss.22, Zhang.24, Moskalenko.21, Moskalenko.22, Zhao.25.1, Zhao.26}. A comparative summary of these architectures and their key characteristics are presented in Table~\ref{tab:architecture_comparison}. In the following, we critically assess these existing paradigms and propose that the double-transmon coupler (DTC) provides a uniquely favorable solution, capable of simultaneously meeting the stringent requirements stated above.

For direct-coupling architectures, the candidate schemes are purely capacitive \cite{Bao.22, Dogan.23, Huang.26}, purely galvanic inductive \cite{Lin.25.1, Lin.25.2, Ma.24}, or a hybrid galvanic-capacitive approach \cite{Chakraborty.25, Nguyen.22}. Purely capacitive coupling is generally insufficient as there exists an always-on residual $ZZ$ interactions \cite{Bao.22, Dogan.23, Huang.26}. By contrast, purely galvanic or hybrid approaches can achieve the sufficiently strong qubit-qubit interactions and high on-to-off ratios. A galvanic connection naturally leverages the large superinductance of fluxonium qubits to realize strong coupling, and such schemes have proven effective for mitigating nearest-neighbor $ZZ$ interactions \cite{Nguyen.22, Ma.24}. By confining both single- and two-qubit operations entirely to the computational subspace, this approach exploits the intrinsically long coherence times of fluxonium and suppresses leakage errors due to its large anharmonicity. However, galvanically connected unit cells present significant design challenges: applying a local flux bias inevitably introduces long-range flux crosstalk across the multi-qubit device \cite{Zhang.24}. Furthermore, the extended galvanic network introduces stray harmonic modes that can strongly couple to the qubits, leading to dephasing via photon shot noise \cite{Ma.24}.

Architectures employing single-mode couplers utilize either fixed-frequency \cite{Rosenfeld.24, Xiong.25} or tunable elements \cite{Zwanenburg.26, Ding.23, Singh.26, Lange.25, Kugut.25, Heunisch.25}. Fixed-frequency couplers, such as linear resonators \cite{Rosenfeld.24, Xiong.25}, offer an extended physical length that compensates for the compact footprint of fluxonium qubits, thereby reducing wiring density. Nevertheless, the always-on effective coupling mediated by the resonator might lead to detrimental spectator-induced crosstalk, especially when the resonator is populated with a large photon number during parallel two-qubit gate operations, which remains unstudied in the literature. Alternatively, tunable couplers, such as flux-tunable transmons \cite{Zwanenburg.26, Ding.23, Singh.26, Lange.25, Kugut.25, Heunisch.25, WangJH.25}, facilitate active switching of the inter-qubit coupling. These tunable transmons are typically configured as either grounded or floating. To suppress spurious $ZZ$ interactions responsible for spectator-induced crosstalk in the operational subspace, grounded couplers rely on a direct inter-qubit capacitance to offset the effective coupling mediated by the coupler itself \cite{Yan.18}. Consequently, this direct capacitive channel introduces a critical trade-off by exacerbating microwave driving crosstalk between adjacent qubits. 
Conversely, while floating couplers can suppress the residual $ZZ$ interaction without a direct capacitive path \cite{Sete.21}, maintaining a large $E_J/E_C$ ratio to mitigate charge noise results in a comparatively weaker effective interaction between the qubits; this is because, with floating-type couplers, the qubit-qubit coupling is inherently mediated by the coupler's differential mode \cite{Ding.23}.

To overcome these limitations, we turn to double-mode couplers as a fundamentally more versatile approach. In particular, we consider a hybrid architecture in which qubits are capacitively coupled to a double-transmon element with intrinsic inductive coupling \cite{Goto.22, Kubo.24, Campbell.23, Li.24, Li.25, Zhao.25.1, Cai.25}. The double-transmon coupler (DTC) operates by tuning to a specific flux bias at which the inter-qubit couplings mediated by its two normal modes are equal in magnitude and opposite in sign, thereby canceling each other~\cite{Li.24}. From a circuit perspective, this cancellation of the two coupling paths is equivalent to turning off the inductive coupling, as the flux-dependent effective inductance of the shared Josephson junction within the DTC diverges to infinity at this bias point. This divergence creates an effective open circuit that prevents current flow through the junction, thereby nullifying the coupling path between the adjacent qubits \cite{Chen.14}. Conversely, the interaction is effectively ``turned on" by flux biasing the DTC away from this cancellation point. In the active regime, the couplings mediated by the two normal modes are no longer equal, and the detuning between the qubit and coupler modes decreases, strongly enhancing the net effective inter-qubit coupling.

The ability to switch the qubit–qubit interaction on and off via flux tuning provides key advantages for scalability in the DTC architecture. Although originally developed for transmon qubits \cite{Goto.22, Kubo.24, Campbell.23, Li.24, Li.25, Cai.25}, the DTC can be repurposed as a highly effective tunable coupler for fluxonium systems \cite{Zhao.25.1}. The alternating capacitive and galvanic connections naturally mitigate the long-range flux crosstalk inherent to purely galvanic designs, while the absence of direct inter-qubit capacitive paths simultaneously suppresses microwave driving crosstalk. In the ideal limit of vanishing parasitic intra-coupler capacitance, the DTC can, in principle, achieve an infinitely large on–off ratio through destructive interference between coupling pathways mediated by its two normal modes. In the activated regime, the large charge-dipole matrix elements of fluxonium plasmon transitions enable strong microwave-driven interactions, providing an efficient mechanism for implementing microwave-activated phase (MAP) gates~\cite{Ding.23}. Finally, the extended physical footprint of the double-transmon increases the spatial separation between fluxonium qubits, further easing wiring constraints. Consequently, this architecture simultaneously fulfills the key criteria outlined above, establishing a promising route toward scalable fluxonium-based quantum processors.

\begin{figure*}[ht]
    \centering
    \includegraphics[width=0.28\linewidth]{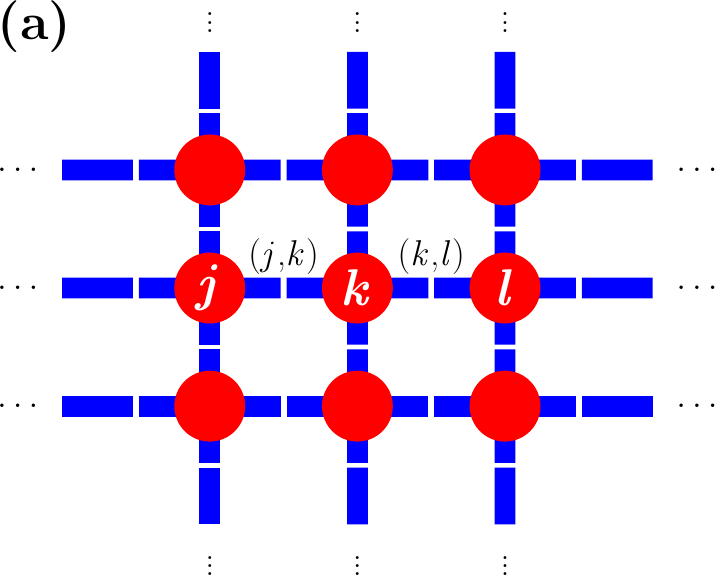}
    \includegraphics[width=0.63\linewidth]{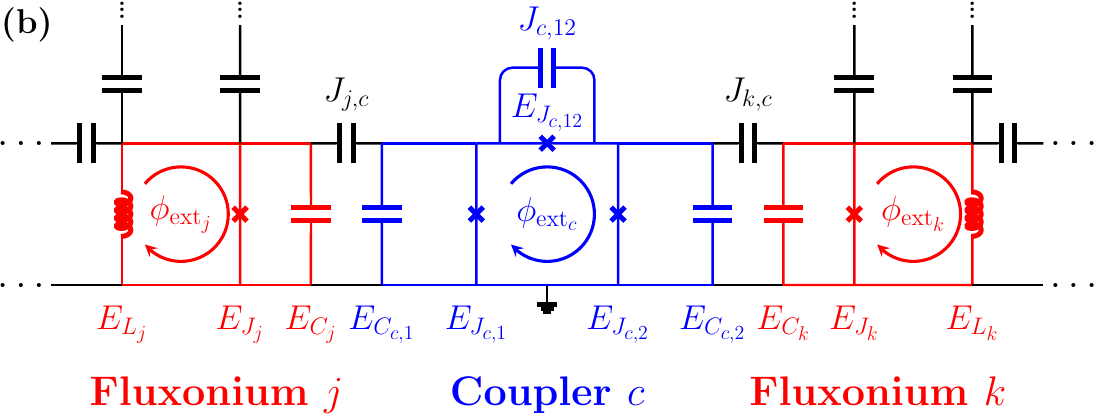}
    \caption{(a) Schematic of a large-scale grid of fluxonium qubits (red circles) connected via double-transmon  couplers (blue). (b) Detailed circuit diagram of a pair of fluxonium qubits, indexed $j$ and $k$, coupled by a DTC, labeled $c=(j,k)$.} 
    \label{fig:grid}
\end{figure*}

While these advantages are substantial, this architecture imposes additional constraints on system-level design. Compared with single grounded transmon coupler schemes, the effective coupling between the qubit mode and DTC's differential mode, relevant for two-qubit gate operations, is generally reduced. Nevertheless, this reduction can be mitigated by suppressing capacitance loading in the fluxonium circuit~\cite{QG2026}. In addition, parameter design in the F–DTC–F system remains challenging due to the complexity of the underlying four-mode energy spectrum, particularly when practical constraints from measurement and fabrication are taken into account. Consequently, developing a systematic and physically transparent framework for parameter design in such systems constitutes the central objective of this work. 

\section{Model and Problem Formulation}\label{sec:ProblemFormulation}

\subsection{System Hamiltonian} \label{subsec:system_H}

We investigate a scalable multi-qubit architecture comprising fluxonium qubits interconnected via double-transmon couplers (DTCs), as illustrated in Fig.~\ref{fig:grid}(a). The total Hamiltonian of the multi-qubit device is modeled as
\begin{equation}
H_\mathrm{total} = H_\mathrm{sys} + H_r + H_d,
\end{equation}
where $H_\mathrm{sys}$ describes the system of fluxonium qubits, DTCs, and their mutual interactions; $H_r$ accounts for the readout and reset resonator modes as well as the interaction with qubits or DTCs; and $H_d$ represents the external microwave drives. In particular, the system Hamiltonian includes the bare component terms and the coupling terms, given by
\begin{equation}
\label{eq:H}
\begin{split}
& H_{\mathrm{sys}} =\sum_j H_{q_j}(E_{C_j}, E_{J_j}, E_{L_j}) \\
& + \sum_{c} H_{c} (E_{C_{c,1}}, E_{C_{c,2}}, E_{J_{c,1}}, E_{J_{c,2}}, E_{J_{c,12}}, J_{c,12}) \\
& + \sum_{j,c} H_{q_j,c}^{\mathrm{int}} (J_{j,c}),
\end{split}
\end{equation}
where $H_{q_j}$, $H_c$, and $H_{q_j,c}^\mathrm{int}$ represent the Hamiltonians of the $j$-th fluxonium, the $c$-th DTC, and the interactions between them, respectively. For clarity, we index the DTCs as pairs $c=(j,k)$, indicating that the $j$-th and $k$-th fluxonium qubits are coupled via the $c$-th DTC, cf.~Figs.~\ref{fig:grid}(a) and (b).
Defined by the charging $E_{C_j}$, Josephson $E_{J_j}$, and inductive $E_{L_j}$ energies, the Hamiltonian of the $j$-th fluxonium is
\begin{equation} \label{eq:H_fluxonium}
H_{q_j}=4E_{C_j}\hat{n}_j^2 - E_{J_j}\cos\hat{\phi}_j+\frac{E_{L_j}}{2} \left(\hat{\phi}_j-\phi_{\mathrm{ext}_j}\right)^2,
\end{equation}
where $\hat{n}_j$ and $\hat{\phi}_j$ are the charge and flux operators while $\phi_{\mathrm{ext}_j}$ is the flux bias through the qubit loop, cf.~Fig.~\ref{fig:grid}(b). 

The Hamiltonian $H_c$ for the $c$-th DTC, defined by the charging energies $E_{C_{c,m}}$, Josephson energies $E_{J_{c,m}}$ of the $m$-th transmon ($m=1,2$), inter-transmon Josephson energy $E_{J_{c,12}}$ and capacitive coupling $J_{c,12}$, is given by
\begin{align}\label{eq:H_DTC}
\begin{split}
H_{c} = & \sum_{m=1}^2\left[4E_{C_{c,m}}\hat{n}_{c,m}^2- E_{J_{c,m}}\cos\left(\hat{\phi}_{c,m}+\widetilde{C}_{c,m}\phi_{\mathrm{ext}_c}\right)\right]
\\ &- E_{J_{c,12}}\cos\left(\hat{\phi}_{c,2} - \hat{\phi}_{c,1}-\widetilde{C}_{c,12}\phi_{\mathrm{ext}_c}\right)
\\ &+ J_{c,12}\hat{n}_{c,1}\hat{n}_{c,2},
\end{split}
\end{align}
where $\hat{n}_{c,m}$ and $\hat{\phi}_{c,m}$ are the charge and phase operators, respectively, and $\phi_{\mathrm{ext}_c}$ is the external flux bias threading the loop formed by the Josephson junctions [cf.~Fig.~\ref{fig:grid}(b)]. The external flux is distributed across the three junctions (associated with $E_{J_{c,12}}$, $E_{J_{c,1}}$, and $E_{J_{c,2}}$) according to the coefficients $\widetilde{C}_{c,m}$ and $\widetilde{C}_{c,12}$, which are defined in Table~\ref{tab:dtc_HO_parameters}. This specific parameterization ensures that the time derivative of the external flux vanishes in the Hamiltonian~\cite{Campbell.23}. Furthermore, it accurately captures the quantum dynamics when a time-dependent external flux is applied~\cite{You.19}, which facilitates the derivation of the flux-noise-induced relaxation rates detailed in Appendix~\ref{sec:DTC_flux_noise}. For simplicity, we assume symmetric Josephson junctions for the DTC throughout this work, setting $E_{J_{c,1}}=E_{J_{c,2}}$. However, the DTC retains its fundamental capability to selectively enhance or nullify effective inter-qubit interactions even in the presence of up to $10\%$ junction asymmetry~\cite{Zhao.25.1}. We explicitly demonstrate this robustness in Secs.~\ref{subsubsec:lea_min} and \ref{subsec:xtalk}, where non-uniformities in $E_{C_{c,m}}$ and $E_{J_{c,m}}$ are introduced via random parameter perturbations.

\begin{table}[!t]
\begin{tabular}{l p{6.5cm}} 
    \hline \hline
    \textbf{Parameter} & \textbf{Description} \\
    \hline
    $|k_j\rangle$ & Bare state $|k\rangle$ of the $j$-th qubit \\
    $E_{k_j}$ & Energy of bare state $|k\rangle$ of the $j$-th qubit \\
    $\omega_{k_j}=E_{k_j}/\hbar$ & Angular frequency of bare state $|k\rangle$ of the $j$-th qubit \\ 
    $\omega_{{k,l}_j}$ & Transition frequency $\omega_{l}-\omega_{k}$ between bare states of the $j$-th qubit \\
    $|jck\rangle$ & Composite eigenstate of fluxonium qubits $j, k$ and the intervening DTC $c$. For example, $|101\rangle$ corresponds to both qubits in $|1\rangle$ and the DTC in $|0\rangle$. \\
    $|j c_1 k c_2 l\rangle$ & Composite eigenstate of three fluxonium qubits ($j,k,l$) and the intervening DTCs ($c_1,c_2$) \\
    $\omega_{k}$ & Angular frequency of eigenstate $|k\rangle$\\
    $\omega_{k,l}$ & Transition frequency $\omega_{l}-\omega_{k}$ between eigenstates $|k\rangle$ and $|l\rangle$ \\
    $\phi_{\mathrm{ext}_j}$ & External flux bias of the $j$-th fluxonium \\
    $\phi_{\mathrm{ext}_c}$ & External flux bias of the $c$-th DTC \\
    $\omega_{-_c}$ & Frequency of the DTC differential mode, which equals $\omega_{0,1_c}$ ($\omega_{0,2_c}$) at $\phi_{\mathrm{ext}_c}>\phi_{\mathrm{off}_c}$ $(<\phi_{\mathrm{off}_c})$\\
    $\omega_{+_c}$ & Frequency of the DTC common mode, which equals to $\omega_{0,2_c}$ ($\omega_{0,1_c}$) at $\phi_{\mathrm{ext}_c}>\phi_{\mathrm{off}_c}$ $(<\phi_{\mathrm{off}_c})$\\
    $\phi_{\mathrm{off}_c}$ & DTC ``turn-off'' flux bias (effective coupling vanishes) \\
    $\phi_{\mathrm{on}_c}$ & DTC ``turn-on'' flux bias (maximum effective coupling); occurs at resonance $\omega_{1,2_j}=\omega_{\mathrm{on}_c}$ \\
    $\omega_{\mathrm{on(off)}_c}$ & Frequency of the DTC mode at flux bias $\phi_{\mathrm{on(off)}_c}$ \\ [1ex]
    \hline \hline
\end{tabular}
\caption{Nomenclature for derived spectral properties and external control parameters used throughout this work. The notations for bare state energies and transition frequencies apply analogously to the $c$-th double-transmon coupler (DTC) by substituting the qubit index $j$ with the coupler index $c$.}
\label{tab:def}
\end{table}

\begin{table*}[t]
\begin{ruledtabular}
\begin{tabular}{lll}
Performance Metric & Definition / Description & Requirement \\
\hline
1. Single-qubit fidelity ($F_{\text{1q}}$) & See Appendix~\ref{appen:defSQF} & $F_{\text{1q}} > F_{\text{1q}}^*$ \\
2. Two-qubit gate characteristics   & Leakage $\eta$ (cf.~Appendix~\ref{subsec:leakage}) and & $\eta < \eta^*$, \\
& two-qubit gate fidelity $F_{\text{2q}}$ (cf.~Appendix~\ref{appen:2QF})
& $F_{\text{2q}} > F_{\text{2q}}^*$
\\
3. Spectator crosstalk ($\zeta$)          & Frequency shifts of transitions relevant  & $\zeta < \zeta^*$ \\
                                          & to the MAP gate due to residual coupling & \\
                                          & with spectator qubits (cf.~Appendix~\ref{sec:defxtalk}) &
                                          \\
4. Qubit readout                          & Dispersive shift $\chi$, critical photon                        & $|\chi| \sim \chi^*$, $n_{\text{crit}} \gtrsim n_{\text{crit}}^*$, \\
                                          & number $n_{\text{crit}}$ and SNR condition $|2\chi|/\kappa$        & $|2\chi|/\kappa \sim (|2\chi|/\kappa)^*$ \\
5. Qubit reset                            & Reset duration $t_{\text{reset}}$ and             & $t_{\text{reset}} < t_{\text{reset}}^*$, \\
                                          & reset fidelity $F_{\text{reset}}$                & $F_{\text{reset}} > F_{\text{reset}}^*$ \\
6. Decoherence induced by resonator mode  & Dephasing via photon shot noise and                   
& $T_\phi^\mathrm{shot}>\left(T_\phi^\mathrm{shot}\right)^*$, \\
  & energy relaxation via the Purcell effect & $T_1^\mathrm{Purcell}>\left(T_1^\mathrm{Purcell}\right)^*$ \\
\end{tabular}
\end{ruledtabular}
\caption{\label{tab:metrics} Summary of multi-qubit performance metrics and target requirements. Asterisks ($^*$) denote the target or optimal parameter values required for high-fidelity operation.}
\label{tab:requirements}
\end{table*}

The interaction between the $j$-th fluxonium and the $c$-th DTC is mediated by their mutual capacitance, described by the Hamiltonian
\begin{equation}\label{eq:H_qc}
H_{q_j,c}^{\mathrm{int}} = J_{j,c} \hat{n}_j \hat{n}_{c,m},
\end{equation}
where $J_{j,c}$ is the capacitive coupling strength, $\hat{n}_j$ denotes the charge operator of the $j$-th fluxonium, and $\hat{n}_{c,m}$ denotes the charge operator of the $m$-th mode of the $c$-th DTC. In this work, we restrict our analysis to nearest-neighbor capacitive couplings for simplicity, neglecting parasitic interactions between non-adjacent DTCs. Consequently, in the setup depicted in Fig.~\ref{fig:grid}(b), the coupling $J_{j,c}$ corresponds specifically to the interaction between the $j$-th fluxonium and the adjacent transmon of the $c$-th DTC.

Qubit readout and active reset are mediated by resonators capacitively coupled to the fluxonium qubits, described by the Hamiltonian
\begin{equation} \label{eq:H_resonator}
H_{r_j}/\hbar = \omega_{r_j} \hat{a}^\dagger_{r_j} \hat{a}_{r_j} + g_{r_j}(\hat{n}_j+\alpha_{j,c}\hat{n}_{c,m}) (\hat{a}^\dagger_{r_j} + \hat{a}_{r_j}),
\end{equation}
where $\omega_{r_j}$ is the resonator frequency and $g_{r_j}$ denotes the capacitive coupling strength between resonator $r_j$ and the $j$-th fluxonium. The resonator field decay is dominated by the coupling quality factor $Q_j$ to the feedline (overcoupled regime), yielding a photon decay rate $\kappa_j = \omega_{r_j}/Q_j$. The coefficient $\alpha_{j,c}$ quantifies the parasitic capacitive coupling between the resonator and the adjacent DTC. To differentiate the readout and reset systems, we reserve unprimed variables for the readout resonators and use primed notation for all active reset parameters, e.g.~$\omega'_{r_j}$ and $g'_{r_j}$ to designate the angular frequency and capacitive coupling strength of the $j$-th reset resonator, respectively.

Having established the static properties, we next proceed to the dynamic analysis of the system. Particularly, single- and two-qubit gates are performed by applying microwave drives to the fluxonium qubits. This driving is modeled by the Hamiltonian

\begin{equation}\label{eq:drive}
H_d(t) = \Omega_{d_j}(t) \left(\hat{n}_j + \gamma_{j,c} \hat{n}_{c,m} \right),
\end{equation}
where $\Omega_{d_j}(t)$ represents the time-dependent microwave pulse applied on the $j$-th fluxonium, and the term $\Omega_{d_j}(t)\gamma_{j,c}$ accounts for the parasitic drive on the adjacent coupler. Note that $\gamma_{j,c}$ denotes the ratio of the microwave-drive mutual capacitances to the DTC and the qubit, and plays a non-negligible role in practical device design. Similar to $\alpha_{j,c}$, this parasitic drive is intrinsic to the device geometry. Yet, rather than being detrimental, a carefully optimized $\gamma_{j,c}$ can enhance gate performance by suppressing leakage transitions. While a detailed optimization is outside the scope of this work, we incorporate realistic values for $\gamma_{j,c}$ derived from electrostatic simulations of the layout.

Having defined the underlying circuit Hamiltonians in terms of fundamental design parameters, specifically the constituent charging, Josephson, and inductive energies, alongside the inter-element coupling strengths, we obtain the relevant energy spectra and eigenstates by diagonalizing these governing equations. To provide a clear logical foundation for the subsequent analysis, Table~\ref{tab:def} formalizes the nomenclature for these derived spectral properties and external controls used throughout this work. Because the characterization of the tunable-coupler architecture involves multiple levels of physical abstraction, we organize these quantities into three distinct categories. First, we define the uncoupled properties of the system, establishing the notation for the bare states $|k_j\rangle$, eigenenergies $E_{k_j}$, and transition frequencies $\omega_{k,l_j}$, extracted from the isolated fluxonium and DTC Hamiltonians. Second, we specify the composite eigenstates ($|jck\rangle$, $|jc_1 k c_2 l\rangle$) and their corresponding eigenvalues ($\omega_k$) and transition frequencies ($\omega_{k,l}$) of the interacting multi-body system, which are essential for describing the targeted two-qubit gate dynamics and tracking parasitic crosstalk channels. Third, to simplify the discussion of the coupler's functionality, we analyze the lowest two excited states of the DTC in terms of its differential and common modes (with frequencies $\omega_{-_c}$ and $\omega_{+_c}$), noting that their assignment as the first or second excited state depends on the applied flux bias. Adopting this mode-based framework is instrumental in deriving the effective inter-qubit interactions mediated by the DTC (see Appendix~\ref{appen:parameter}), yielding crucial insights for mitigating spectator-induced crosstalk.
Finally, we detail the external control variables, encompassing the operational flux biases ($\phi_{\mathrm{on}_c}$, $\phi_{\mathrm{off}_c}$) and their corresponding mode frequencies ($\omega_{\mathrm{on}_c}$, $\omega_{\mathrm{off}_c}$) for the DTC ``turn-on'' and ``turn-off'' configurations that dictate the effective inter-qubit coupling network.

Notably, fluxonium qubits exhibit a highly structured energy spectrum that enables a natural separation between protected computational states and strongly coupled excited transitions. In this work, we operate at the half-flux sweet spot ($\phi_{\mathrm{ext},j} = \pi$), where the computational subspace is defined by the two lowest levels, $\{|0\rangle, |1\rangle\}$. At this bias point, charge matrix elements within the computational subspace are strongly suppressed, resulting in reduced dielectric loss and correspondingly enhanced energy relaxation times ($T_1$)~\cite{Long.19}. In contrast, higher-lying plasmon transitions retain large, transmon-like dipole matrix elements. We exploit this separation by utilizing the plasmon transitions to mediate capacitive coupling to the double-transmon couplers (DTCs), as well as to enable efficient interactions with readout and reset resonators.

\subsection{Criteria for Evaluation Based on DiVincenzo's Requirements } \label{subsec:criteria}

To benchmark the performance of our multi-qubit device, we define a set of criteria inspired by DiVincenzo’s standards \cite{DiVincenzo.00}. For a viable general-purpose processor, the composite system formed by fluxonium qubits and DTCs should demonstrate the benchmarking metrics listed in Table \ref{tab:requirements}.

While specific target thresholds depend on the intended application, ranging from near-term algorithms to fault-tolerant error correction, we adopt the following rigorous baseline standards for this work: single- and two-qubit gate fidelities of $F_\mathrm{1q}^*>99.99\%$ and $F_\mathrm{2q}^*>99.9\%$; reset metrics of $t^*_\mathrm{reset}<300$~ns and $F_\mathrm{reset}^*>99\%$; a leakage probability of $\eta^* <10^{-3}$; readout parameters of $2\chi^*\sim2$~MHz, $\left(|2\chi_j|/\kappa_j\right)^*\sim1$, and $n_\mathrm{crit}^*\gtrsim50$; resonator-induced decoherence limits of $\left(T_\phi^\mathrm{shot}\right)^*>10~\mu\mathrm{s}$ and $\left(T_1^\mathrm{Purcell}\right)^*>10~\mu\mathrm{s}$; and a spectator crosstalk limit of $\zeta^*<10$~kHz. These thresholds are carefully chosen to guarantee high-fidelity operation. Specifically, the leakage limit ($\eta^* < 10^{-3}$) strictly bounds the leakage-induced gate infidelity; as shown analytically in Appendix~\ref{appen:leaF}, this leakage rate contributes an error of $2.5\times10^{-4}$, safely within the overall two-qubit gate error budget of $1-F_\mathrm{2q}^* = 10^{-3}$. The readout requirements are established based on the optimization criteria detailed in Sec.~\ref{subsec:readout}. To ensure that resonator modes do not excessively degrade coherence, the Purcell relaxation target is set to $\left(T_1^\mathrm{Purcell}\right)^*>10~\mu\mathrm{s}$, which is sufficient to maintain both single-qubit and MAP-gate fidelities above $99.9\%$ (cf.~Appendices~\ref{appen:defSQF} and \ref{ref:rel_2QFid}). Although a direct analytical estimate for the MAP gate fidelity under photon shot noise is unavailable, we use the single-qubit gate fidelity as an indicative benchmark: the target dephasing time $\left(T_\phi^\mathrm{shot}\right)^*>10~\mu\mathrm{s}$ corresponds to a negligible shot-noise-induced single-qubit infidelity of $10^{-5}$ ($F_\mathrm{1q}=99.999\%$). Finally, the acceptable limit for spectator crosstalk ($\zeta^* < 10$~kHz) is derived from qubit dynamics simulations that evaluate the resulting two-qubit gate performance (see Appendix~\ref{subsec:xtalk_fid}). These simulations demonstrate that coherent errors induced by a $10$~kHz crosstalk strength remain negligible, on the order of $10^{-5}$.

In contrast to prior studies that often isolate specific gate metrics or assume ideal operating conditions, our approach explicitly incorporates practical device-level realities. Specifically, we account for crucial measurement and calibration trade-offs, for instance, balancing fast, high signal-to-noise ratio (SNR) readout against the preservation of qubit coherence, alongside fabrication-induced parameter variations across the device. Satisfying the comprehensive criteria outlined in Table~\ref{tab:requirements} under these realistic experimental limitations makes identifying a globally optimal parameter set, which encompasses 16 distinct circuit parameters, significantly more challenging. By addressing these multi-objective constraints simultaneously, we provide a robust and experimentally viable framework for scalable fluxonium quantum processors.

\subsection{Problem Formulation}

The characterization of the multi-qubit architecture relies on a comprehensive set of bare circuit elements, operational parameters, and inter-element couplings. For notational convenience, we organize these into specific parameter vectors corresponding to their respective Hamiltonians: the fluxonium parameters from Eq.~\eqref{eq:H_fluxonium}, $\mathbf{x}_q=\{E_{C_j}, E_{J_j}, E_{L_j}\}$; the DTC circuit parameters from Eq.~\eqref{eq:H_DTC}, $\mathbf{x}_c=\{E_{C_{c,m}}, E_{J_{c,m}}, E_{J_{c,12}}, J_{c,12}\}$; the fluxonium-DTC coupling parameters from Eq.~\eqref{eq:H_qc}, $\mathbf{x}_{qc} = \{J_{j,c}\}$; the resonator parameters from Eq.~\eqref{eq:H_resonator}, $\mathbf{x}_r = \{\omega_{r_j}, g_{r_j}, \alpha_{j,c}, Q_j\}$; and the external drive parameters from Eq.~\eqref{eq:drive}, $\mathbf{x}_d = \{\gamma_{j,c}\}$. 
To bridge these formal definitions with practical physical intuition, we refine our treatment of two specific subsets. First, while the DTC is technically governed by the bare elements in $\mathbf{x}_c$, it is physically more intuitive to characterize its operation using derived frequencies. Therefore, we alternatively parameterize the DTC set as $\mathbf{x}_c=\{\omega_{\mathrm{on}_c}, \omega_{\mathrm{off}_c}\}$, corresponding to the ``on" and ``off" configurations defined in Table~\ref{tab:def}. Second, regarding the external drive, we explicitly exclude the time-dependent microwave drive $\Omega_{d_j}(t)$ from the parameter space. Because $\Omega_{d_j}(t)$ encompasses both the applied pulse envelope and the carrier frequency, its rigorous optimization entails advanced quantum optimal control and pulse shaping techniques (such as DRAG pulses~\cite{Motzoi.13}) that fall outside the primary architectural scope of this study.

Consequently, the overall performance of the multi-qubit architecture is fully characterized by the aggregate parameter set $\mathbf{x} = \bigcup_{i\in\{q,c,qc,r,d\}}\mathbf{x}_i$. 
Given $\mathbf{x}$, the full system Hamiltonian can be constructed. From this, both the static and dynamical properties of the system are obtained, enabling the evaluation of the performance metrics defined above.
Our objective is to identify an optimal configuration $\mathbf{x}^*$ that satisfies the performance metrics outlined in Table~\ref{tab:requirements}, thereby enabling the DTC-based fluxonium device to execute fast, high-fidelity parallel gates as well as efficient reset and readout operations. Furthermore, acknowledging that fabrication imperfections are inevitable, we investigate the robustness of key performance metrics against parameter variations. This analysis imposes additional constraints on $\mathbf{x}$, ensuring that device performance remains robust in the presence of fabrication uncertainties.

\begin{figure*}[ht!]
    \centering
    \includegraphics[width=0.6\linewidth]{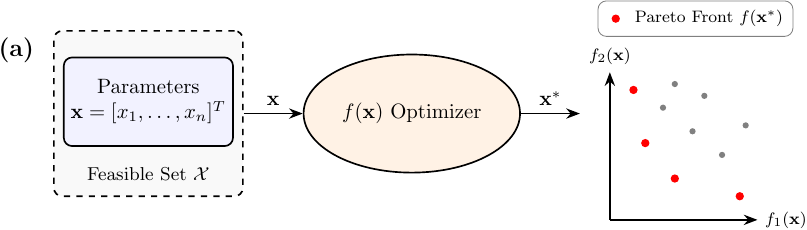}
    \vspace{10pt}
    \\
    \includegraphics[width=1.0\linewidth]{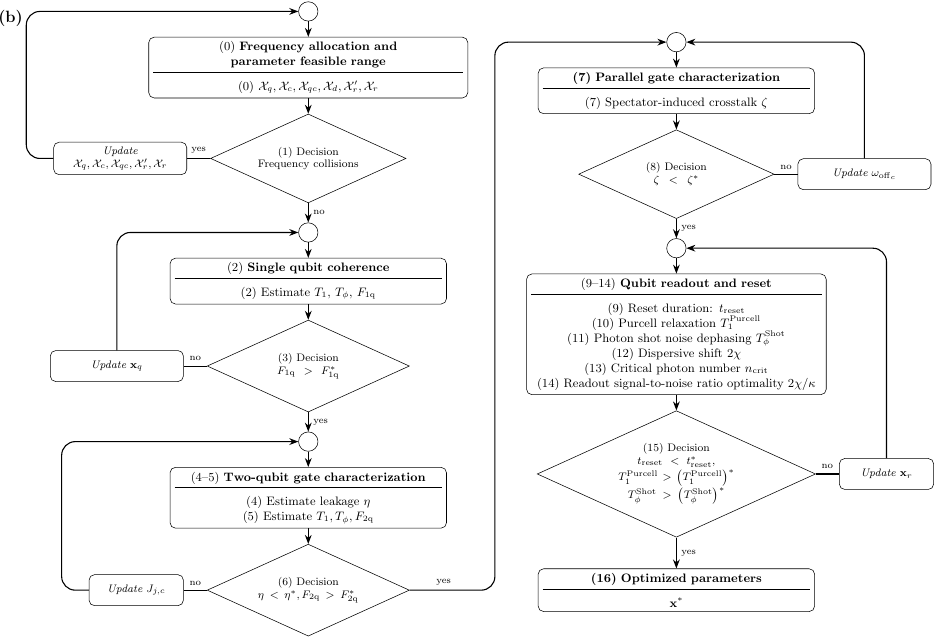}
    \caption{(a) Schematic of a multi-objective optimization workflow mapping a feasible set of input parameters $\mathbf{x}$ to an optimal Pareto front via an optimizer evaluating multiple objective functions $f(\mathbf{x})$. (b) Optimization workflow for the DTC-based fluxonium multi-qubit architecture. In the decision blocks, starred parameters (e.g., $F_\mathrm{1q}^*,\ \eta^*$) indicate the threshold values determining whether the design parameters meet the target specifications. We explicitly evaluate the robustness of key characteristics exhibiting non-trivial sensitivity to fabrication variances, including frequency allocation (Step 0), leakage $\eta$ (Step 4), and spectator crosstalk $\zeta$ (Step 7).}
    
    \label{fig:optimization}
\end{figure*}

For a given circuit topology, e.g., Fig.~\ref{fig:grid}(a), a specific hardware configuration is represented by the aggregate parameter vector $\mathbf{x} \in \mathcal{X}$, where $\mathcal{X}$ denotes the allowable parameter space. The system's performance is evaluated against $M$ distinct metrics, which are summarized in Table~\ref{tab:requirements}. To simplify the problem formulation, we assume without loss of generality that the objective is to minimize every metric. Consequently, performance targets that inherently require maximization, such as single- and two-qubit gate fidelities ($F_\mathrm{1q}$, $F_\mathrm{2q}$), are recast as infidelities ($1-F$). We define a vector-valued objective function $f(\mathbf{x})$ that maps a parameter configuration to its corresponding metric values, formulated as the following multi-objective optimization problem:
\begin{equation}
\min_{\mathbf{x}\in \mathcal{X}} f(\mathbf{x}) = \left[ f_1(\mathbf{x}), \dots, f_M(\mathbf{x}) \right]^\mathrm{T},
\end{equation}
where each scalar function $f_i(\mathbf{x})$ evaluates the value of the $i$-th performance metric. Because it is generally impossible to find a single parameter set $\mathbf{x}$ that simultaneously minimizes all competing metrics, our goal is to identify a Pareto-optimal parameter vector $\mathbf{x}^*$, cf.~Fig.~\ref{fig:optimization}(a). Formally, $\mathbf{x}^*$ is Pareto optimal if there exists no $\mathbf{x} \in \mathcal{X}$ that satisfies $f_i(\mathbf{x}) \leq f_i(\mathbf{x}^*)$ for all $i \in \{1, \dots, M\}$, with strict inequality for at least one such $i$ \cite{ZhuMT.26}. Physically, this implies that improving any single performance metric necessarily comes at the expense of degrading another.

Ideally, identifying the optimal parameter vector $\mathbf{x}^*$ would rely on a fully automated, global optimization routine requiring minimal user intervention. However, evaluating specific multi-qubit performance metrics stated above, such as spectator-induced crosstalk, driven leakage probabilities, and comprehensive system decoherence rates, is highly computationally intensive. This immense computational overhead severely restricts the sampling rate, rendering brute-force searches or high-dimensional global optimizations computationally intractable across the vast available parameter space $\mathcal{X}$.

To circumvent this bottleneck, we have developed a systematic workflow (Fig.~\ref{fig:optimization}(b)) that semi-decouples the optimization of these various metrics, despite their inherent physical interdependence on shared subsets of $\mathbf{x}$. 
Particularly, we adopt a two-phase optimization strategy. First, frequency allocation is performed to delineate the feasible parameter regime at the level of the system’s energy spectrum. Building upon this, the search space is further reduced by ordering parameter selection according to the underlying physical hierarchy. This hierarchical approach progressively lowers the effective dimensionality of the optimization problem. In the next section, we detail this protocol and demonstrate its systematic convergence to a robust, high-fidelity operating regime.

\section{Practical Parameter Design: Metrology and Implementation}\label{sec:MetrologyandImplementation}

To systematically generate a set of optimal parameters $\mathbf{x}^*$, we introduce a metrology and workflow that iteratively refines key parameters until specific design criteria are met, as illustrated in Fig.~\ref{fig:optimization}(b). Steps 0--1 introduce a frequency allocation scheme and set feasible ranges for parameters, ensuring that frequency collisions do not occur. Steps 2--3 optimize the single-qubit parameters $\mathbf{x}_q$ according to the criteria defined in Performance Metric 1 of Table~\ref{tab:requirements}.
Next, Steps 4--6 optimize $\mathbf{x}_{qc}$ by characterizing two-qubit gate properties, specifically leakage $\eta$ and gate fidelity $F_\mathrm{2q}$, to ensure the targets in Performance Metric 2 of Table~\ref{tab:requirements} are met. Following this, Steps 7--8 update the DTC parameters in $\mathbf{x}_c$ (i.e.~$\omega_{\mathrm{off}_c}$) to minimize spectator-induced crosstalk $\zeta$, as required by Performance Metric 3. Finally, Steps 9--14 verify the properties of the readout and reset resonators to satisfy the requirements listed in Performance Metrics 4--6 of Table~\ref{tab:requirements}.

The workflow in Fig.~\ref{fig:optimization}(b) is structured hierarchically, prioritizing the definition of circuit parameters, which requires navigating complex inter-dependencies, before targeting isolated features where parameter tuning is decoupled from the rest of the system. Specifically, Steps 0--1 establish the baseline configuration and valid frequency landscape, while Steps 2--15 refine each parameter subset through detailed simulations and analysis. In the following subsections, we detail the execution of this workflow to derive the optimal parameter configuration $\mathbf{x}^*$.

\subsection{Frequency allocation scheme and parameter feasible range} \label{subsec:freq_alloc}

Leveraging the rich spectral structure of fluxonium qubits, we propose a frequency-allocation scheme ordered from lowest to highest operating frequencies. The hierarchy and underlying rationale for each spectral band are as follows: (1) The single-qubit transition satisfies $\omega_{0,1_j} < \omega'_{r_j}$ at half qubit flux ($\phi_{\mathrm{ext}_j}=\pi$), ensuring that active reset can be executed by flux-biasing the qubit into resonance with the reset resonator. (2) The qubit-reset mode occupies the spectral gap $\omega_{0,1_j} < \omega'_{r_j} < \omega_{1,2_j}$; this placement prevents the reset resonator from degrading the two-qubit MAP gate fidelity via photon shot noise or Purcell relaxation. (3) The DTC ``turn-on'' frequency is tuned to resonance, $\omega_{\mathrm{on}_c} = \omega_{1,2_j}$, facilitating strong effective coupling between neighboring fluxonium qubits for fast MAP gate operations. (4) The DTC ``turn-off'' frequency is biased such that $\omega_{1,2_j} < \omega_{\mathrm{off}_c} < \omega_{0,3_j}$, which maintains sufficient detuning between the "on" and ``off" states to suppress spectator crosstalk, while avoiding excessive detuning that would increase flux-noise sensitivity at the operational point. (5) Finally, the readout frequency is positioned in the upper spectrum, $\omega_{0,3_j} < \omega_{r_j} < \omega_{1,4_j}$, enabling qubit measurement via dispersive coupling between the resonator mode $\omega_{r_j}$ and the $\omega_{0,3_j}$ and $\omega_{1,4_j}$ transitions. Crucially, this highest frequency band is explicitly chosen to avoid spectral overlap with all other control operations. By assigning distinct frequency bands to separate functional roles, this strategically interleaved allocation significantly reduces parameter interdependence. In large-scale quantum processors, this isolation benefits both calibration and fabrication by enabling relatively independent tuning procedures and enhancing robustness against fabrication-induced variations. Next, we introduce feasible parameter ranges as bounds for the optimization process shown in Fig.~\ref{fig:optimization}(b). The ranges identified in the subsequent analysis are summarized in Table~\ref{tab:param}.

The initial stage in the device design is determining the feasible set of fluxonium parameters $\mathcal{X}_q$, as these values fundamentally govern single-qubit coherence and gate performance. We adopt a baseline charging energy of $E_{C_j}/h \approx 1$~GHz, consistent with standard literature \cite{Ding.23, Nguyen.19, Xiong.22}. We then adopt typical literature values for the Josephson and inductive energies, setting $E_{J_{j}}/h \in [2, 9]$~GHz and $E_{L_{j}}/h \in [0.5, 1]$~GHz \cite{Long.19, Gaurav.25, Ding.23, Lin.18, Earnest.18, Manucharyan.09, Ficheux.21, Lin.25.2}.

\begin{table}[!b]
\begin{tabular}{p{1.8cm}p{2.0cm}p{2.2cm}p{1.9cm}}
\hline \hline
\textbf{Circuit} \newline \textbf{element} & \textbf{Parameter} & \textbf{Feasible} \newline \textbf{range} $\mathcal{X}$ & \textbf{Optimized} \newline \textbf{value} $\mathbf{x}^*$ \\
\midrule
Fluxonium & $E_{C_j}/h$ & 1.10 & 1.10 \\
& $E_{J_j}/h$ & [2.00, 9.00] & 4.65 \\
& $E_{L_j}/h$ & [0.50, 1.00] & 0.65 \\
\midrule
DTC  & $E_{C_{c,m}}/h$ & [0.15, 0.35] & 0.30 \\
(bare \newline parameters)  & $E_{J_{c,m}}/h$ & $E_{J_{c,m}}/E_{C_{c,m}}$ \newline $>30$ & 11.30 \\
& $E_{J_{c,12}}/h$ & [0, $E_{J_{c,m}}$] & 1.30 \\
& $J_{c,12}/h$ & [0, 0.20] & 0.14 \\
\midrule
DTC 
& $\omega_{\mathrm{on}_c}/2\pi$ & $\omega_{1,2_j}$ & 4.23 \\
(derived
\newline parameters) & $\omega_{\mathrm{off}_c}/2\pi$ & $\omega_{0,1_c}=\omega_{0,2_c}$ \newline at $\phi_{\mathrm{off}_c}$ & 5.79 \\
\midrule
Drive & $\gamma_{j,c}$ & 0.35 & 0.35 \\
\midrule
Coupling & $J_{q,c}/h$ & [0, 0.70] & 0.50 \\
\midrule
Readout & $\omega_{r_j}/2\pi$ & [7.20, 8.20] & 7.50 \\
Resonator & $g_{r_j}/h$ & [0.03, 0.20] & 0.13 \\
& $Q_{j}$ & [2000, 5000] & 3000 \\
\midrule
Reset & $\omega'_{r_j}/2\pi$ & [2.00, 4.00] & 3.45 \\
Resonator & $g'_{r_j}/h$ & $<0.20$ & 0.05 \\
& $Q'_j$ & [100, 3000] & 400 \\[1ex]
\hline \hline
\end{tabular}
\caption{\label{tab:param} Summary of feasible parameter ranges and optimized values for the F-DTC-F architecture. All parameters are expressed in units of GHz, with the exception of the dimensionless resonator quality factors $Q_j$ and $Q'_j$. Note that the DTC operational frequencies are derived quantities: $\omega_{\mathrm{on}_c}$ is strictly determined by the optimized fluxonium plasmon frequency, whereas $\omega_{\mathrm{off}_c}$ is derived from the coupler circuit parameters at the turn-off flux $\phi_{\mathrm{off}_c}$.}
\end{table}

Next, we determine the parameter ranges for the DTC, $\mathcal{X}_c$. Leveraging transmon anharmonicity and sensitivity to charge noise, the individual charging energies are chosen from common literature values, yielding $E_{C_{c,m}}/h \in [0.15, 0.35]$~GHz \cite{Arute.19, Koch.07, Eric.22}. To mitigate sensitivity to charge noise, the individual Josephson-junction energies of the transmons in the DTC are constrained such that $E_{J_{c,m}}/E_{C_{c,m}} > 30$ \cite{Koch.07}. The inductive coupling between the two transmons in the DTC is set to be smaller than their individual Josephson-junction energies. This maintains a proper DTC energy spectrum, which comprises a flux-tunable differential mode and a nearly constant common mode. Optimizing $\mathbf{x}_c$ requires satisfying the following condition: at the coupler turn-on flux, the coupler frequency is resonant with the fluxonium plasma frequency at the flux sweet spot ($\phi_{\mathrm{ext}_j}=\pi$), such that $\omega_{\mathrm{on}_c} = \omega_{1,2_j}$. This resonance mediates a strong effective fluxonium-fluxonium coupling, enabling a fast, high-fidelity two-qubit MAP gate (see Appendix~\ref{appen:MAP}). To suppress sensitivity to flux noise, we design this interaction to occur near the coupler's flux sweet spot ($\phi_{\mathrm{ext}_c}=\pi$), where the transition frequency is first-order insensitive to flux fluctuations. We designate this operating point as the ``on" state, $\phi_{\mathrm{on}_c}$, with a corresponding coupler frequency $\omega_{\mathrm{on}_c}$. Note that while one could alternatively set $\phi_{\mathrm{on}_c}=0$, that regime is beyond the scope of this work. At $\phi_{\mathrm{ext}_c}=\phi_{\mathrm{off}_c}$, the coupler ``off" frequency $\omega_{\mathrm{off}_c}$, where the effective inter-qubit coupling vanishes in the limit $J_{c,12}=0$, is higher than $\omega_{\mathrm{on}_c}$ and coincides with a degeneracy between two DTC normal modes. To provide a more physically intuitive framework for the parameter optimization workflow depicted in Fig.~\ref{fig:optimization}(b), rather than directly tuning the bare circuit elements ($E_{C_{c,m}}$, $E_{J_{c,m}}$, $E_{J_{c,12}}$, and $J_{c,12}$), we optimize the derived DTC operational frequencies ($\omega_{\mathrm{on}_c}$ and $\omega_{\mathrm{off}_c}$) to optimize the multi-qubit characteristics. Nevertheless, we note that this frequency-based approach is mathematically and physically equivalent to optimizing the underlying bare circuit parameters of the coupler. For completeness, both of them are included in Table~\ref{tab:param}.

\begin{figure}[t]
\centering
\includegraphics[width=\linewidth]{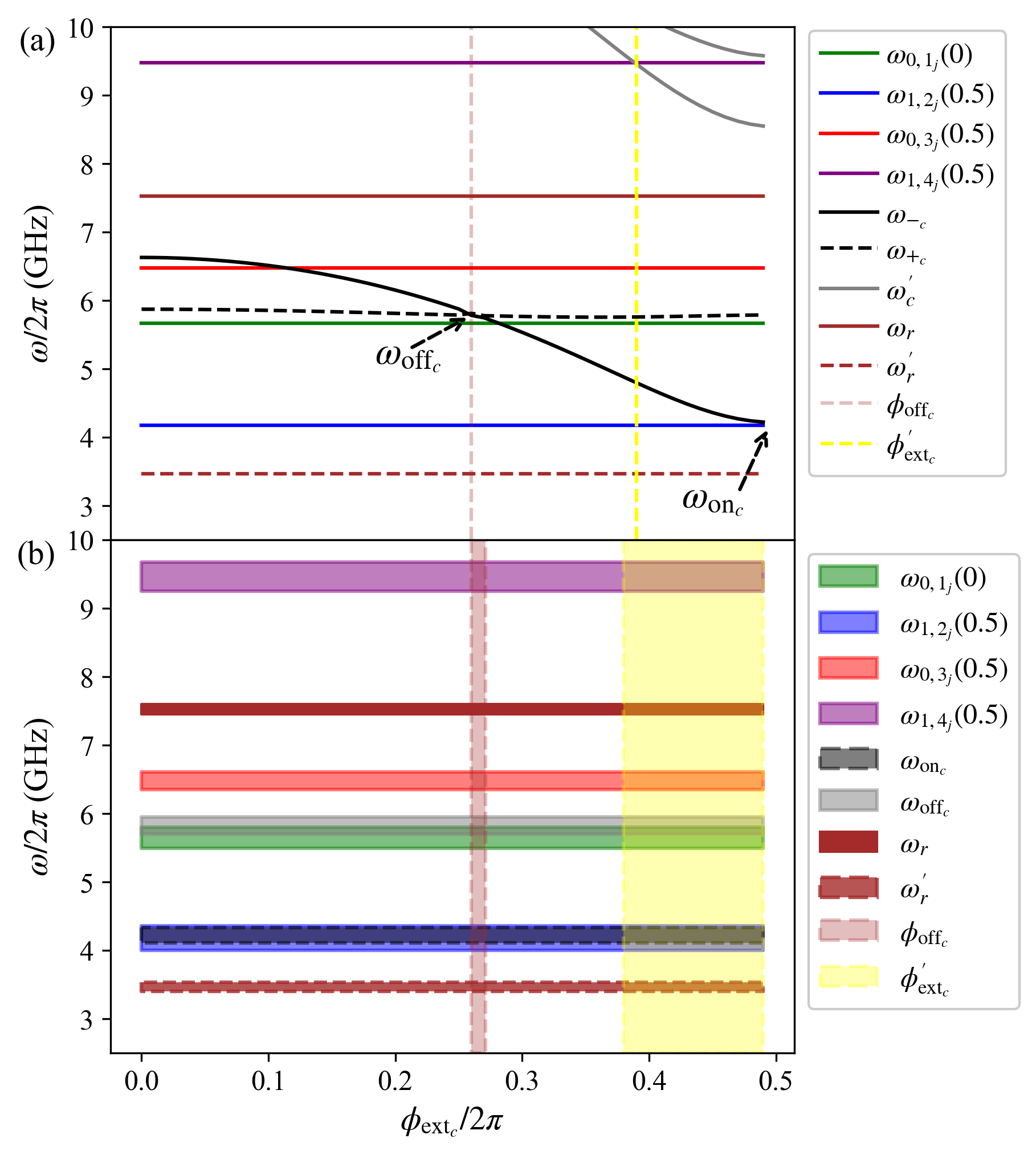}
\caption{(a) Frequency allocation scheme: bare frequencies of the fluxonium qubit, biased at half and integer qubit flux $\phi_{\mathrm{ext}_j}$, and DTC transitions as a function of the coupler flux bias $\phi_{\mathrm{ext}_c}$. (b) Robustness analysis: statistical distribution of bare frequencies under stochastic parameter fluctuations (100 randomized samples). Sensitivity is modeled using a 5\% perturbation for fluxonium $E_J$, $E_L$, and DTC $E_J/h < 5$~GHz, and a 2\% perturbation for all $E_C$ and DTC $E_J/h > 5$~GHz. The notation $\omega_{k,l_\alpha}(\phi_{\mathrm{ext},\alpha})$ refers to the transition frequency between states $|k\rangle$ and $|l\rangle$ at flux bias $\phi_{\mathrm{ext}_\alpha}$ for component $\alpha$ (where $\alpha=j$ for the fluxonium and $\alpha=c$ for the DTC). $\omega'_c$ denotes the higher energy manifold of the DTC. $\omega_r$ and $\omega'_{r_j}$ indicate the readout and reset resonator frequencies, respectively. Vertical colored regions indicate key operating points: $\phi_{\mathrm{off}_c}$ is the bias where effective coupling vanishes, and $\phi'_{\mathrm{ext}_c}$ denotes the bias where the resonance between the coupler's higher-excited manifold and the fluxonium $\omega_{1,4_j}$ transition occurs. Results are computed based on optimized parameters listed in Table~\ref{tab:param}.}
\label{fig:bare_robust}
\end{figure}

\begin{figure*}[!ht]
    \centering
    \includegraphics[width=0.8\linewidth]{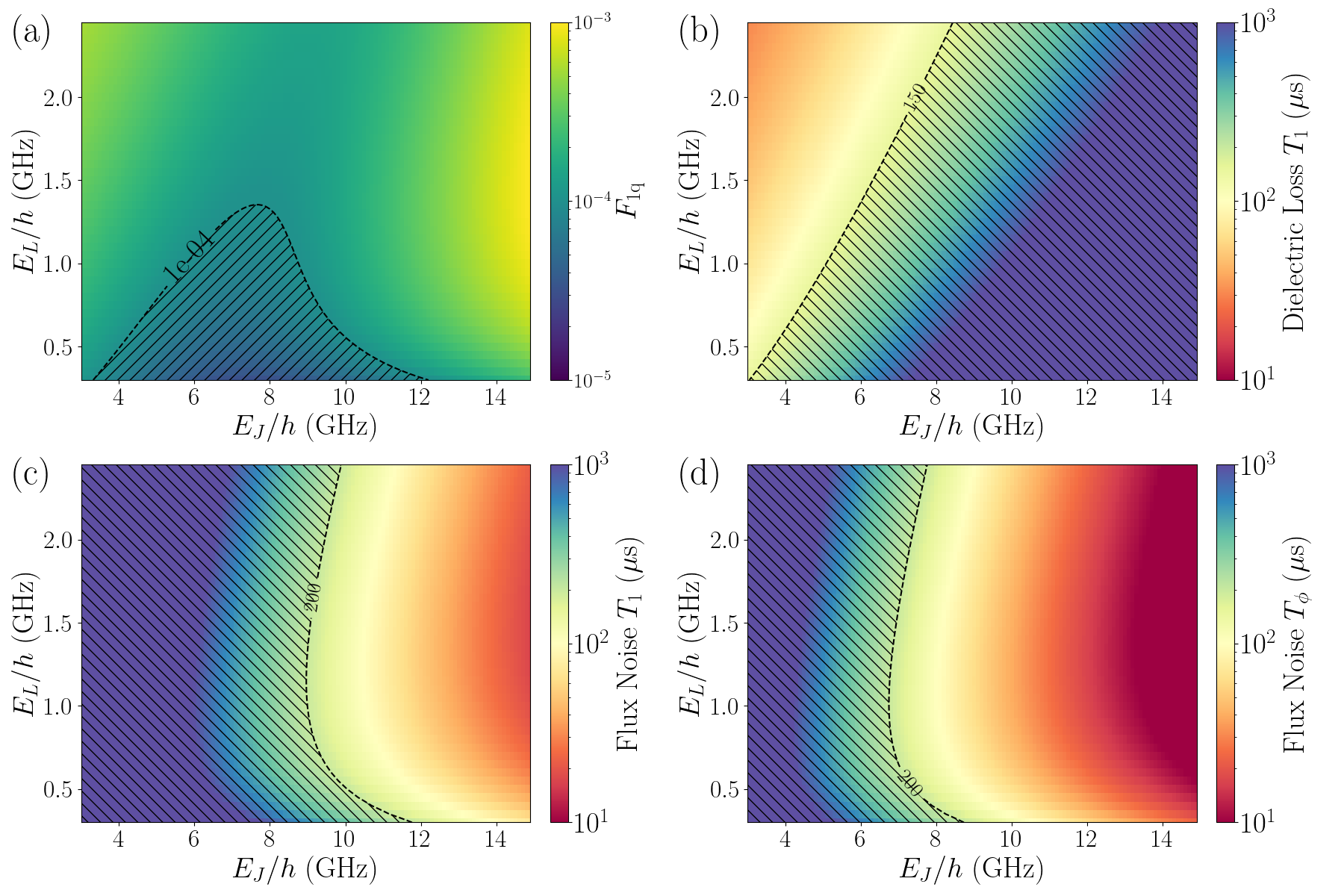}
    \caption{Single-qubit coherence and gate fidelity landscape as a function of $E_L$ and $E_J$, with fixed $E_C/2\pi = 1.1$~GHz. (a) Average single-qubit gate fidelity estimated with gate time $t_g=50$ ns. The region where fidelity exceeds $99.99\%$ is demarcated by black contours. (b–c) Relaxation times $T_1$ limited by (b) dielectric loss and (c) flux noise. (d) Dephasing time $T_\phi$ limited by flux noise. In panels (b–d), regions with coherence times above specific thresholds are highlighted.} 
    \label{fig:sq_t1t2}
\end{figure*}

For the resonator parameters, we delegate qubit readout and reset operations to dedicated resonators. The feasible ranges for these resonator frequencies, denoted as $\mathcal{X}_r$ and $\mathcal{X}'_r$, respectively, depend on the final optimized fluxonium parameters $\mathbf{x}_q^*$. To demonstrate the execution of the frequency-allocation scheme defined above, we use a representative set of fluxonium parameters ($E_{C_j}/h = 1.1$~GHz, $E_{J_j}/h = 4.65$~GHz, $E_{L_j}/h = 0.65$~GHz), which yields the energy spectrum shown in Fig.~\ref{fig:bare_robust}(a). This allows us to identify optimal ranges for the readout resonator frequency, $\omega_{r_j}/2\pi \in [7.2, 8.2]$~GHz, and the reset resonator frequency, $\omega'_{r_j}/2\pi \in [2, 4]$~GHz, which satisfy the spectral hierarchy established earlier. Note that for the reset resonator, the bounding condition $\omega'_{r_j} < \omega_{0,1_j}$ must hold specifically at zero qubit flux ($\phi_{\mathrm{ext}_j}=0$).
Additionally, we aim to bias the DTC at the turn-off flux $\phi_{\mathrm{off}_c}$ during qubit readout to minimize calibration overhead. However, at a specific collision flux $\phi'_{\mathrm{ext}_c}$ (indicated by the yellow vertical line in Fig.~\ref{fig:bare_robust}(a)), higher-energy transitions of the DTC, specifically $\omega_{{0,3}_c}$, $\omega_{{0,4}_c}$, and $\omega_{{0,5}_c}$, may cross the upper fluxonium transition $\omega_{{1,4}_j}$, potentially destabilizing the readout.  Therefore, further tuning of $\mathbf{x}_c$ is required to ensure $\phi_{\mathrm{off}_c} \neq \phi'_{\mathrm{ext}_c}$. For the readout resonator, the acceptable parameter ranges for the quality factor $Q_j$ and the qubit-resonator coupling strength $g_{r_j}$ are defined to ensure the resonator remains in the dispersive regime without severely sacrificing readout speed. This guarantees an optimal signal-to-noise ratio (SNR) via a sufficient photon loss rate. Similarly, the valid parameter ranges for the reset resonator's quality factor $Q'_j$ and coupling strength $g'_{r_j}$ are established to facilitate sufficiently fast qubit reset without degrading qubit coherence.

Regarding the drive parameters, active optimization typically involves advanced control strategies, such as selectively canceling the effective drive on unwanted transitions within the MAP gate scheme~\cite{Ding.23} or actively mitigating microwave crosstalk~\cite{Nuerbolati.22}. However, because such dynamic drive optimization falls outside the primary architectural scope of this work, we do not treat $\gamma_{j,c}$ as a free parameter. Instead, we adopt a fixed, realistic parasitic drive ratio of $\gamma_{j,c}=0.35$, which is extracted directly from electrostatic simulations.

Finally, we constrain the fluxonium-DTC coupling strength $J_{j,c}$ to the maximum value permitted by the specific device geometry shown in Fig.~\ref{fig:grid}(a). Based on electrostatic simulations of this architecture, we estimate this upper bound to be $J_{j,c}/h \sim 700$~MHz.

Notably, during parameter optimization, we perform a robustness check on the candidate parameter set $\mathbf{x}$ to ensure the absence of frequency collisions prior to evaluating the metrics defined in Table~\ref{tab:requirements}. This verification is essential for practical device design, as fabrication-induced variations are unavoidable. Specifically, the following conditions are verified:

\begin{itemize}
    \item $\omega'_{r_j} < \omega_{0,1_j}$ at $\phi_{\mathrm{ext}_j}=0$;
    \item No frequency collisions occur within the set $\{\omega_{0,3_j}, \omega_{1,4_j}, \omega_{1,2_j}, \omega_{\mathrm{off}_c}, \omega_{r_j}, \omega'_{r_j}\}$;
    \item The operating range for the ``off'' bias $\phi_{\mathrm{off}_c}$ does not overlap with any collision flux $\phi'_{\mathrm{ext}_c}$ where $\omega_{0,m_c} = \omega_{1,4_j}$ for $m \in \{3, 4, 5\}$. 
\end{itemize}

To illustrate this frequency-allocation scheme and the associated robustness checks, we use the final optimized parameter set $\mathbf{x}^*$ as a representative example. Figure~\ref{fig:bare_robust}(a) details the relevant frequencies, while Fig.~\ref{fig:bare_robust}(b) shows the frequency dispersion under the parameter variations specified in the caption. It is evident that the optimized parameters listed in Table~\ref{tab:param} satisfy the criteria outlined above across the considered range of variations. The successful completion of this check corresponds to Steps 0--1 in the optimization workflow depicted in Fig.~\ref{fig:optimization}(b). Although not explicitly illustrated in the flowchart, we emphasize that whenever a parameter is updated, such as during the rejection iterations in Steps 3, 6, 8, 15 in Fig.~\ref{fig:optimization}(b), the robustness checks of Step 2 are strictly re-evaluated to guarantee the absence of frequency collisions. We also note that the proposed frequency-allocation scheme is not unique. Alternative architectural choices, such as utilizing a single shared resonator for both qubit readout and reset, would result in a different parameter landscape and necessitate reordering the frequency-allocation workflow.

Beyond avoiding spectral collisions, a critical advantage of this preliminary frequency allocation is that it systematically decouples the physical design constraints. By assigning distinct spectral bands to all essential quantum operations, the architecture structurally simplifies the device design. Consequently, what would otherwise constitute a highly complex, multi-parameter, multi-target simultaneous optimization problem is reduced to a tractable sequential workflow. This spectral partition ensures that the parameter subsets governing specific operations, such as qubit reset, single and parallel MAP gate executions, and readout, can be optimized largely independently, minimizing the risk of competing constraints.
In the following subsections, we detail the optimization workflow through which the optimal solution $\mathbf{x}^*$ is obtained from the feasible parameter space $\mathcal{X}$.

\subsection{Qubit coherence and single qubit gate}

As a first step, we examine the average single-qubit gate fidelity (see Appendix~\ref{appen:SQF} for details) as a function of $E_J$ and $E_L$, in the presence of flux noise and dielectric loss, for $E_{C_j}/h = 1.1$~GHz, as shown in Fig.~\ref{fig:sq_t1t2}(a). These calculations assume realistic noise parameters: a $1/f$ flux-noise amplitude of $A_\Phi=3.7\,\mu\Phi_0$ and a dielectric loss tangent of $\tan \delta_c = 1.84\times 10^{-6}$ (evaluated at $\omega/2\pi=300$~MHz using the scaling $\tan \delta_c(\omega)=4\times10^{-6}\times [\omega/(2\pi \times 6~\text{GHz})]^{0.26}$ \cite{Ateshian.07}). To ensure high-fidelity operation ($F_{1q} > 99.99\%$), we restrict the remaining parameters to $E_{J_j}/h \in [4.5, 9.5]$~GHz and $E_{L_j}/h \in [0.3, 0.65]$~GHz. The relaxation times limited by dielectric loss and flux noise are primarily governed by the qubit transition frequency $\omega_{{0,1}_j}$~\cite{Wang.25}, which is jointly determined by $E_J$ and $E_L$, as shown in Fig.~\ref{fig:sq_t1t2}(b)-(d).
Provided the fluxonium parameters remain within these bounds, different combinations of the full parameter set $\mathbf{x}$ can lead to equally optimal single-qubit performance metrics, corresponding to different solutions along the Pareto front shown schematically in Fig.~\ref{fig:optimization}(a). Consequently, we select a priori a set of fluxonium parameters $\mathbf{x}_q^* = \{E^*_{C_j}, E^*_{J_j}, E^*_{L_j}\}$ based on the coherence constraints shown in Fig.~\ref{fig:sq_t1t2}(a) and detailed in Table~\ref{tab:param}. This analysis corresponds to Steps 2–3 of the optimization workflow in Fig.~\ref{fig:optimization}(b).

\subsection{Two-qubit gate characterization} \label{subsec:2Q_g}
\subsubsection{Leakage minimization}\label{subsubsec:lea_min}
In addition to decoherence, the primary limitation of MAP gate operation arises from driven leakage transitions, particularly when short gate times are used to mitigate decoherence. For a given parameter set $\mathbf{x}_q \cup \mathbf{x}_c \cup \mathbf{x}_{qc}$, $\eta_{j\beta,k\alpha}$  denotes the probability of a leakage transition $|k\rangle \to |\alpha\rangle$ induced while driving the target MAP transition $|j\rangle \to |\beta\rangle$ (see Appendix~\ref{subsec:leakage} for details). To quantify MAP gate's robustness against leakage, we define the metric
\begin{equation}
    \widetilde{\eta} = \min_{ j,\beta}\left(\max_{ k\neq j, \alpha}\eta_{j\beta,k\alpha}\right).
\end{equation}
This quantity represents the worst-case leakage probability associated with the best available MAP transition; essentially, it identifies the optimal parameter set by minimizing the maximum leakage. Ideally, we seek to minimize this leakage probability for a fixed gate time $t_g$.

The most straightforward strategy is to engineer $\mathbf{x}_{c}$ such that the DTC ``turn-on'' frequency is resonant with the fluxonium plasmon frequencies, i.e., $\omega_{\mathrm{on}_c} = \omega_{1,2_j} = \omega_{1,2_k}$, as implemented in Table~\ref{tab:param} (cf.~Fig.~\ref{fig:bare_robust}(a)) \cite{ACEngineer}. Under this on-resonance condition, the DTC mode and fluxonium plasmon transitions are strongly hybridized, resulting in the strong state-dependent noncomputational transitions necessary for MAP gate operation. However, due to inevitable fabrication variances, perfect resonance is impossible to guarantee. Nevertheless, the impact of frequency detuning caused by parameter variations can be mitigated by increasing the fluxonium-DTC coupling strength $J_{j,c}$.

\begin{figure}[t]
    \centering
    \includegraphics[width=\linewidth]{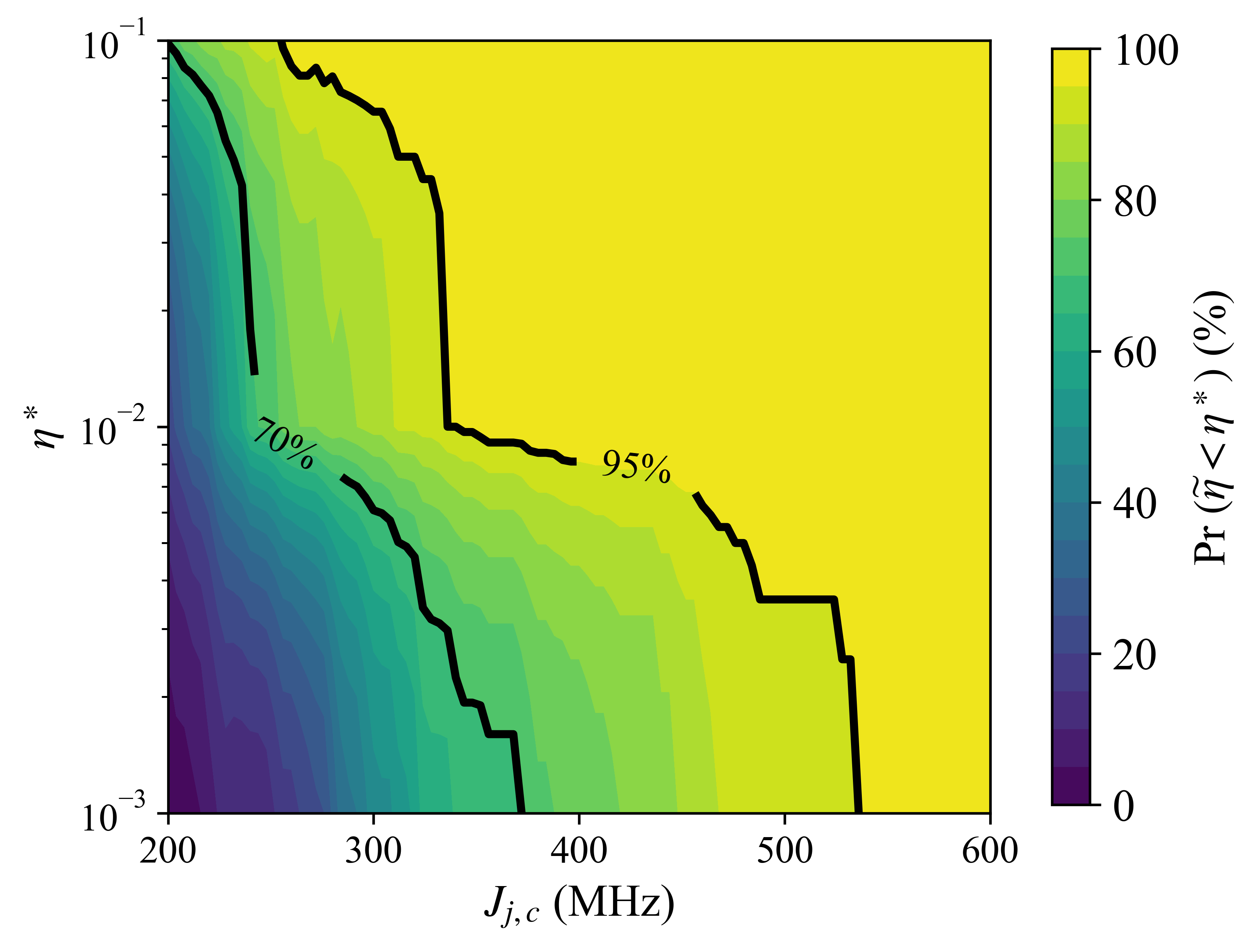}
    \caption{Robustness of MAP gate leakage estimates ($\widetilde{\eta}$) against fabrication uncertainty. Simulations use the parameters from Table~\ref{tab:param} with a gate time of $t_g=50$~ns. Parameter variations are modeled as follows: 5\% for fluxonium $E_J, E_L$ and DTC junctions with $E_J/h < 5$~GHz; and 2\% for all $E_C$ and DTC junctions with $E_J/h > 5$~GHz. The color scale indicates the percentage of samples (out of 100 randomized instances) yielding a leakage $\widetilde{\eta}$ below the threshold $\eta^\ast$.}
    \label{fig:leakage_robust}
\end{figure}

As derived in Appendix~\ref{appen:MAP_prot}, a larger coupling $J_{j,c}$ increases the detuning between the target MAP transition and nearby leakage transitions. Furthermore, the analysis in Appendix~\ref{subsec:leakage} demonstrates that the leakage probability $\eta_{j\beta,k\alpha}$ scales inversely with these detunings. This theoretical prediction is confirmed by the robustness analysis in Fig.~\ref{fig:leakage_robust}, which presents the statistics of $\widetilde{\eta}$ under parameter variations. Notably, Fig.~\ref{fig:leakage_robust} indicates that to ensure $\widetilde{\eta} < 10^{-3}$ across all sampled variations of $\mathbf{x}$, the coupling strength $J_{j,c}/h$ must be at least 500~MHz. Conversely, if a weaker coupling is chosen (e.g., $J_{j,c}/h \approx 300$~MHz), the majority of sample instances yield leakage probabilities an order of magnitude higher, reaching levels of $\sim 10^{-2}$. This analysis concludes Steps 4 and 6 of the optimization workflow outlined in Fig.~\ref{fig:optimization}(b). It is important to note that coupling strengths of $J_{j,c}/h \sim$ 500~MHz are attainable within realistic device designs. Achieving this target, however, requires deliberate suppression of capacitance loading, which is often regarded as a challenge for fluxonium qubits due to their limited capacitance budget, particularly in high-connectivity architectures. Nevertheless, our recent study shows that this constraint does not pose a fundamental obstacle to extending fluxonium systems to two-dimensional lattices~\cite{QG2026}.

\subsubsection{Two-qubit gate fidelities}
\begin{figure}[t]
    \centering
    \includegraphics[width=0.95\linewidth]{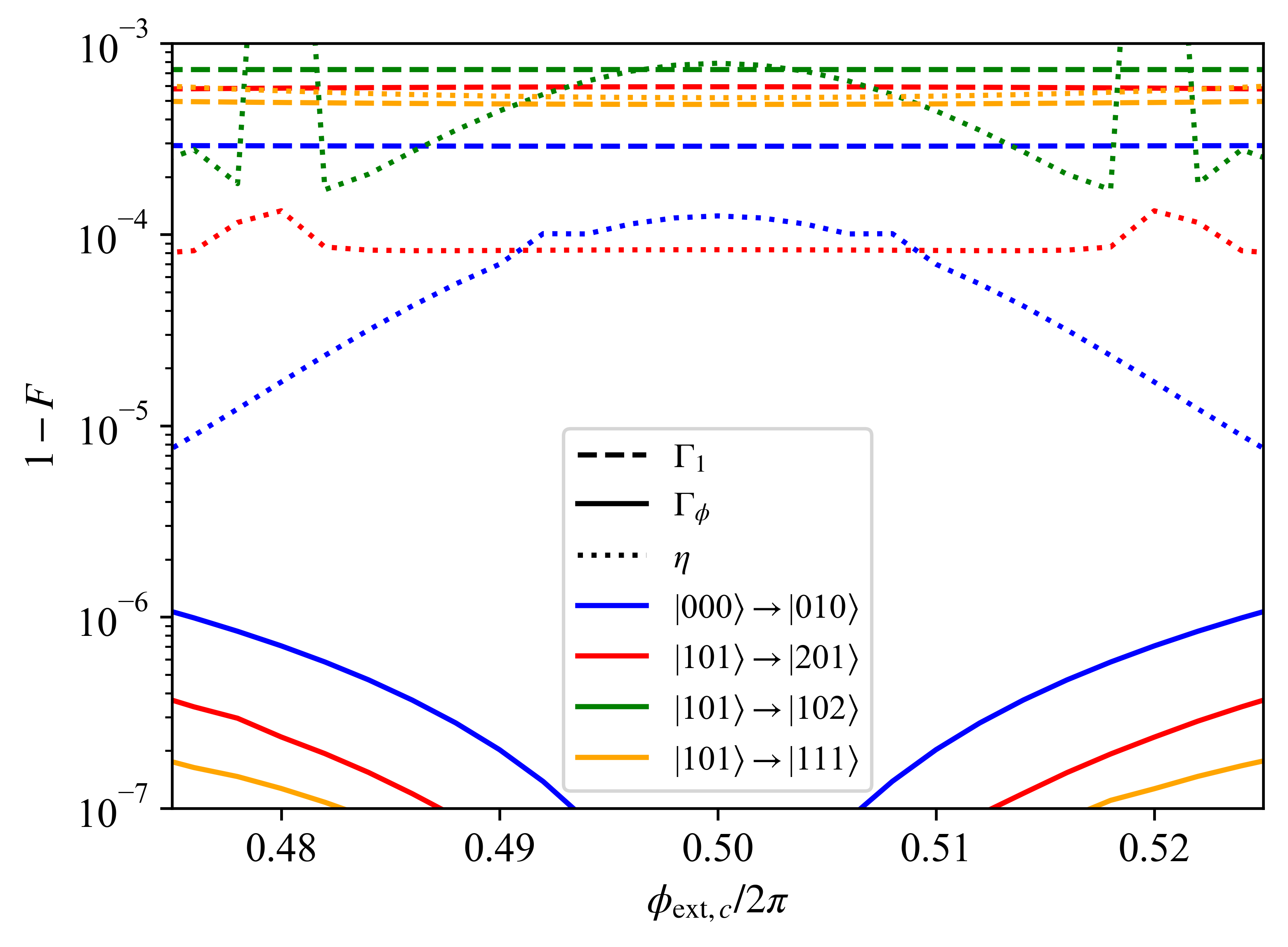}
    \caption{Estimated MAP gate infidelity as a function of the DTC flux bias $\phi_{\mathrm{ext}_c}$, showing individual contributions from higher-level dephasing $\Gamma_\phi$ (solid lines), higher-level relaxation $\Gamma_1$ (dashed lines), and leakage out of the computational subspace $\eta$ (dotted lines). Calculations are performed using the parameters from Table~\ref{tab:param} with a gate time $t_g = 50$~ns. Blue, red, green, and orange curves correspond to MAP gates implemented by driving the $|000\rangle \rightarrow |010\rangle$, $|101\rangle \rightarrow |201\rangle$, $|101\rangle \rightarrow |102\rangle$, and $|101\rangle \rightarrow |111\rangle$ transitions, respectively. The simulation assumes realistic noise parameters: a $1/f$ flux-noise amplitude of $A_\Phi=3.7\,\mu\Phi_0$ and a dielectric loss tangent of $\tan \delta_c = 3.8 \times 10^{-6}$, evaluated at $\omega/2\pi=4$~GHz using the scaling $\tan \delta_c(\omega)=4\times10^{-6}\times [\omega/(2\pi \times 6~\text{GHz})]^{0.26}$ \cite{Ateshian.07}.}
    \label{fig:map_fid}
\end{figure}
During MAP gate operation, the device is subject to various noise channels, primarily flux noise and dielectric loss. The theoretical framework for estimating gate fidelities is detailed in Appendix~\ref{sec:fid_multi}. Figure~\ref{fig:map_fid} presents the estimated MAP gate fidelities with various selected energy transitions calculated using the parameters from Table~\ref{tab:param}.

The results indicate that the gate infidelity is dominated by energy relaxation. An analysis of the specific decay rates reveals that the primary contributor to infidelity for the $|000\rangle \to |010\rangle$ transition is dielectric loss from both fluxonium qubits and the DTC. Conversely, for the remaining transitions, dielectric loss in the fluxonium qubits is the dominant mechanism. This distinction arises because the excited state of the former corresponds to an excitation of the DTC mode, whereas the latter involve excitations of the fluxonium plasmon modes.

Furthermore, the contribution from pure dephasing increases as the operating point moves away from the DTC sweet spot at $\phi_{\mathrm{ext}_c}=\pi$. This trend is attributed to the increased sensitivity to flux noise caused by the steeper dispersion of the DTC's flux-tunable mode, as illustrated in Fig.~\ref{fig:MAP_t}. Notably, the dephasing-induced infidelities presented here are estimated using an undriven (Ramsey-like) model (see Appendix~\ref{appen:dephaseHigherLvl}), which does not fully capture the dynamics of the driven transitions during the MAP gate operation. Because driven systems generally exhibit enhanced resilience to low-frequency dephasing noise~\cite{Guo.18}, and an exact analytical formulation for this driven multi-level evolution is highly complex, we expect the true dephasing-induced gate infidelities to be lower than the values plotted in Fig.~\ref{fig:map_fid}. Consequently, our undriven model serves as a conservative upper bound for the expected error.  Finally, within the range of $\phi_{\mathrm{ext}_c}$ shown in Fig.~\ref{fig:map_fid}, leakage-induced infidelity remains negligible, consistent with the conclusions of the preceding subsection.

In summary, the parameters listed in Table~\ref{tab:param} yield two-qubit gate fidelities exceeding $99.9\%$. This result confirms the successful completion of Steps 5-6 in the optimization workflow outlined in Fig.~\ref{fig:optimization}(b).

\subsection{Parallel gates: spectator-induced crosstalk}\label{subsec:xtalk}

To ensure the high-fidelity execution of parallel gate operations, target gates must remain strictly robust against crosstalk induced by spectator qubits. The double-transmon coupler (DTC) emerges as a highly promising architecture to achieve this requisite isolation; its  ``turn-off'' mechanism (detailed in Appendix~\ref{sec:turn_off}) theoretically yields an infinite on-off ratio in the ideal limit of vanishing inter-transmon capacitive coupling.

\begin{figure}[t]
    \includegraphics[width=\linewidth]{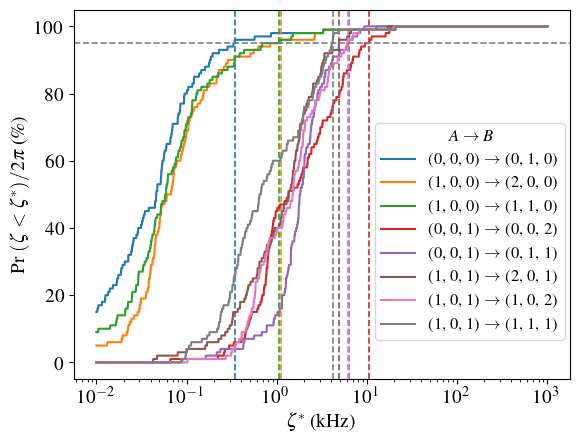}
    \caption{Robustness analysis of spectator-induced crosstalk $\zeta$ under fabrication uncertainty. Curves of different colors correspond to distinct selected energy transitions. Simulations use the parameters from Table~\ref{tab:param} with random variations modeled as follows: 5\% for fluxonium $E_J, E_L$ and DTC junctions with $E_J/h < 5$~GHz; and 2\% for all $E_C$ and DTC junctions with $E_J/h > 5$~GHz. Statistics are compiled from 100 randomized samples. The colored vertical lines indicate the 95th percentile (the threshold below which 95\% of the samples fall).}
    \label{fig:xtalk_robust}
\end{figure}

\begin{figure*}[t]
    \centering
    \includegraphics[width=0.34\linewidth]{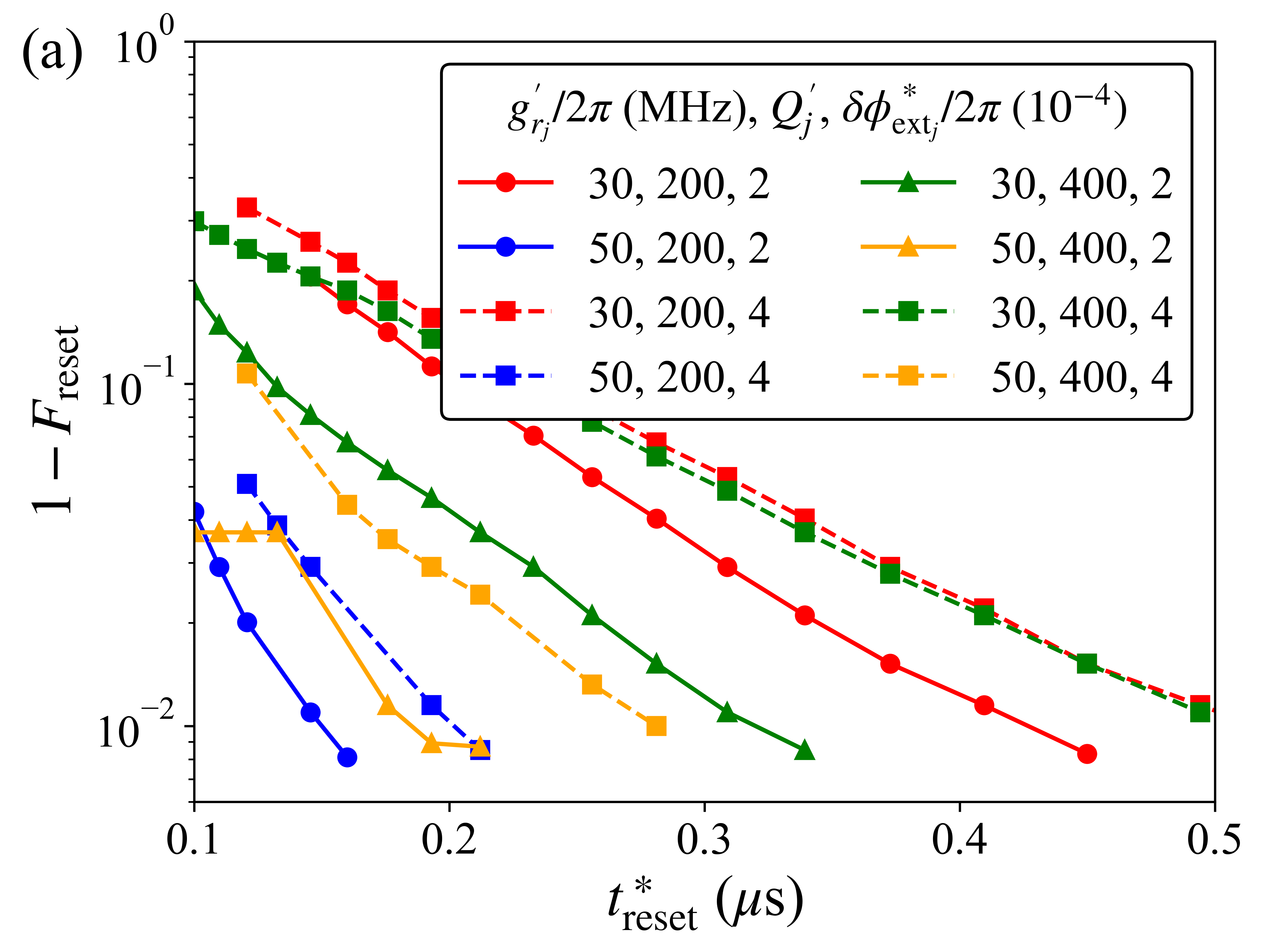}
    \includegraphics[width=0.64\linewidth]{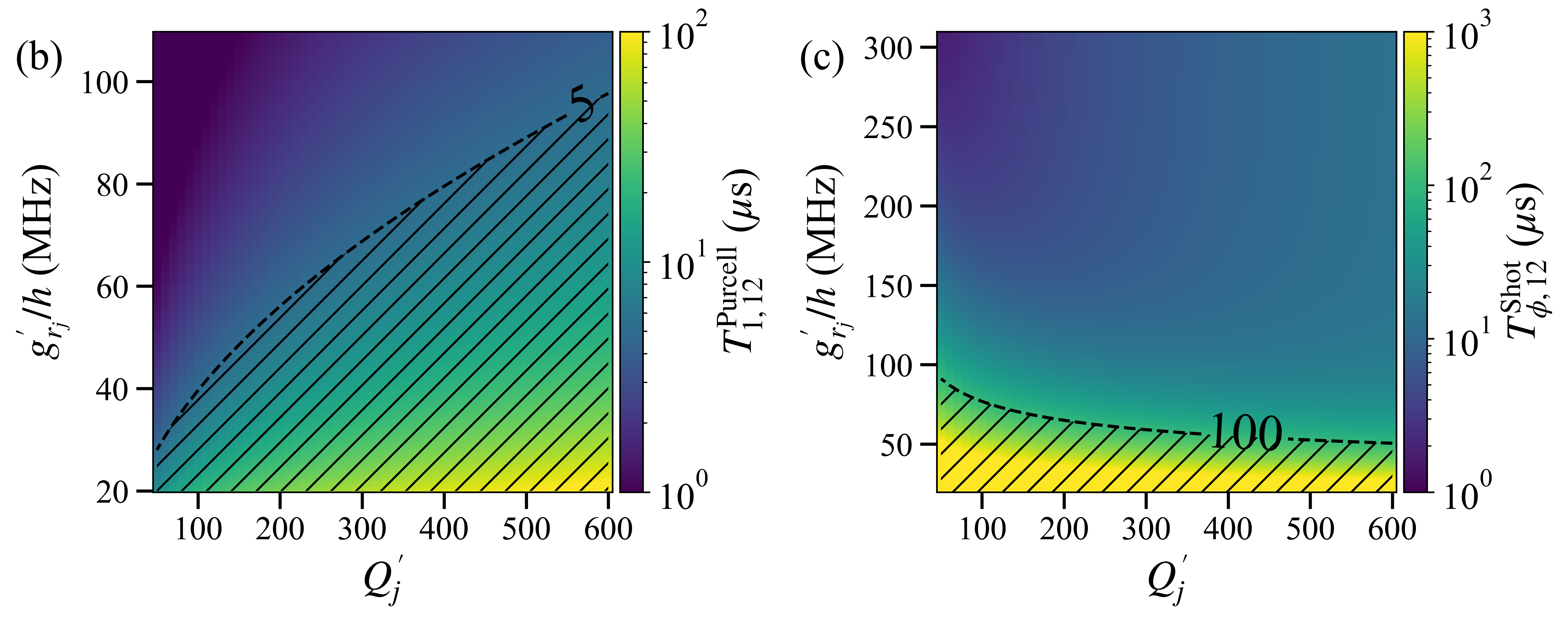}
    \caption{(a) Reset infidelity $1-F_\mathrm{reset}$ as a function of reset duration $t_\mathrm{reset}$. The marker symbols and colors indicate the corresponding fluxonium-resonator coupling $g'_{r_j}$, resonator quality factor $Q'_j$, and $\delta\phi^\ast_{\mathrm{ext},j}$ \cite{FluxSens}. (b) Purcell relaxation time and (c) photon shot-noise dephasing time for the fluxonium plasmon transition $\omega_{1,2_j}$, plotted as functions of $Q'_j$ and $g'_{r_j}$.In panels (b–c), regions with coherence times above specific thresholds are highlighted.}
    \label{fig:reset}
\end{figure*}

Realizing this ideal isolation, however, presents a significant parameter-allocation challenge when accounting for realistic fabrication uncertainties.
Guided by the approximate form of the effective fluxonium-fluxonium coupling (detailed in Appendix~\ref{appen:parameter}), a robust design strategy must ensure a large detuning between the DTC mode frequency at the turn-off flux, $\omega_{\mathrm{off}_c}$, and the fluxonium plasmon frequencies $\omega_{1,2_{j}}, \omega_{1,2_{k}}$. Physically, the effective inter-qubit coupling arises from two distinct pathways: inductive coupling and parasitic capacitive coupling. While the inductive contribution can be dynamically tuned off, the capacitive coupling remains inherently fixed by the static system parameters. Consequently, residual interactions arising from this parasitic capacitive coupling cannot be eliminated through tuning the inductive coupling and must instead be suppressed by increasing the detuning between the qubit's plasmon and DTC modes.
Satisfying the condition $g_{c,\mathrm{ind}} \ll |\omega_{1,2_{j,k}} - \omega_{\mathrm{off}_c}|$ is particularly critical given the strong fluxonium-DTC coupling ($J_{q,c}/h \sim 500$~MHz) required to execute the MAP gate (see Sec.~\ref{subsec:2Q_g}). Without sufficient detuning, inevitable parameter variations across the device can lead to unintended resonances between $\omega_{\mathrm{off},c}$ and $\omega_{{1,2}_j},\omega_{{1,2}_k}$, thereby compromising the DTC ``turn-off'' mechanism and degrading the isolation of the architecture.

The primary observable consequence of this degraded isolation is spectator-induced crosstalk, which occurs when the effective inter-qubit coupling is not perfectly nulled. We quantitatively define this crosstalk as the frequency shift of a target MAP transition conditioned on the quantum states of adjacent spectator qubits (see Appendix~\ref{sec:turn_off}).
Physically, this spectral shift renders the targeted two-qubit operation inadvertently dependent on the spectator dynamics, manifesting as an unwanted controlled-controlled-phase operation or conditioned leakage acting on the active qubit pair.
Such residual entanglement fundamentally limits the fidelity of simultaneous gate executions. To evaluate the efficacy of our proposed parameter allocation, Fig.~\ref{fig:xtalk_robust} presents a robustness analysis of spectator-induced crosstalk during a MAP gate, utilizing the optimized system parameters detailed in Table~\ref{tab:param}.

We observe that the designed detuning of $(\omega_{\mathrm{off}_c} - \omega_{1,2_{j}})/2\pi = (\omega_{\mathrm{off}_c} - \omega_{1,2_{k}})/2\pi \sim 1.8$~GHz, as shown in Fig.~\ref{fig:bare_robust}(a), is sufficient to maintain crosstalk below 10~kHz across all sampled parameter variations. This 10~kHz threshold serves as a strict benchmark, selected to guarantee that the MAP gate infidelity induced by spectator crosstalk is bounded to $\lesssim 10^{-5}$, cf.~Appendix~\ref{subsec:xtalk_fid}. Conversely, we demonstrate that a large frequency separation is essential for robust isolation: as illustrated in Fig.~\ref{fig:param_1}, reducing the detuning to $\sim 0.8$~GHz results in an order-of-magnitude increase in crosstalk due to the increased likelihood of frequency collisions. This procedure enables the optimization of key DTC parameters, namely, $\omega_{\mathrm{off}_c}$.
This analysis concludes Steps 7--8 of the optimization workflow outlined in Fig.~\ref{fig:optimization}(b).

Notably, the above discussion focuses on nulling spectator-induced frequency shifts on transitions relevant to MAP gate operations. Equally important is the suppression of such shifts in qubit frequencies within the computational subspace, which is essential for high-fidelity parallel single-qubit gates. However, this particular type of crosstalk is inherently mitigated in the MAP gate scheme due to the large ratio of the charge dipoles in the plasmon to the fluxon transitions, i.e.,~$\langle 1|\hat{n}_j|2\rangle \gg \langle 0|\hat{n}_j|1\rangle$ and large detuning between coupler frequencies in the activated regime and fluxon frequencies, i.e.~$\omega_{\mathrm{on}_c}\gg \omega_{{0,1}_j}$. Consequently, the $ZZ$ and $XX$ interactions within the qubit computational subspace remain negligible, obviating the need for active mitigation protocols.

\subsection{Qubit reset}\label{subsec:reset}

Due to the long coherence times of fluxonium qubits in the computational subspace and their relatively low transition frequencies (hundreds of MHz), active reset is crucial to prevent thermalization. In this work, we implement active reset using dedicated low-frequency resonators ($\sim 3.5$~GHz). The resonator frequency is chosen to be high enough that the reset fidelity $F_\mathrm{reset}$ \cite{ResetFid} is not limited by the thermal photon population, yet low enough to minimize the impact of the Purcell effect and photon shot noise on the MAP gate fidelity.

The coupling strength $g'_{r_j}$ between the qubit and the resonator must balance the robustness of the reset operation against flux variations (characterized by the notation $\delta\phi^\ast_{\mathrm{ext}_j}$ \cite{FluxSens}) with the decoherence induced on the qubit plasmon transitions. A larger coupling strength $g'_{r_j}$ enhances robustness against flux noise but reduces both the Purcell relaxation time $T_{1,12}^\mathrm{Purcell}$ and the photon shot-noise dephasing time $T_{\phi,12}^\mathrm{shot}$ of the plasmon transitions. Similarly, the resonator's coupling quality factor $Q'_j$ must balance the reset duration $t_\mathrm{reset}$ against induced decoherence. A lower quality factor $Q'_j$ enables faster reset but increases the decoherence rate on the plasmon transitions. Consequently, determining the optimal reset parameters $\mathbf{x}_r = \{\omega'_{r_j}, g'_{r_j}, Q'_j\}$ is a multi-objective optimization task targeting five benchmark metrics: $\{t_\mathrm{reset}, \delta\phi^\ast_{\mathrm{ext},j}, F_\mathrm{reset}, T_{1,12}^\mathrm{Purcell}, T_{\phi,12}^\mathrm{shot}\}$.

We define the reset fidelity $F_\mathrm{reset}$ as the probability of the fluxonium qubit returning to its ground state $|0\rangle$ after applying a flux bias to bring the qubit transition frequency $\omega_{01_j}$ into resonance with the reset resonator frequency $\omega_r'$. Figure~\ref{fig:reset}(a) presents the calculated reset fidelity $F_\mathrm{reset}$, obtained by simulating the dynamics of the fully coupled fluxonium-DTC-resonator system (governed by the Hamiltonian introduced in Sec.~\ref{subsec:system_H}).
Figure~\ref{fig:reset}(a) demonstrates that a reset fidelity of $F_\mathrm{reset} > 99\%$ within a duration of $t_\mathrm{reset} \sim 300$~ns can be achieved using $g'_{r_j}/2\pi = 50$~MHz and $Q'_j = 400$. At this operating point, the reset protocol demonstrates robustness against flux variations up to a value $\delta \phi_{\mathrm{ext}_j}^\ast/2\pi \approx 4 \times 10^{-4}$, ensuring reliable operation under realistic experimental conditions. Furthermore, the reset duration can be accelerated to $t_\mathrm{reset} \sim 200$~ns by decreasing the quality factor to $Q'_j = 200$ while maintaining the identical effective coupling and flux-noise robustness. Furthermore, this configuration maintains sufficiently long coherence time for MAP gate fidelity $>99.9\%$, with estimated limits of $T_{1,12}^\mathrm{Purcell} > 5~\mu$s and $T_{\phi,12}^\mathrm{Shot} > 100~\mu$s, as shown in Fig.~\ref{fig:reset}(b) and (c). This concludes Steps 9, 10, 11 and 15 in the optimization flow in Fig.~\ref{fig:optimization}(b).

\subsection{Qubit readout}\label{subsec:readout}

Qubit readout plays a central role in superconducting quantum processors and requires careful balancing measurement speed, state distinguishability, and the preservation of qubit coherence. Key system variables, such as the dispersive shift $|2\chi|$, the cavity decay rate $\kappa$, and the readout photon number, must be simultaneously engineered to maximize the signal-to-noise ratio (SNR) while mitigating measurement back-action. While the ultimate choice of parameters is strongly dictated by specific experimental architectures and objectives, here we establish a generalized baseline that ensures robust readout performance, deferring highly specialized optimizations for ultra-fast or ultra-high-fidelity integration to application-specific designs.

\begin{figure}[t]
    \centering
    \includegraphics[width=\linewidth]{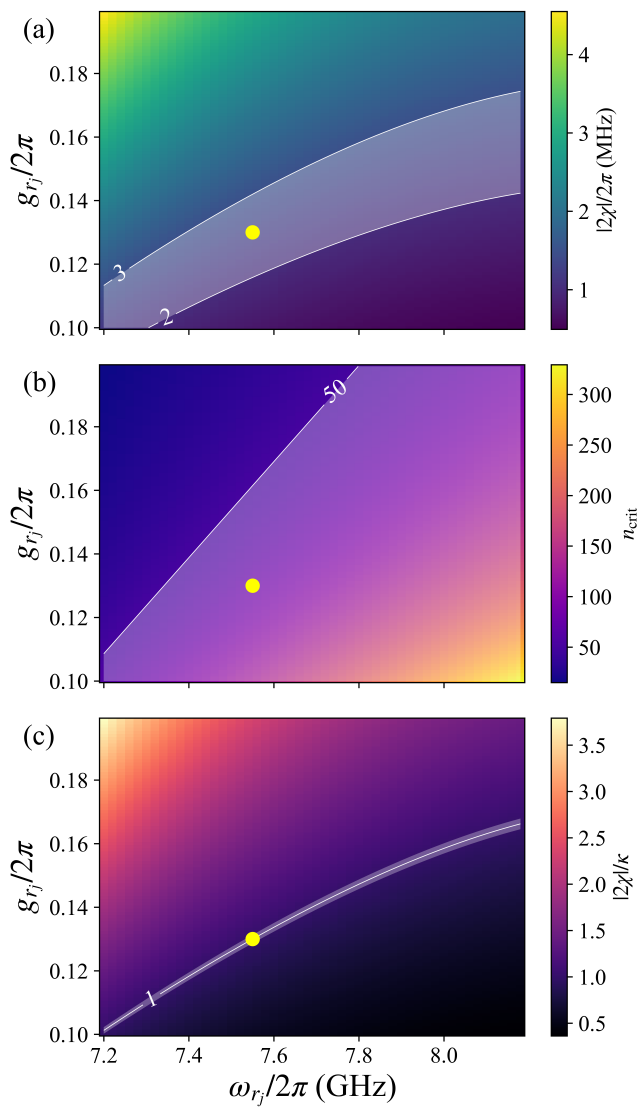}
    \caption{(a) The dispersive shift $|2\chi|/2\pi$, (b) the critical photon number $n_\mathrm{crit}$, and (c) the signal-to-noise ratio (SNR) condition $|2\chi|/\kappa$ as a function of the readout resonator frequency $\omega_r/2\pi$ and the qubit-resonator coupling strength $g_{r_j}$. The resonator decay rate is defined as $\kappa = \omega_{r_j}/Q_j$, with the quality factor fixed at $Q_j=3000$. White contour lines and shaded bands delineate the boundaries of the target requirements (e.g., $|2\chi|/2\pi \sim 2$~MHz, $n_\mathrm{crit} \geq 50$, $|2\chi|/\kappa \approx 1$). The yellow dot marks a specific optimized parameter instance at $(\omega_{r_j}/2\pi, g_{r_j}/2\pi) = (7.55~\mathrm{GHz}, 130~\mathrm{MHz})$ that simultaneously satisfies all three performance metrics.}
    \label{fig:readout}
\end{figure}

Specifically, we target a parameter regime characterized by a dispersive shift of $|2\chi|/2\pi \sim 1$~MHz, a ratio of $|2\chi|/\kappa \approx 2$, and a critical photon number $n_\mathrm{crit} \gtrsim 50$. The first condition facilitates sufficiently fast state discrimination without unnecessarily compromising the intrinsic qubit coherence. The second condition strictly maximizes the readout SNR \cite{Gambetta.08}, yielding a high readout fidelity that is fundamentally independent of all other system parameters. Finally, maintaining a large $n_\mathrm{crit} \gtrsim 50$ ensures projective, near–quantum-nondemolition (QND) readout by avoiding the resonance between multi-photon transitions of the resonator and noncomputational transitions of the fluxonium. Enforcing this bound ensures that a nominal steady-state readout drive of $n_\mathrm{drive} \gtrsim 10$ ($n_\mathrm{drive} \approx n_\mathrm{crit}/5$) safely restricts measurement-induced transition probability to $<1\%$ \cite{Walter.17}.

With the optimized fluxonium parameters detailed in Table~\ref{tab:param}, Fig.~\ref{fig:readout} demonstrates that a qubit-resonator coupling of $g_{r_j}/2\pi=130$~MHz and a resonator frequency of $\omega_{r_j}/2\pi \approx 7.55$~GHz successfully secure this target parameter regime. To evaluate the cavity decay rate $\kappa$, we assume a realistic quality factor of $Q_j = 3000$. Specifically, this configuration yields a dispersive shift of $|2\chi|/2\pi \approx 2.55$~MHz, a decay ratio of $|2\chi|/\kappa = 1$, and a critical photon number of $n_\mathrm{crit} \approx 75$ (indicated by the yellow dot in Fig.~\ref{fig:readout}), comfortably satisfying the dispersive readout conditions established above.

Beyond dispersive readout performance, Purcell relaxation arising from the qubit-resonator coupling constitutes a prominent decoherence channel. We quantify its impact on two-qubit MAP gate fidelity by analyzing the extent to which Purcell decay damps the higher-level auxiliary states involved during gate operation.
While plasmon transitions in typical fluxonium systems can couple strongly to the readout resonator and lead to enhanced relaxation of highly excited states, our architecture strategically mitigates this effect. In our engineered parameter regime, the state-dependent dispersive shift is primarily governed by the $\omega_{{0,3}_{j}}$ and $\omega_{{1,4}_{j}}$ transitions, whereas the plasmon transition $\omega_{{1,2}_{j}}$ used for the MAP gate remains far detuned from the readout resonator frequency. As a result, Purcell-induced decay of the relevant gate states is strongly suppressed, rendering its impact on two-qubit gate fidelity negligible. The analyses presented above conclude Steps 12--15 of our systematic parameter optimization flow in Fig.~\ref{fig:optimization}(b).

\section{CONCLUSIONS AND OUTLOOK}\label{sec:Conclusion}
In this work, we have presented a comprehensive design strategy for scalable fluxonium-based quantum processors using double-transmon couplers. By leveraging the rich energy structure of fluxonium qubits, we introduce a frequency-allocation scheme that separates qubit transitions, coupler modes, and resonator bands into distinct spectral regions. A central result of this work is that such frequency separation transforms an otherwise highly complex, multi-parameter, multi-objective optimization problem into a structured, sequential design workflow. By assigning single- and two-qubit control, readout, and reset operations to largely independent spectral bands, the corresponding parameter sets can be optimized in a nearly decoupled manner. This significantly reduces the effective dimensionality and computational complexity of the design problem, while preserving global consistency across system-level performance metrics.
Building on this principle, we have developed a systematic design framework that identifies feasible parameter regimes for fluxonium-based, DTC-coupled multi-qubit quantum processors.  
Specifically, we identified key benchmarking metrics to guide this optimization: single-qubit gate fidelity ($F_{\mathrm{1q}}$); leakage ($\eta$) and environmental noise-induced decoherence for two-qubit gates; the dispersive shift ($\chi$), critical photon number ($n_\mathrm{crit}$), and SNR condition ($|2\chi|/\kappa$) for qubit readout; reset fidelity ($F_{\mathrm{reset}}$) and duration ($t_{\mathrm{reset}}$); and spectator-induced crosstalk ($\zeta$) during parallel operations. Optimization based on these benchmarks enables high-fidelity, shallow-depth circuit execution. 
Furthermore, the framework incorporates robustness analysis to mitigate sensitivity to fabrication-induced parameter variations, a primary limitation to performance uniformity in large-scale quantum processors. Consequently, our work bridges microscopic Hamiltonian design and system-level performance in a quantitative and predictive manner.

Although this study focuses on the DTC architecture, the proposed framework is readily generalizable to other coupling schemes in fluxonium-based multi-qubit architectures, such as tunable transmon couplers \cite{Chakraborty.25, Zwanenburg.26, Ding.23, Singh.26, Lange.25, Kugut.25, Heunisch.25} and resonator-based couplers \cite{Rosenfeld.24, Xiong.25}, provided that analogous benchmarking criteria are established. More broadly, this work represents a step toward systematic, and ultimately automated, parameter design for quantum devices. By ensuring that architectural choices are driven by rigorous performance metrics, we can significantly accelerate the development of robust quantum processors, paving the way for automated, model-driven quantum processor design.

\appendix
\section{Effective system Hamiltonian} \label{appen:parameter}

\subsection{DTC Hamiltonian in the harmonic oscillator basis}
\begin{table}[t]
\renewcommand{\arraystretch}{2}
\begin{ruledtabular}
\begin{tabular}{lc}
Term & Value \\
\hline
    $\phi_{\mathrm{zpf},c,m}$
&
    $\frac{1}{\sqrt{2}}
    \sqrt{\hbar \omega_{c,m}/\left(E'_{J_{c,m}}+E'_{J_{c,12}}\right)}$
\\
    $n_{\mathrm{zpf},c,m}$
&
    $
    \frac{1}{\sqrt{2}}
    \sqrt{\hbar \omega_{c,m}/(8E_{C_{c,m}})}$
\\
    $E'_{J_{c,12}}$
&
    $E_{J_{c,12}} \cos \phi_{\mathrm{ext}_c}$
\\
    $E_{J_{c,m}}'$
&
    $\sqrt{E_{J_{c,m}}^2-E_{J_{c,12}}^2\sin^2\left(\phi_{\mathrm{ext}_c}\right)}$
\\
    $E_{C_{c,m}}$
&
    $e^2/(2C_{\Sigma_{c,m}})$
\\
    $C_{\Sigma_{c,m}}$
&
    $\frac{C_{c}^2}{C_{c,12}+C_{c,2}}$
\\
    $C_{c}^2$
&
    $
    C_{c,1}C_{c,2}
    +
    C_{c,1}C_{c,12} 
    + C_{c,2}C_{c,12}
    $
\\
    $
    \widetilde{C}_{c,m}
    $
&
    $
    C_{c,m}C_{c,12}/C_c^2
    $
\\
    $g_{c,\mathrm{cap}}$
&
    $\frac{C_{c,12}\sqrt{\omega_{c,1}\omega_{c,2}}}
    {2\sqrt{\left(C_{c,1}+C_{c,12}\right)
    \left(C_{c,2}+C_{c,12}\right)
    }}
    $
\\
    $g_{c,\mathrm{ind}}$
&
    $\frac{4E'_{J_{c,12}}\sqrt{E_{C_{c,1}}E_{C_{c,2}}}}
    {\hbar^2\sqrt{\omega_{c,1}\omega_{c,2}}}
    $
\\
    $\nu_{c,m}^{(4)}$
&
    $E_{C_{c,m}}/(12\hbar)$
\end{tabular}
\caption{DTC Hamiltonian parameter definitions. $C_{c,m}$ denotes the capacitance of $m$-th transmon, $C_{c,12}$ is the mutual capacitance between 1-st and 2-nd transmon.}
\label{tab:dtc_HO_parameters}
\end{ruledtabular}
\end{table}

\begin{figure}[t]
    \centering
    \includegraphics[width=0.8\linewidth]{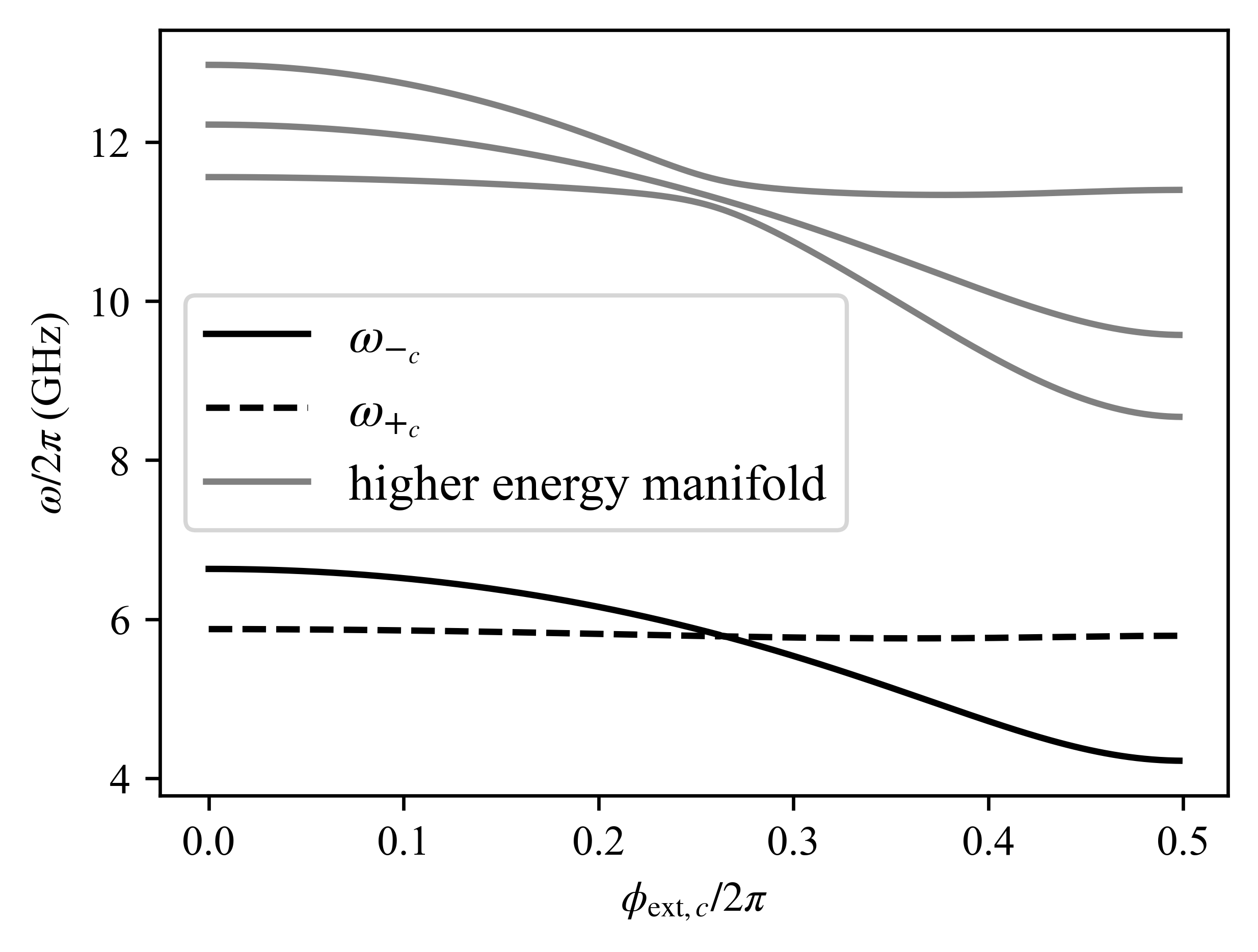}
    \caption{Normal modes of a DTC vs.~flux bias $\phi_{\mathrm{ext}_c}$. The DTC parameters are taken from Table \ref{tab:param}.}
    \label{fig:dtc_mode}
\end{figure}

The DTC Hamiltonian $H_{c}$ can be alternatively expressed in the harmonic oscillator basis by introducing the operators \cite{Campbell.23}
\begin{subequations}
\begin{align}
    \hat{\phi}_j&=-i \phi_{\mathrm{zpf},j} (\hat{a}_j-\hat{a}_j^\dagger),
    \\
    \hat{n}_j&=n_{\mathrm{zpf},j} (\hat{a}_j+\hat{a}_j^\dagger),
\end{align}
\end{subequations}
with the definitions of the phase (number) zero-point fluctuations $\phi_{\mathrm{zpf}}$ ($n_{\mathrm{zpf}}$) given in Table~\ref{tab:dtc_HO_parameters}. 
This results in the approximated DTC Hamiltonian (in the small $E_{J_{c,12}}/E_{J_m}$ limit)
\begin{align} \label{eq:DTC_H_HO}
\begin{split}
    H_{c}/\hbar&\approx 
    \sum_{n=1}^2 \left[ \omega_{c,n} 
    \hat{a}_{c,n}^\dagger \hat{a}_{c,n}
    - \nu^{(4)}_{c,n}\left(\hat{a}_{c,n}-\hat{a}_{c,n}^\dagger\right)^{4}
    \right]
    \\
    & + g_{c,\mathrm{ind}}\left(\hat{a}_{c,1}-\hat{a}_{c,1}^\dagger\right)\left(\hat{a}_{c,2}-\hat{a}_{c,2}^\dagger\right)
    \\
    & + g_{c,\mathrm{cap}}\left(\hat{a}_{c,1}+\hat{a}_{c,1}^\dagger\right)\left(\hat{a}_{c,2}+\hat{a}_{c,2}^\dagger\right),
\end{split}
\end{align}
where higher-order inductive couplings and qubit nonlinearities $\nu^{(p)}$ for $p\geq 6$ have been neglected. In Eq.~\eqref{eq:DTC_H_HO}, $\omega_{c,n}$ ($\nu^{(4)}_{c,n}$) denotes the frequency (nonlinearity) of mode $n$, and $g_{c,\mathrm{ind}}$ ($g_{c,\mathrm{cap}}$) represents the inductive (capacitive) coupling between modes 1 and 2 of the $c$-th DTC, cf.~Table \ref{tab:dtc_HO_parameters} for the explicit form of the parameters. The DTC eigenenergies as a function of flux bias $\phi_{\mathrm{ext}_c}$, calculated using the parameters in Table~\ref{tab:param}, are shown in Fig.~\ref{fig:dtc_mode}. The lowest eigenenergy manifold comprises a nearly constant-frequency mode and a flux-tunable mode, with frequencies denoted as $\omega_{+_c}$ and $\omega_{-_c}$, respectively. The same spectral feature is also evident in the DTC energy spectrum shown in Fig.~\ref{fig:bare_robust} of the main text.

\subsection{Effective fluxonium-fluxonium interaction}\label{subsubsec:f-f}

It is instructive to derive the effective interaction between neighboring fluxonium qubits, indexed $j$ and $k$, mediated by the intervening DTC $c=(j,k)$. Given the strong anharmonicity of the fluxonium spectrum, we restrict our analysis to the relevant two-level subspaces of the qubits and the DTC, assuming couplings to other energy levels are negligible. While approximate, this projection onto specific target subspaces yields valuable physical insight into the effective inter-qubit coupling mechanism.

Consider the relevant energy levels $\{|a\rangle, |b\rangle\}$ for the $j$-th and $k$-th fluxoniums and the two modes of the DTC, with frequencies $\omega_{a,b_j}$, $\omega_{a,b_k}$, $\omega_{c,1}$, and $\omega_{c,2}$, respectively. For simplicity, we assume degenerate modes in the DTC, i.e.~$\omega_{c,n}=\omega_c$. Neglecting higher-order terms, the subsystem Hamiltonian is given by
\begin{align}
\begin{split}
    H_{{a,b}_{j,k}}&\approx \left[\sum_{m\in\{j,k\}}
    (\omega_{{a,b}_m}/2) \hat{\sigma}_{{a,b}_m}^z 
    + \sum_{n=1}^2\omega_{c} \hat{a}_{c,n}^\dagger \hat{a}_{c,n} 
    \right.
    \\
    & \left.+ \sum_{m\in\{j,k\}}\sum_{n=1}^2 g_{{a,b}_m,c_n} \left(\hat{\sigma}_{{a,b}_m}^\dagger\hat{a}_{c,n} + \mathrm{h.c.}\right)
    \right]
    \\
    &+ g_{c,\mathrm{ind}}(\hat{a}_{c,1}- \hat{a}_{c,1}^\dagger)(\hat{a}_{c,2}- \hat{a}_{c,2}^\dagger)
    \\
    &+ g_{c,\mathrm{cap}}(\hat{a}_{c,1}+ \hat{a}_{c,1}^\dagger)(\hat{a}_{c,2}+ \hat{a}_{c,2}^\dagger),
\end{split}
\end{align}
where $\hat{\sigma}^z_{{a,b}_m} = |b_m\rangle \langle b_m| - |a_m\rangle \langle a_m|$ and $\sigma_{a,b_m}=|a_m\rangle\langle b_m|$ ($\sigma^\dagger_{a,b_m}=|b_m\rangle\langle a_m|$) are the Pauli-$z$ operator and lowering (raising) operator in the $m$-th qubit subspace spanned by $\{|a_m\rangle, |b_m\rangle\}$, respectively, $\hat{a}_{c,n}$ $(\hat{a}^\dagger_{c,n})$ is the annihilation (creation) operator for the $n$-th mode of the $c$-th coupler whose frequency is $\omega_{c,n}$, and $g_{{a,b}_m,c_n} = \langle a_m| \langle 0_{c,n}| H_\mathrm{sys} |b_m\rangle |1_{c,n}\rangle$ is the coupling matrix element. 

An effective Hamiltonian can be derived by performing a Schrieffer-Wolff transformation to eliminate the direct fluxonium-DTC coupling. Assuming the detuning $\Delta_{a,b_m} = \omega_{a,b_m} - \omega_{c}$ is much larger than the coupling strength $g_{{a,b}_m,c_n}$, the effective Hamiltonian becomes \cite{Campbell.23, Zhao.25.1}
\begin{align}
\begin{split}
    H_{a,b_{j,k}}^\mathrm{eff} 
    &= \sum_{m\in\{j,k\}}\left[ \omega_{{a,b}_m} \hat{\sigma}_{{a,b}_m}^\dagger \hat{\sigma}_{{a,b}_m}\right] 
    \\
    &+ g_{a,b}^\mathrm{eff}\left({\hat{\sigma}^{\dagger}_{a,b_j}} \hat{\sigma}_{{a,b}_k} + \text{h.c.}\right),
\end{split}
\end{align}
where the effective coupling strength is expressed as
\begin{align}\label{eq:effC}
\begin{split}
g_{a,b}^\mathrm{eff} \approx 
\frac{g_{a,b_j, c_1}g_{a,b_k,c_2}}{2}\sum_{m=j,k}
\left[\frac{g_{c,\mathrm{cap}}-g_{c,\mathrm{ind}}}{\Delta_{a,b_m}^2}\right.
\\
\left.- \frac{g_{c,\mathrm{cap}}+g_{c,\mathrm{ind}}}{\omega_c\Delta_{a,b_m}}
\right].
\end{split}
\end{align}
In the absence of parasitic capacitive coupling, i.e., $g_{c,\mathrm{cap}} = 0$, the effective interaction can be fully suppressed by tuning the external flux of the DTC such that the inductive contribution vanishes, $g_{c,\mathrm{ind}} = 0$. At this operating point, the cancellation is exact and independent of both the qubit parameters and the specific transitions involved, reflecting perfect destructive interference between the common and differential modes of the DTC.
In the presence of finite capacitive coupling, the total effective coupling includes an additional contribution that cannot be nullified for all potential transitions through tuning the flux alone. As a result, the location of the turn-off point depends on both the qubit parameters and the selected energy levels.

\section{Microwave-activated-phase (MAP) gate realization}\label{appen:MAP}
\subsection{DTC ``turn-on'' mechanism and MAP gate protocol} \label{appen:MAP_prot}
The effective interaction between neighboring fluxonium qubits, $j$ and $k$, is mediated by the intervening DTC $c=(j,k)$. Since the physical coupling between the fluxoniums and the DTC is capacitive, achieving a significant effective coupling strength requires exploiting the large charge matrix elements associated with the fluxonium plasmon transitions. Consequently, strong inter-qubit coupling is realized by tuning the flux-dependent mode of the DTC into resonance with the fluxonium plasmon frequencies, i.e., $\omega_{1,2_j} \approx \omega_{1,2_k} \approx \omega_{-_c}$. Near this degeneracy point, the interaction shifts from a virtual, dispersive mediation to a resonant hybridization, effectively ``turn-on'' the coupler. We denote the DTC flux bias $\phi_{\mathrm{ext}_c}$ at which this degeneracy occurs as $\phi_{\mathrm{on}_c}$, and the corresponding DTC mode frequency as $\omega_{\mathrm{on}_c}$. When the DTC is biased to this ``on'' state, the effective fluxonium-fluxonium coupling $g_{1,2}^\mathrm{eff}$ induces a level repulsion between the plasmon states. This interaction manifests as distinct shifts in the transition frequencies (cf.~Table \ref{tab:def} for definition of $\omega_k$) 
\begin{subequations}\label{eq:geff}
\begin{align}
    2\left[(\omega_{201} - \omega_{101}) - (\omega_{200} - \omega_{100})\right] &\approx \Delta - \tilde{g}_\mathrm{eff},\label{eq:t1}
    \\
    2\left[(\omega_{102} - \omega_{101}) - (\omega_{002} - \omega_{001})\right] &\approx -\Delta + \tilde{g}_\mathrm{eff},\label{eq:t2}
    \\
    \left[(\omega_{111} - \omega_{101}) - (\omega_{200} - \omega_{100})\right] &\approx \omega_{-_c}-\omega_{{1,2}_j},
\end{align}
\end{subequations}
where 
\begin{subequations}
\begin{align}
    \tilde{g}_\mathrm{eff} &= \sqrt{4(g_{1,2}^\mathrm{eff})^2+\Delta^2},
    \\
    \Delta&=\left(\omega_{1,2_j} - \omega_{1,2_k}\right).
\end{align}
\end{subequations} 
From Eq.~\eqref{eq:geff}, it can be deduced that for degenerate plasmon frequencies and DTC normal mode, specifically when $\Delta=0$ and $\omega_{-_c}=\omega_{1,2_j}=\omega_{1,2_k}$, the effective coupling $g_{1,2}^\mathrm{eff}$ leads to splitting between the three eigenstates $|201\rangle$, $|102\rangle$, and $|111\rangle$.
The resulting splitting between these transition frequencies enables frequency-selective addressing. By applying a microwave drive resonant with a specific transition, while remaining off-resonant to others, we can selectively rotate the system. A controlled-phase (CZ) gate is realized by applying a drive pulse of duration $t_g$ sufficient to perform a $2\pi$ rotation on the targeted pair of states. This cyclic evolution returns the population to the computational subspace while accumulating a geometric conditional phase of $\pi$ \cite{Ding.23}. 

\subsection{Leakage estimation} \label{subsec:leakage}
\begin{figure}
    \centering
    \includegraphics[width=\linewidth]{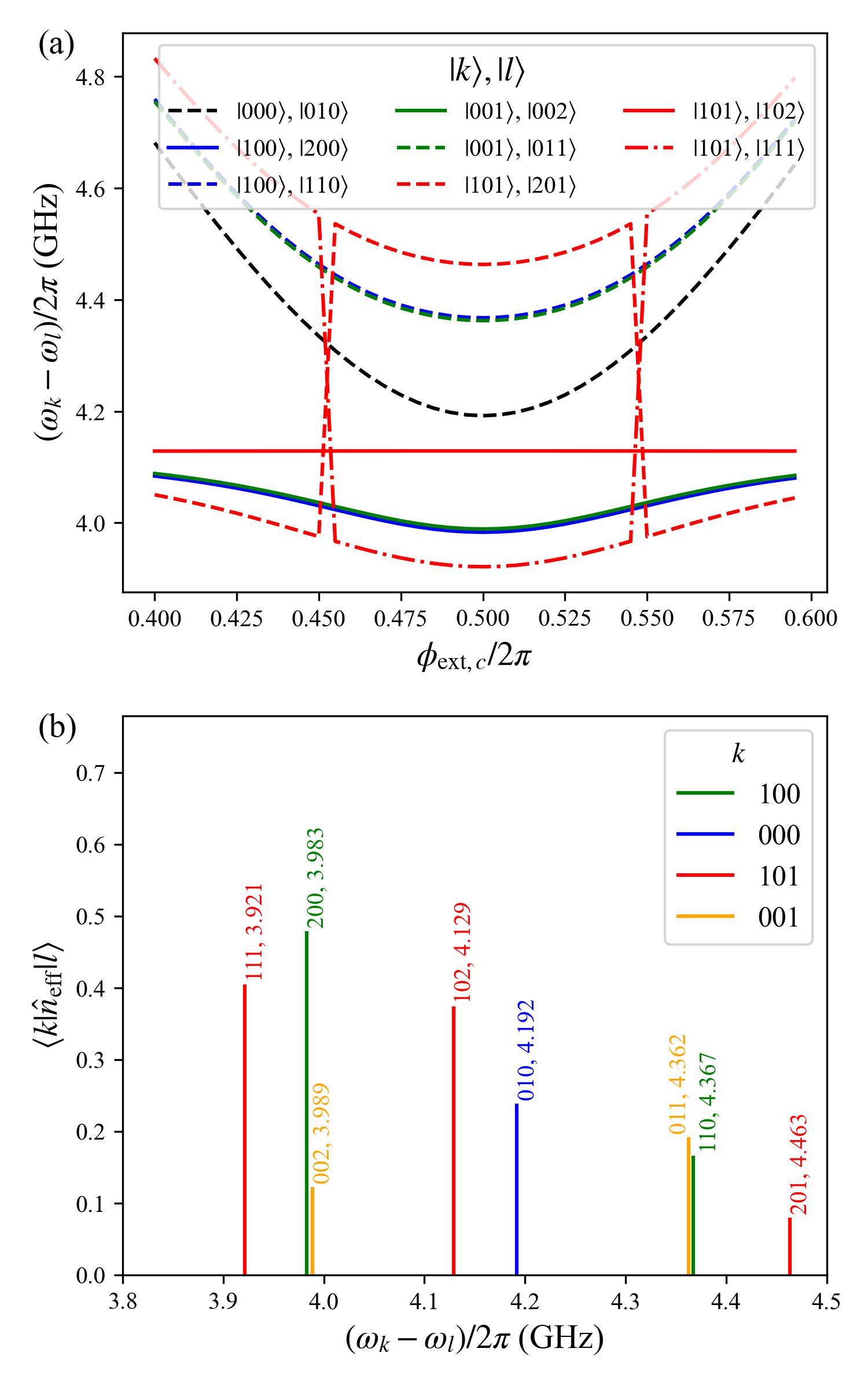}
    \caption{(a) Transition frequencies of all eight candidate transitions for the MAP gate operation. (b) The effective drive strength of the transition $|k\rangle\rightarrow |l\rangle$, denoted as $\langle k|\hat{n}_\mathrm{eff}|l\rangle$ at coupler sweet spot, i.e.~$\phi_{\mathrm{on}_c}=\pi$. The label of each vertical line denotes the final state of the transition $|k\rangle$ and the corresponding transition frequency. The results are computed with parameters in Table \ref{tab:param}.}
    \label{fig:MAP_t}
\end{figure}

Here, we adopt the parameters from Table.~\ref{tab:param} as an example to illustrate the discussion on leakage in the MAP gate scenario. 
At the operating point of the MAP gate ($\phi_{\mathrm{on}_c}\approx \pi$), a manifold of eight transitions arises, as shown in Fig.~\ref{fig:MAP_t}(a). Since any of these eight transitions can be utilized to perform the MAP gate operation, we must select the one that yields the smallest leakage error. To estimate this error, we employ time-dependent perturbation theory and assume the leakage probability during the gate operation significantly smaller than one. Consider a cosine envelope microwave drive whose frequency $\omega_d$ is in resonance with the transition frequency of $|j\rangle\rightarrow|\beta\rangle$, i.e.~$\omega_d=\omega_{j,\beta}$, if there exists another transition $|k\rangle \rightarrow |\alpha\rangle$ whose frequency $\omega_{k,\alpha}$ is close to $\omega_{j,\beta}$, a leakage transition will be driven. The probability of the leakage state $|\beta\rangle$, denoted as $\eta_{j \beta,k\alpha}$, at the end of the pulse is given by 
\begin{equation}\label{eq:leakage}
    \eta_{j\beta,k\alpha} = \left|\frac{1}{t_g}\frac{n_{k\alpha}}{n_{j\beta}}\frac{2\pi}{\Delta_{j\beta,k\alpha}}\frac{1}{1-[t_g\Delta_{j\beta,k\alpha}/(2\pi)]^2}\right|^2.
\end{equation}
Here, $t_g$ is the gate duration, while $n_{j\beta}=\langle j|\hat{n}_\mathrm{eff}|\beta\rangle$ and $n_{k\alpha}=\langle k|\hat{n}_\mathrm{eff}|\alpha\rangle$ represent the microwave drive strengths on the ``intended" and ``leakage" transitions, respectively. The effective charge transition operator is defined as $\hat{n}_\mathrm{eff}=\hat{n}_j+\gamma_{j,c} \hat{n}_{c,1}$ (see Eq.~\eqref{eq:drive} for the definition of $\gamma_{j,c}$), and $\Delta_{j\beta,k\alpha} = |\omega_{\alpha} - \omega_{k}|-|\omega_{\beta} - \omega_{j}|$ denotes the detuning between the frequencies of the intended and leakage transitions.

Using Eq.~\eqref{eq:leakage}, one can judiciously choose the ``best" transition to perform the MAP gate. For instance, as shown in Fig.~\ref{fig:MAP_t}(b), although the transition $|101\rangle \rightarrow |201\rangle$ yields the largest detuning from the nearest leakage state ($\Delta_{101;201} \approx 96\,\mathrm{MHz}$), the transition $|101\rangle \rightarrow |102\rangle$, despite having a smaller detuning ($63\,\mathrm{MHz}$), may yield lower overall leakage. This is due to a more favorable ratio of matrix elements, specifically a relatively larger intended charge transition $\langle 101|\hat{n}_\mathrm{eff}|102\rangle$ compared to the values of $\langle k|\hat{n}_\mathrm{eff}|\alpha\rangle$ of its nearest leakage transitions. The relationship between the leakage probability and the resulting leakage-induced MAP gate fidelity is presented explicitly in Appendix~\ref{appen:leaF}.

\section{Spectator induced crosstalk}\label{sec:turn_off}
\subsection{Definition and ``turn-off'' mechanism}\label{sec:defxtalk}
To quantify the isolation between qubits, we first define a metric for spectator-induced crosstalk. We consider a composite system of three fluxonium qubits ($q_1, q_2, q_3$) connected via DTCs. We focus on a scenario where a MAP gate is applied to the active pair $(q_2, q_3)$ by driving a transition between states $|A\rangle$ and $|B\rangle$, while $q_1$ serves as a spectator. The magnitude of the frequency shift induced by the spectator is tunable via the flux bias of the intervening coupler, $\phi_{\mathrm{ext}_{c=(1,2)}}$. We define the frequency shift $\zeta_{ij, A\to B}$ as the variation in the target transition energy ($A \to B$) when the spectator $q_1$ changes from state $i$ to state $j$
\begin{equation}\zeta_{ij, A\to B}(\phi_{\mathrm{ext},12}) = \left| (E_{j0B} - E_{j0A}) - (E_{i0B} - E_{i0A}) \right|,
\end{equation}
where $E_{k0\Lambda}$ denotes the eigenenergy of the system state adiabatically connected to the bare state $|k\rangle_{q_1} \otimes |0\rangle_{c_{12}} \otimes |\Lambda\rangle_{\mathrm{active}}$. Here, $k \in \{i, j\}$ represents the state of the spectator $q_1$, the coupler $c_{12}$ remains in its ground state, and the active subsystem corresponds to $|\Lambda\rangle_{\mathrm{active}} \in \{|A\rangle, |B\rangle\}$. 

For robust parallel gate operation, the spectator qubit may occupy any state in the subspace $\{|0\rangle, |1\rangle, |2\rangle\}$. Consequently, we define the system crosstalk $\zeta$ as the worst-case frequency shift, minimized over the DTC flux bias
\begin{equation}\label{eq:xtalk_def}
\zeta = \min_{\phi_{\mathrm{ext},12}} \left( \max_{\substack{i,j \in {0,1,2} \ i \neq j}} \zeta_{ij, A\to B}(\phi_{\mathrm{ext},12}) \right).
\end{equation}

Minimizing this crosstalk requires effectively isolating the qubit pairs. In architectures using DTC-coupled fluxonium qubits, this isolation is achieved by biasing the coupler at a specific ``turn-off'' flux $\phi_{\mathrm{off}_c}$, where the two normal modes of the DTC are degenerate ($\omega_{-_c} = \omega_{+_c}$). At this operating point, the effective inductive coupling $g_{c,\mathrm{ind}}$ becomes zero (cf.~Eq.~\eqref{eq:geff}). Consequently, in the ideal limit of vanishing capacitive coupling between the transmons within the DTC (i.e., $J_{c,12}=0$), the total effective coupling between neighboring fluxonium qubits vanishes. This ensures that the state of a spectator qubit induces zero frequency shifts on its neighbors.

\begin{figure} [t]
    \centering
    \includegraphics[width=\linewidth]{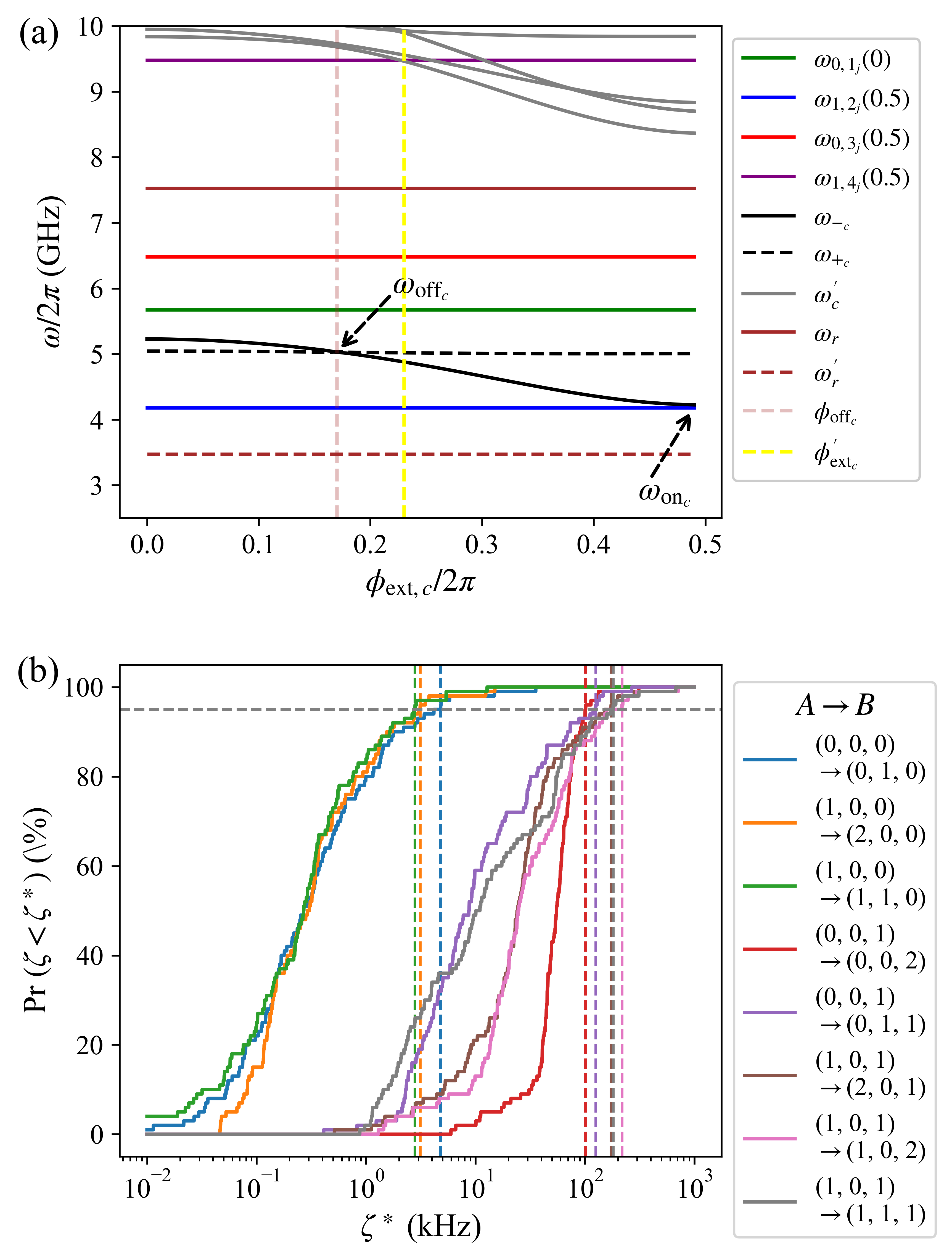}
    \caption{(a) Frequency allocation and (b) robustness analysis of spectator-induced crosstalk $\zeta$. The results are obtained using the optimized parameters from Table~\ref{tab:param}, but with modified DTC junction energies: $E_{J,c_m}/h=11.3$~GHz and $E_{J,c_{12}}/h=1.3$~GHz.}
    \label{fig:param_1}
\end{figure}
When a non-zero capacitive coupling ($J_{c,12} \neq 0$) is present, the cancellation of the total effective coupling becomes frequency-dependent. Specifically, ``turn-off'' the interaction for transitions at different frequencies requires slightly different coupler flux biases. For non-degenerate qubits, this frequency dependence prevents the simultaneous cancellation of all interaction terms, giving rise to residual crosstalk. Nevertheless, for small capacitive couplings, an inspection of Eq.~\eqref{eq:geff} reveals that this residual crosstalk can be effectively mitigated through parameter engineering. By ensuring a large detuning $\Delta$ between the fluxonium plasmon frequencies and the DTC normal mode frequencies at the turn-off flux, such that $\Delta \gg g_{c,\mathrm{cap}}$, the frequency shifts induced by the spectator are suppressed. 

This design consideration is particularly critical when evaluating crosstalk robustness against parameter variations. In the presence of fabrication uncertainties, insufficient detuning can lead to specific variation instances where the DTC's nearly flux-independent mode $\omega_{+_c}$ collides with the fluxonium plasmon frequency, thereby amplifying residual crosstalk. For example, Fig.~\ref{fig:param_1} illustrates a scenario where the worst-case crosstalk $\zeta$ is approximately an order of magnitude larger than that obtained with our optimal parameter set $\mathcal{P^*}$ (cf.~Fig.~\ref{fig:xtalk_robust}). This degradation arises obviously because the smaller design value of $\Delta$ in Fig.~\ref{fig:param_1} increases the likelihood of frequency collisions under random parameter variations.

\subsection{Effect of crosstalk on MAP gate fidelities} \label{subsec:xtalk_fid}
\begin{figure}
    \centering
    \includegraphics[width=0.9\linewidth]{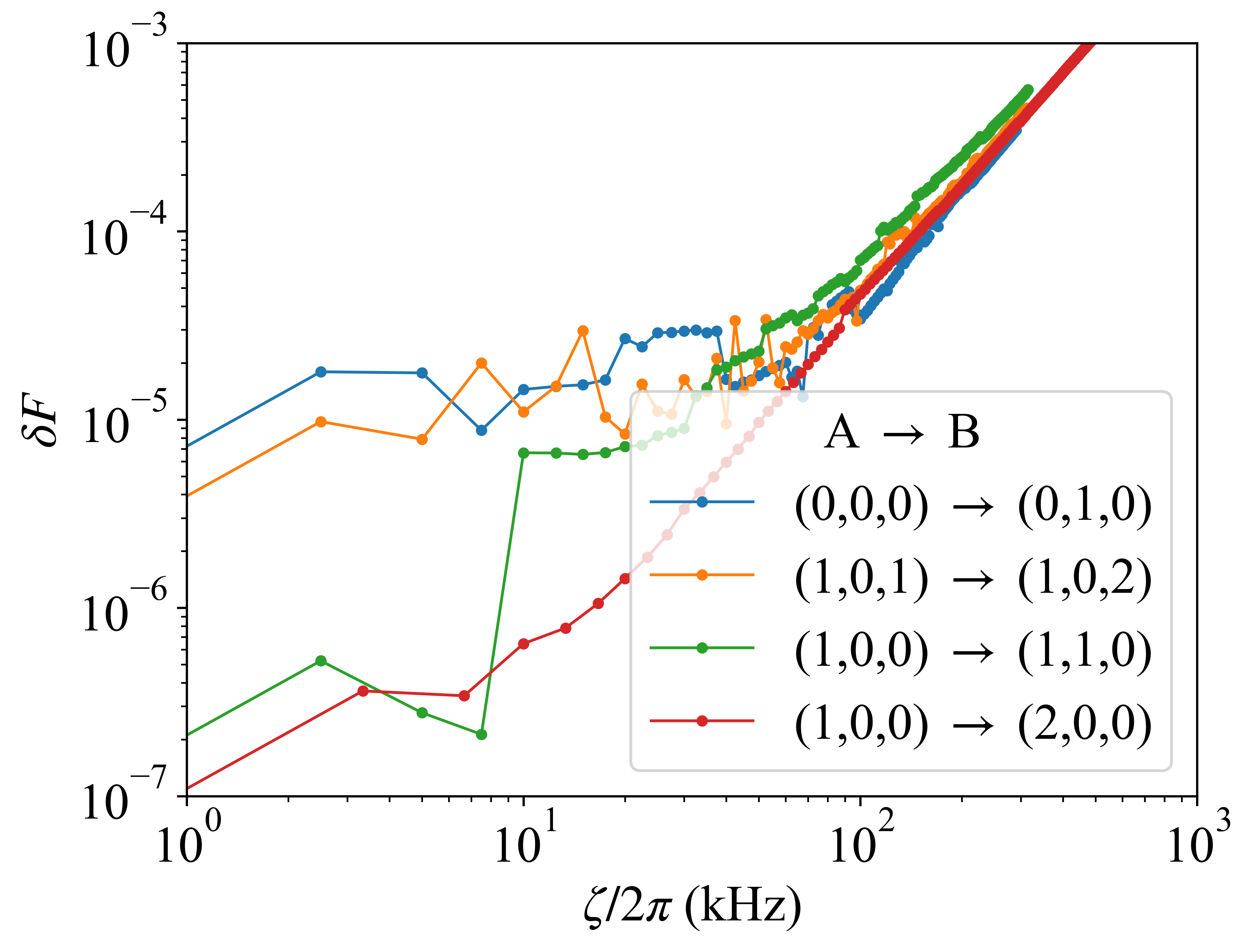}
    \caption{Crosstalk-induced reduction in MAP gate fidelity, denoted as $\delta F$, evaluated in the absence of relaxation and pure dephasing. The fidelity reduction is plotted as a function of the crosstalk strength, $\zeta$. Results are calculated using the optimized parameters from Table~\ref{tab:param}, with an asymmetry introduced between the interacting fluxonium Josephson energies ($E_{J_1}/h=4.65$~GHz and $E_{J_2}/h=4.5$~GHz).}
    \label{fig:xtalk_fid}
\end{figure}

The impact of crosstalk on two-qubit gate fidelity is highly sensitive to the specific system parameters and the chosen gate scheme. To determine the maximum tolerable crosstalk that our DTC-coupled fluxonium architecture can withstand before MAP gate performance degrades, we evaluate the system using the optimized parameters detailed in Table~\ref{tab:param}. Figure~\ref{fig:xtalk_fid} illustrates the crosstalk-induced infidelity, $\delta F$, as a function of the crosstalk strength $\zeta$. We observe that the architecture demonstrates strong resilience: for a crosstalk strength of $\zeta/2\pi \lesssim 10$~kHz ($100$~kHz), the induced infidelity remains tightly bounded at $\delta F \lesssim 10^{-5}$ ($\lesssim 10^{-4}$).

\section{Single-qubit coherence time and average gate fidelity}\label{appen:SQF}
\subsection{Relaxation rate \texorpdfstring{$\Gamma_1$}{Gamma 1}}
Consider a Hamiltonian parameterized by an external parameter $\lambda$, $\hat{H}(\lambda)$, we assume that the overall parameter can be decomposed into an externally controlled value and a small noise term, i.e.~$\lambda=\lambda_e + \lambda_n$ with $\lambda_n \ll \lambda_e$. A Taylor expansion of the Hamiltonian yields \cite{Koch.07}
\begin{equation}
    \hat{H} \rightarrow \hat{H} + \lambda_n \hat{H}^{(1)}_\lambda,
\end{equation}
where
\begin{equation}
    \hat{H}^{(1)}_\lambda = \left.\frac{\partial \hat{H}}{\partial \lambda}\right|_{\lambda^e}.
\end{equation}
The depolarization rate is then 
\begin{equation}\label{eq:rel_rate}
    \Gamma_{1,kl}^\lambda = \frac{2}{\hbar^2}\left|\langle k | \hat{H}^{(1)}_\lambda | l \rangle \right|^2S_\lambda(\omega_{k,l}),
\end{equation}
where $\omega_{k,l}=(E_l - E_k)/\hbar$ is the frequency difference between state $|k\rangle$ and $|l\rangle$ and $S_\Phi(\omega)$ is the noise spectrum. For $1/f$ noise, $S_\Phi(\omega)=2\pi (A_\Phi^2/\omega)$.

\subsection{Pure dephasing rate \texorpdfstring{$\Gamma_\phi$}{Gamma phi}}
For $1/f$ noise spectrum, the first and second-order dephasing rate are \cite{Nguyen.19, Ithier.05}
\begin{subequations}\label{eq:T_phi}
\begin{align}
    \Gamma^\lambda_{\phi,kl} &= A_\lambda \frac{\partial \omega_{k,l}}{\partial \Phi},
    \\
    \Gamma^{\lambda^2}_{\phi,kl} & = A_\lambda^2 \frac{\partial^2 \omega_{k,l}}{\partial \Phi^2},
\end{align}
\end{subequations}

\subsection{Estimates for fluxonium single-qubit relaxation (\texorpdfstring{$T_1$}{T 1}) and pure dephasing time (\texorpdfstring{$T_\phi$}{T phi})}

\subsubsection{Flux noise} 
Assuming the flux noise exhibits a $1/f$ character, the relaxation rate (at the half-integer flux) between state $|i\rangle$ and $|j\rangle$ of the fluxonium qubit are \cite{Ateshian.07, Nguyen.19}
\begin{equation}\label{eq:flux_T1_q}
    \Gamma_{1,kl}^{\Phi} = 8 \left(\frac{\pi E_L}{\hbar \Phi_0}\right)^2 
    \left|\langle k |\hat{\phi}|l\rangle\right|^2 S_\Phi (\omega_{k,l}).
\end{equation}
The dephasing rate directly follows from Eq.~\eqref{eq:T_phi}.

\subsubsection{Dielectric loss}
The relaxation rate attributed to dielectric loss in the fluxonium qubit is given by
\cite{Ateshian.07}
\begin{equation}\label{eq:diel_q}
    \Gamma^\mathrm{diel}_{1,kl} = \frac{1}{4 E_C}
    \left|\langle k| \hat{\phi}|l\rangle\right|^2 \hbar \omega_{k,l}^2 \tan \delta_c \coth\left(\frac{\hbar \omega_{k,l}}{2 k_B T_\mathrm{eff}}\right),
\end{equation}
where $\tan \delta_c$ is the effective loss tangent, and $T_\mathrm{eff}$ is the effective temperature. 

\subsubsection{Purcell relaxation}
Consider the $j$-th fluxonium is dispersively coupled to a resonator of frequency $\omega_{r_j}$ with coupling strength $g_{r_j}$, the Purcell relaxation between two state $|k_j\rangle$ and $|l_j\rangle$ is \cite{Sete.14}
\begin{equation}
\Gamma_{1,kl}^\mathrm{Purcell} = \frac{\kappa}{2}-\frac{\sqrt{2}}{2}\sqrt{-A+\sqrt{A^2+(\kappa \Delta)^2}},
\end{equation}
where $A=\Delta^2+4g_{r_j}^2-\kappa^2/4$ and $\Delta = \omega_{k,l_j}-\omega_{r_j}$.

\subsubsection{Photon shot noise}
The fluxonium qubits in this architecture are susceptible to dephasing from photon number fluctuations in the reset resonator. This susceptibility is exacerbated by the reset resonator's relatively low frequency, $\omega'_{r_j}$, which leads to a relatively higher thermal population. For the undriven reset resonator, the dephasing rate between states $|k_j\rangle$ and $|l_j\rangle$ is \cite{Krantz.19}
\begin{equation}
\Gamma_{\phi,kl} = \eta\frac{4\chi^2}{\kappa}\bar{n},
\end{equation}
defined by the dispersive shift $\chi$, $\eta=\kappa^2/(\kappa^2+4\chi^2)$ and the mean thermal photon number $\bar{n} = (\exp(\hbar \omega'_r/k_B T_\mathrm{eff})-1)^{-1}$.

\subsection{Estimates for DTC relaxation (\texorpdfstring{$T_1$}{T 1}) and pure dephasing time (\texorpdfstring{$T_\phi$}{T phi})}
\subsubsection{Flux noise}\label{sec:DTC_flux_noise}
From Eq.~\eqref{eq:H_DTC}, the Hamiltonian derivative with respect to external flux is
\begin{align}\label{eq:flux_T1_dtc}
\begin{split}
    \hat{H}_\Phi^{(1)}=\frac{2\pi}{\Phi_0}\left[\sum_{m=1}^2 E_{J_{c,m}}\sin\left(\hat{\phi}_{c,m}+\widetilde{C}_{c,m}\phi_{\mathrm{ext},c}\right)\right.
    \\
    \left.
    E_{J_{c,12}}\sin\left(\hat{\phi}_{c,2}-\hat{\phi}_{c,1}-\widetilde{C}_{c,12}\phi_{\mathrm{ext}_c}\right)\right].
\end{split}
\end{align}
The relaxation rate can then be estimated with Eq.~\eqref{eq:rel_rate}.

\subsubsection{Dielectric loss}
The MAP gate operation exploits the coupling between the flux-tunable mode of the DTC, $\omega_{c_-}$, and the fluxonium plasmon modes, $\omega_{1,2_j}$. In this configuration, the primary relaxation channel arising from dielectric loss is associated with the flux-tunable mode, which corresponds to the differential mode of the DTC. To quantify this loss, we first evaluate the effective capacitance connected to the differential mode.

Expressing the DTC Hamiltonian in terms of node voltages $\hat{\dot{\phi}}_{c,m}$ and node fluxes $\hat{\phi}_{c,m}$, we write $H=T+U$, where
\begin{align}
\begin{split}
    T &=\left(\frac{\Phi_0}{2\pi}\right)^2
    \left[\sum_{m=1}^2\frac{C_{c,m}}{2}\hat{\dot{\phi}}_{c,m}^2 + \frac{C_{c,12}}{2}\left(\hat{\dot{\phi}}_{c,1}-\hat{\dot{\phi}}_{c,2}\right)^2\right],
    \\
     U&= -\sum_{m=1}^2 E_{J_{c,m}}\cos\left(\hat{\phi}_{c,m}+\widetilde{C}_{c,m}\phi_{\mathrm{ext}_c}\right)
\\ &- E_{J_{c,12}}\cos\left(\hat{\phi}_{c,2} - \hat{\phi}_{c,1}-\widetilde{C}_{c,12}\phi_{\mathrm{ext}_c}\right).
\end{split}
\end{align}
Assuming symmetric junction parameters $E_{J_{c,1}} = E_{J_{c,2}}$ and capacitances $C_{c,1}=C_{c,2}$, we rewrite the Hamiltonian in the common and differential bases, defined as $\hat{\phi}_{c,\mathrm{c}}=(\hat{\phi}_{c,1}+\hat{\phi}_{c,2})/2$ and $\hat{\phi}_{c,\mathrm{d}}=(\hat{\phi}_{c,1}-\hat{\phi}_{c,2})/2$, respectively. This yields
\begin{align}
    \begin{split}
        U &= - 2 E_{J_{c,1}} \cos \left(\hat{\phi}_{c,\mathrm{c}}+\widetilde{C}_{c,1}\phi_{\mathrm{ext}_c}\right) \\
        &-E_{J_{c,12}} \cos\left(2\hat{\phi}_{c,\mathrm{d}}-\widetilde{C}_{c,12}\phi_{\mathrm{ext}_c}\right),
        \\
        T &= \left(\frac{\Phi_0}{2\pi}\right)^2 \left[\left(\frac{C_{c,1}+C_{c,2}}{2}\right)\hat{\phi}_{c,\mathrm{c}}^2\right.
        \\&+\left.\left(\frac{C_{c,1}+C_{c,2}}{2}+2C_{c,12}\right)\hat{\phi}_{c,\mathrm{d}}^2\right],
    \end{split}
\end{align}
where $\Phi_0=h/2e$. From the kinetic energy term, we identify the effective capacitance of the differential mode as $C_\mathrm{eff}=(C_{c,1}+C_{c,2})/2+2C_{c,12}$. Consequently, the dielectric loss rate is estimated as
\begin{equation}
    \Gamma_{1,kl}^\mathrm{diel}=\frac{C_\mathrm{eff}}{(2e)^2}\left|\langle k|\hat{\phi}_{c,\mathrm{d}}|l\rangle\right|^2\hbar \omega_{k,l}^2  \tan \delta_c \coth\left(\frac{\hbar \omega_{k,l}}{2 k_B T_\mathrm{eff}}\right)
\end{equation}

\subsection{Average gate fidelity} \label{appen:defSQF}
Given a pure dephasing time $T_\phi$ (with $1/f$ noise spectrum) and relaxation time $T_1$, the average gate fidelity is \cite{Malley.15}
\begin{equation}
    F_{\mathrm{1q}} = 1-\left[\frac{t_g}{3 T_1}+\frac{1}{3}\left(\frac{t_g}{T_\phi}\right)^2\right]
\end{equation}

\section{Multi-qubit coherence time and average two-qubit gate fidelity}\label{sec:fid_multi}

\subsection{Relaxation rate $\Gamma_1$ and pure dephasing rate $\Gamma_\phi$}\label{subsec:multilevel_t1t2}
The relaxation rate for the multi-qubit composite system is derived by generalizing Eq.~\eqref{eq:rel_rate}. This derivation involves replacing the bare states $|i\rangle$ and $|j\rangle$ with the dressed eigenstates of the composite system and promoting the local operators, such as the phase operator $\hat{\phi}$ in Eqs.~\eqref{eq:flux_T1_q}, \eqref{eq:diel_q}, and \eqref{eq:flux_T1_dtc}, to the full composite Hilbert space. Similarly, the pure dephasing rate is determined using Eq.~\eqref{eq:T_phi}, where $\omega_{k,l}$ represents the transition frequencies of the composite system.

\subsection{Gate error modeling} \label{appen:2QF}
In this section, we present analytical estimates of MAP gate fidelities in the presence of incoherent errors. We model the operation of a two-qubit gate as a completely positive trace-preserving map $\mathcal{K}$ acting on an input density state $\rho$, such that $\rho'=\mathcal{K}(\rho)$. To derive the gate fidelity, we express $\mathcal{K}$ in the Kraus representation, $\mathcal{K}(\rho)=\sum_k K_k \rho K_k^\dagger$, satisfying the completeness relation $\sum_k K_k^\dagger K_k=I$. The average state fidelity is given by
\begin{equation}\label{eq:fid}
    F_\mathrm{2q} = \frac{1}{n(n+1)}\left[
    \mathrm{Tr}\left(\sum_k M_k^\dagger M_k\right) + \sum_k \left|\mathrm{Tr}(M_k)\right|^2
    \right],
\end{equation}
where $M_k=P U_0^\dagger K_k P$. Here, $U_0$ represents the ideal unitary operator, $P$ is the projection operator onto the computational subspace, and $n$ is the dimension of the subspace.

\subsubsection{Relaxation involving higher-energy levels} \label{ref:rel_2QFid}
In the MAP gate scheme, the dynamics involve five relevant states: the four computational states $\mathcal{C}=\{|000\rangle, |001\rangle, |100\rangle, |101\rangle\}$ and a higher-energy auxiliary state $|\beta\rangle$. Following the formalism in Ref.~\cite{Ding.23}, we model $|j\rangle$ and $|\beta\rangle$ decay into each other at the same rates $\Gamma_1$ under driven evolution and set the ideal evolution as $U_0=I$. The fidelity is given by
\begin{equation}
    F = \frac{16+e^{-\Gamma_1t_g} + 3 e^{-\Gamma_1t_g/2}}{20}.
\end{equation}

\subsubsection{Dephasing of higher energy levels} \label{appen:dephaseHigherLvl}
For the dephasing analysis, we confine the dynamics to the subspace spanned by the four computational states and the auxiliary state. The pure dephasing rate $\Gamma_{\phi,kl}$ between a pair of states $|k\rangle$ and $|l\rangle$ is estimated as described in Sec.~\ref{subsec:multilevel_t1t2}. This dephasing induces a quantum operation $\rho'=\mathcal{K}_\phi(\rho)$, where the density matrix elements $\rho_{k,l}=\langle k|\rho|l\rangle$ evolve as
\begin{subequations}
    \begin{align}
        \rho'_{k,l\neq k} &= \exp\left[-(\Gamma_{\phi,kl}t)^2\right]\rho_{k,l},\\
        \rho'_{k,k} &= \rho_{k,k}.
    \end{align}
\end{subequations}
Here, the Gaussian decay profile arises from the assumption of $1/f$ noise. The evolution can be alternatively expressed as a Hadamard product, $\rho' = \rho \circ D$, where $D$ is the decoherence matrix with off-diagonal elements $D_{kl}=\exp[-(\Gamma_{\phi,kl}t)^2]$ and diagonal elements $D_{kk}=1$. To obtain the Kraus representation of $\mathcal{K}_\phi$, we diagonalize $D$ such that $D = \sum_{m}\lambda_m v_m v_m^\dagger$, where $\lambda_m$ and $v_m$ are the eigenvalues and eigenvectors, respectively. The Kraus operators are then given by
\begin{equation}
    K_{\phi,m} =\mathrm{diag}(\sqrt{\lambda_m}v_m).
\end{equation}
As in the relaxation analysis, we simplify the fidelity estimation by assuming an effective identity operation ($U_0=I$) and applying Eq.~\eqref{eq:fid}.

\subsubsection{Leakage infidelity} \label{appen:leaF}
In the MAP gate scheme, the dynamics involve the four computational states $\mathcal{C}=\{|000\rangle, |001\rangle, |100\rangle, |101\rangle\}$ and the higher-energy auxiliary states $|\beta\rangle$ and $|\alpha\rangle$. The gate is realized by driving the transition between $|101\rangle$ and $|\beta\rangle$ to induce a geometric phase. To model the gate error, we consider a representative scenario where a parasitic leakage transition occurs between $|100\rangle$ and the auxiliary state $|\alpha\rangle$. We note that while we focus on this specific transition for concreteness, the derivation below can be generalized to any analogous leakage path originating from a computational state.

At the end of the operation, the state $|101\rangle$ acquires the target $\pi$ phase, while a non-negligible leakage population is mixed into $|\alpha\rangle$. Expressed in the basis $\{|000\rangle, |001\rangle, |100\rangle, |101\rangle, |\alpha\rangle\}$, the effective unitary operator is
\begin{equation}
U = \begin{pmatrix}
1 & 0 & 0 & 0 & 0 \\
0 & 1 & 0 & 0 & 0 \\
0 & 0 & \sqrt{1-\eta} & 0 & -i \sqrt{\eta} \\
0 & 0 & 0 & -1 & 0 \\
0 & 0 & -i \sqrt{\eta}  & 0 & \sqrt{1-\eta}
\end{pmatrix},
\end{equation}
Treating this leakage as a coherent error channel, the process is described by a single Kraus operator $K=U$.

We define the ideal CZ unitary operator, extended to include the identity operation on the leakage subspace, as
\begin{align}
\begin{split}
    U_0&=|000\rangle\langle000|
    +|001\rangle\langle001|
    +|100\rangle\langle100|
    \\
    &
    -|101\rangle\langle101|
    +|\alpha\rangle\langle \alpha|,
\end{split}
\end{align}
Using Eq.~\eqref{eq:fid}, we obtain the gate fidelity estimate
\begin{equation}
F\simeq 1-\frac{1}{4}\eta - \frac{3}{80}\eta^2.
\end{equation}
While multiple leakage paths may exist in a realistic device, we consider only the dominant path yielding the largest leakage estimate.

\section*{Acknowledgements}
This research was supported by the Guangdong Provincial Quantum Science Strategic Initiative (Grant No. GDZX2407001). X.M acknowledges insightful discussions with Feng Wu and Hui-Hai Zhao. L.J acknowledges insightful discussions with Rui Li.

\nocite{apsrev41Control}
\bibliography{ref}

@CONTROL{apsrev41Control,title=""}

@unpublished{QG2026,
  author = {Guan, Quan and Chan, {Guo Xuan} and Dou, Xu and Deng,Chunqing and Jin,Lijing},
  note   = {Manuscript in preparation},
}

@misc{Zhan.26,
      title={Scalable Fluxonium Quantum Processors via Tunable-Coupler Architecture}, 
      author={Ze Zhan and Zishuo Li and Fei Wang and Wangwei Lan and Xianchuang Pan and Liang Xiang and Xu Dou and Ran Gao and Guicheng Gong and Yanbo Guo and Quan Guan and Lijuan Hu and Ruizhi Hu and Honghong Ji and Lijing Jin and Yongyue Jin and Chengyao Li and Kannan Lu and Lu Ma and Xizheng Ma and Hongcheng Wang and Jiahui Wang and Huijuan Zhan and Tao Zhou and Xing Zhu and Chunqing Deng and Tenghui Wang},
      eprint={2604.13363},
      archivePrefix={ArXiv}
}

@misc{Zhao.25.1,
      title={Scalable fluxonium qubit architecture with tunable interactions between non-computational levels}, 
      author={Peng Zhao and Guming Zhao and Shaowei Li and Chen Zha and Ming Gong},
      eprint={2504.09888},
      archivePrefix={ArXiv},
}

@article{Campbell.23,
  title = {Modular Tunable Coupler for Superconducting Circuits},
  author = {Campbell, Daniel L. and Kamal, Archana and Ranzani, Leonardo and Senatore, Michael and LaHaye, Matthew D.},
  journal = {Phys. Rev. Appl.},
  volume = {19},
  issue = {6},
  pages = {064043},
  numpages = {17},
  year = {2023},
  month = {Jun},
  publisher = {American Physical Society},
  url = {https://link.aps.org/doi/10.1103/PhysRevApplied.19.064043}
}

@article{Ithier.05,
  title = {Decoherence in a superconducting quantum bit circuit},
  author = {Ithier, G. and Collin, E. and Joyez, P. and Meeson, P. J. and Vion, D. and Esteve, D. and Chiarello, F. and Shnirman, A. and Makhlin, Y. and Schriefl, J. and Sch\"on, G.},
  journal = {Phys. Rev. B},
  volume = {72},
  issue = {13},
  pages = {134519},
  numpages = {22},
  year = {2005},
  month = {Oct},
  publisher = {American Physical Society},
  url = {https://link.aps.org/doi/10.1103/PhysRevB.72.134519}
}

@article{Koch.07,
  title = {Charge-insensitive qubit design derived from the Cooper pair box},
  author = {Koch, Jens and Yu, Terri M. and Gambetta, Jay and Houck, A. A. and Schuster, D. I. and Majer, J. and Blais, Alexandre and Devoret, M. H. and Girvin, S. M. and Schoelkopf, R. J.},
  journal = {Phys. Rev. A},
  volume = {76},
  issue = {4},
  pages = {042319},
  numpages = {19},
  year = {2007},
  month = {Oct},
  publisher = {American Physical Society},
  url = {https://link.aps.org/doi/10.1103/PhysRevA.76.042319}
}

@article{You.19,
  title = {Circuit quantization in the presence of time-dependent external flux},
  author = {You, Xinyuan and Sauls, J. A. and Koch, Jens},
  journal = {Phys. Rev. B},
  volume = {99},
  issue = {17},
  pages = {174512},
  numpages = {10},
  year = {2019},
  month = {May},
  publisher = {American Physical Society},
  url = {https://link.aps.org/doi/10.1103/PhysRevB.99.174512}
}

@misc{Cai.25,
      title={Multiplexed double-transmon coupler scheme in scalable superconducting quantum processor}, 
      author={Tianqi Cai and Chitong Chen and Kunliang Bu and Sainan Huai and Xiaopei Yang and Zhiwen Zong and Yuan Li and Zhenxing Zhang and Yi-Cong Zheng and Shengyu Zhang},
      eprint={2511.02249},
      archivePrefix={ArXiv},
}

@article{Xiong.22,
  title = {Arbitrary controlled-phase gate on fluxonium qubits using differential ac Stark shifts},
  author = {Xiong, Haonan and Ficheux, Quentin and Somoroff, Aaron and Nguyen, Long B. and Dogan, Ebru and Rosenstock, Dario and Wang, Chen and Nesterov, Konstantin N. and Vavilov, Maxim G. and Manucharyan, Vladimir E.},
  journal = {Phys. Rev. Res.},
  volume = {4},
  issue = {2},
  pages = {023040},
  numpages = {16},
  year = {2022},
  month = {Apr},
  publisher = {American Physical Society},
  url = {https://link.aps.org/doi/10.1103/PhysRevResearch.4.023040}
}

@article{Long.19,
  title = {High-Coherence Fluxonium Qubit},
  author = {Nguyen, Long B. and Lin, Yen-Hsiang and Somoroff, Aaron and Mencia, Raymond and Grabon, Nicholas and Manucharyan, Vladimir E.},
  journal = {Phys. Rev. X},
  volume = {9},
  issue = {4},
  pages = {041041},
  numpages = {14},
  year = {2019},
  month = {Nov},
  publisher = {American Physical Society},
  url = {https://link.aps.org/doi/10.1103/PhysRevX.9.041041}
}

@manual{ACEngineer,
	note={More sophisticated strategies exist that exploit multiple drive lines or parasitic coupling to the coupler to suppress leakage transitions, e.g., Refs.~\cite{Nuerbolati.22,Ding.23}. However, such comprehensive drive optimization is beyond the scope of this work.}
}

@manual{FluxSens,
    note={Explicitly, we quantify the robustness of the reset operation against flux variations as $\delta \phi_{\mathrm{ext}_j}^\ast$. This metric is defined as the width of the flux interval over which the reset time $t_{\mathrm{reset}}$ remains below a specified threshold $t_{\mathrm{reset}}^\ast$: $\delta \phi_{\mathrm{ext},j}^\ast = \max \{ \phi_{\mathrm{ext}_j} \mid t_{\mathrm{reset}} < t_{\mathrm{reset}}^\ast \} - \min \{ \phi_{\mathrm{ext}_j} \mid t_{\mathrm{reset}} < t_{\mathrm{reset}}^\ast \}.$
    }
}

@manual{ResetFid,
    note={Reset fidelity is defined as the probability of qubit in ground state after the reset operation. Denote the qubit wavefunction at reset duration $t_\mathrm{reset}$ as $|\psi(t_\mathrm{reset})\rangle$, $F_\mathrm{reset} = \left|\langle \psi(t_\mathrm{reset}) | 0 \rangle \right|^2$.}
}

@article{Sete.14,
  title = {Purcell effect with microwave drive: Suppression of qubit relaxation rate},
  author = {Sete, Eyob A. and Gambetta, Jay M. and Korotkov, Alexander N.},
  journal = {Phys. Rev. B},
  volume = {89},
  issue = {10},
  pages = {104516},
  numpages = {13},
  year = {2014},
  month = {Mar},
  publisher = {American Physical Society},
  url = {https://link.aps.org/doi/10.1103/PhysRevB.89.104516}
}

@article{Krantz.19,
    author = {Krantz, P. and Kjaergaard, M. and Yan, F. and Orlando, T. P. and Gustavsson, S. and Oliver, W. D.},
    title = {A quantum engineer's guide to superconducting qubits},
    journal = {Appl. Phys. Rev.},
    volume = {6},
    number = {2},
    pages = {021318},
    year = {2019},
    month = {06},
    issn = {1931-9401},
    url = {https://doi.org/10.1063/1.5089550},
}

@article{Nuerbolati.22,
    author = {Nuerbolati, Wuerkaixi and Han, Zhikun and Chu, Ji and Zhou, Yuxuan and Tan, Xinsheng and Yu, Yang and Liu, Song and Yan, Fei},
    title = {Canceling microwave crosstalk with fixed-frequency qubits},
    journal = {Appl. Phys. Lett.},
    volume = {120},
    number = {17},
    pages = {174001},
    year = {2022},
    url = {https://doi.org/10.1063/5.0088094},
}

@article{Malley.15,
  title = {Qubit Metrology of Ultralow Phase Noise Using Randomized Benchmarking},
  author = {O'Malley, P. J. J. and Kelly, J. and Barends, R. and Campbell, B. and Chen, Y. and Chen, Z. and Chiaro, B. and Dunsworth, A. and Fowler, A. G. and Hoi, I.-C. and Jeffrey, E. and Megrant, A. and Mutus, J. and Neill, C. and Quintana, C. and Roushan, P. and Sank, D. and Vainsencher, A. and Wenner, J. and White, T. C. and Korotkov, A. N. and Cleland, A. N. and Martinis, John M.},
  journal = {Phys. Rev. Appl.},
  volume = {3},
  issue = {4},
  pages = {044009},
  numpages = {11},
  year = {2015},
  month = {Apr},
  publisher = {American Physical Society},
  url = {https://link.aps.org/doi/10.1103/PhysRevApplied.3.044009}
}

@article{Ding.23,
  title = {High-Fidelity, Frequency-Flexible Two-Qubit Fluxonium Gates with a Transmon Coupler},
  author = {Ding, Leon and Hays, Max and Sung, Youngkyu and Kannan, Bharath and An, Junyoung and Di Paolo, Agustin and Karamlou, Amir H. and Hazard, Thomas M. and Azar, Kate and Kim, David K. and Niedzielski, Bethany M. and Melville, Alexander and Schwartz, Mollie E. and Yoder, Jonilyn L. and Orlando, Terry P. and Gustavsson, Simon and Grover, Jeffrey A. and Serniak, Kyle and Oliver, William D.},
  journal = {Phys. Rev. X},
  volume = {13},
  issue = {3},
  pages = {031035},
  numpages = {24},
  year = {2023},
  month = {Sep},
  publisher = {American Physical Society},
  url = {https://link.aps.org/doi/10.1103/PhysRevX.13.031035}
}

@misc{Ateshian.07,
       author = {{Ateshian}, Lamia and {Hays}, Max and {Rower}, David A. and {Zhang}, Helin and {Azar}, Kate and {Assouly}, R{\'e}ouven and {Ding}, Leon and {Gingras}, Michael and {Stickler}, Hannah and {Niedzielski}, Bethany M. and {Schwartz}, Mollie E. and {Orlando}, Terry P. and {{\^I}-j. Wang}, Joel and {Gustavsson}, Simon and {Grover}, Jeffrey A. and {Serniak}, Kyle and {Oliver}, William D.},
        title = {Temperature and Magnetic-Field Dependence of Energy Relaxation in a Fluxonium Qubit},
        archivePrefix={ArXiv},
      eprint={2507.01175}
}

@article{Nguyen.19,
  title = {High-Coherence Fluxonium Qubit},
  author = {Nguyen, Long B. and Lin, Yen-Hsiang and Somoroff, Aaron and Mencia, Raymond and Grabon, Nicholas and Manucharyan, Vladimir E.},
  journal = {Phys. Rev. X},
  volume = {9},
  issue = {4},
  pages = {041041},
  numpages = {14},
  year = {2019},
  month = {Nov},
  publisher = {American Physical Society},
  url = {https://link.aps.org/doi/10.1103/PhysRevX.9.041041}
}

@article{Li.25,
  title = {Capacitively shunted double-transmon coupler realizing bias-free idling and a high-fidelity CZ gate},
  author = {Li, Rui and Kubo, Kentaro and Ho, Yinghao and Yan, Zhiguang and Inoue, Shinichi and Nakamura, Yasunobu and Goto, Hayato},
  journal = {Phys. Rev. Appl.},
  volume = {23},
  issue = {6},
  pages = {064069},
  numpages = {11},
  year = {2025},
  month = {Jun},
  publisher = {American Physical Society},
  url = {https://link.aps.org/doi/10.1103/l8tq-7sb3}
}

@article{Li.24,
  title = {Realization of High-Fidelity CZ Gate Based on a Double-Transmon Coupler},
  author = {Li, Rui and Kubo, Kentaro and Ho, Yinghao and Yan, Zhiguang and Nakamura, Yasunobu and Goto, Hayato},
  journal = {Phys. Rev. X},
  volume = {14},
  issue = {4},
  pages = {041050},
  numpages = {30},
  year = {2024},
  month = {Nov},
  publisher = {American Physical Society},
  url = {https://link.aps.org/doi/10.1103/PhysRevX.14.041050}
}

@article{DiVincenzo.00,
author = {DiVincenzo, David P.},
title = {The Physical Implementation of Quantum Computation},
journal = {Fortschritte der Physik},
volume = {48},
number = {9-11},
pages = {771-783},
url = {https://onlinelibrary.wiley.com/doi/abs/10.1002/1521-3978%28200009%2948%3A9/11%3C771%3A%3AAID-PROP771%3E3.0.CO%3B2-E},
year = {2000}
}

@article{Yan.18,
  title = {Tunable Coupling Scheme for Implementing High-Fidelity Two-Qubit Gates},
  author = {Yan, Fei and Krantz, Philip and Sung, Youngkyu and Kjaergaard, Morten and Campbell, Daniel L. and Orlando, Terry P. and Gustavsson, Simon and Oliver, William D.},
  journal = {Phys. Rev. Appl.},
  volume = {10},
  issue = {5},
  pages = {054062},
  numpages = {9},
  year = {2018},
  month = {Nov},
  publisher = {American Physical Society},
  url = {https://link.aps.org/doi/10.1103/PhysRevApplied.10.054062}
}

@article{ZhuMT.26,
author = {Zhu, Mutian and Hassanpourghadi, Mohsen and Zhang, Qiaochu and Chen, Mike Shuo-Wei and Levi, A.F.J. and Gupta, Sandeep},
title = {An Active Learning Framework for Analog Circuit Multi-objective Customization},
year = {2026},
publisher = {Association for Computing Machinery},
address = {New York, NY, USA},
issn = {1084-4309},
url = {https://doi.org/10.1145/3789263},
journal = {ACM Trans. Des. Autom. Electron. Syst.},
}

@misc{Zhao.26,
      title={Long-range tunable coupler for modular fluxonium quantum processors}, 
      author={Peng Zhao and Peng Xu and Zheng-Yuan Xue},
      year={2026},
      eprint={2604.12261},
      archivePrefix={ArXiv},
}

@Article{Moskalenko.22,
author={Moskalenko, Ilya N.
and Simakov, Ilya A.
and Abramov, Nikolay N.
and Grigorev, Alexander A.
and Moskalev, Dmitry O.
and Pishchimova, Anastasiya A.
and Smirnov, Nikita S.
and Zikiy, Evgeniy V.
and Rodionov, Ilya A.
and Besedin, Ilya S.},
title={High fidelity two-qubit gates on fluxoniums using a tunable coupler},
journal={npj Quantum Inf.},
year={2022},
month={Nov},
day={08},
volume={8},
number={1},
pages={130},
url={https://doi.org/10.1038/s41534-022-00644-x}
}

@article{Somoroff.23,
  title = {Millisecond Coherence in a Superconducting Qubit},
  author = {Somoroff, Aaron and Ficheux, Quentin and Mencia, Raymond A. and Xiong, Haonan and Kuzmin, Roman and Manucharyan, Vladimir E.},
  journal = {Phys. Rev. Lett.},
  volume = {130},
  issue = {26},
  pages = {267001},
  numpages = {6},
  year = {2023},
  month = {Jun},
  publisher = {American Physical Society},
  url = {https://link.aps.org/doi/10.1103/PhysRevLett.130.267001}
}

@article{Wang.25,
  title = {High-coherence fluxonium qubits manufactured with a wafer-scale-uniformity process},
  author = {Wang, Fei and Lu, Kannan and Zhan, Huijuan and Ma, Lu and Wu, Feng and Sun, Hantao and Deng, Hao and Bai, Yang and Bao, Feng and Chang, Xu and Gao, Ran and Gao, Xun and Gong, Guicheng and Hu, Lijuan and Hu, Ruizi and Ji, Honghong and Ma, Xizheng and Mao, Liyong and Song, Zhijun and Tang, Chengchun and Wang, Hongcheng and Wang, Tenghui and Wang, Ziang and Xia, Tian and Xu, Hongxin and Zhan, Ze and Zhang, Gengyan and Zhou, Tao and Zhu, Mengyu and Zhu, Qingbin and Zhu, Shasha and Zhu, Xing and Shi, Yaoyun and Zhao, Hui-Hai and Deng, Chunqing},
  journal = {Phys. Rev. Appl.},
  volume = {23},
  issue = {4},
  pages = {044064},
  numpages = {14},
  year = {2025},
  month = {Apr},
  publisher = {American Physical Society},
  url = {https://link.aps.org/doi/10.1103/PhysRevApplied.23.044064}
}

@article{Manucharyan.09,
author = {Vladimir E. Manucharyan  and Jens Koch  and Leonid I. Glazman  and Michel H. Devoret },
title = {Fluxonium: Single Cooper-Pair Circuit Free of Charge Offsets},
journal = {Science},
volume = {326},
number = {5949},
pages = {113-116},
year = {2009},
URL = {https://www.science.org/doi/abs/10.1126/science.1175552}}

@article{Nguyen.22,
  title = {Blueprint for a High-Performance Fluxonium Quantum Processor},
  author = {Nguyen, Long B. and Koolstra, Gerwin and Kim, Yosep and Morvan, Alexis and Chistolini, Trevor and Singh, Shraddha and Nesterov, Konstantin N. and J\"unger, Christian and Chen, Larry and Pedramrazi, Zahra and Mitchell, Bradley K. and Kreikebaum, John Mark and Puri, Shruti and Santiago, David I. and Siddiqi, Irfan},
  journal = {PRX Quantum},
  volume = {3},
  issue = {3},
  pages = {037001},
  numpages = {38},
  year = {2022},
  month = {Aug},
  publisher = {American Physical Society},
  url = {https://link.aps.org/doi/10.1103/PRXQuantum.3.037001}
}

@article{Lin.25.1,
url = {https://doi.org/10.1088/1367-2630/adb77b},
year = {2025},
month = {mar},
publisher = {IOP Publishing},
volume = {27},
number = {3},
pages = {033012},
author = {Lin, Wei-Ju and Cho, Hyunheung and Chen, Yinqi and Vavilov, Maxim G and Wang, Chen and Manucharyan, Vladimir E},
title = {Verifying the analogy between transversely coupled spin-1/2 systems and inductively-coupled fluxoniums},
journal = {New J. Phys.}
}

@misc{Chakraborty.25,
       author = {{Chakraborty}, Abhishek and {Bhandari}, Bibek and {Brise{\~n}o-Colunga}, D. Dominic and {Stevenson}, Noah and {Pedramrazi}, Zahra and {Liu}, Chuan-Hong and {Santiago}, David I. and {Siddiqi}, Irfan and {Dressel}, Justin and {Jordan}, Andrew N.},
        title = {Tunable Superconducting Quantum Interference Device Coupler for Fluxonium Qubits},
        eprint = {2508.16907},
        archivePrefix={ArXiv},
}

@misc{Zwanenburg.26,
      title={Crosstalk in Multi-Qubit Fluxonium Architectures with Transmon Couplers}, 
      author={Martijn F. S. Zwanenburg and Christian Kraglund Andersen},
      eprint={2603.09870},
      archivePrefix={ArXiv},
}

@article{Bao.22,
  title = {Fluxonium: An Alternative Qubit Platform for High-Fidelity Operations},
  author = {Bao, Feng and Deng, Hao and Ding, Dawei and Gao, Ran and Gao, Xun and Huang, Cupjin and Jiang, Xun and Ku, Hsiang-Sheng and Li, Zhisheng and Ma, Xizheng and Ni, Xiaotong and Qin, Jin and Song, Zhijun and Sun, Hantao and Tang, Chengchun and Wang, Tenghui and Wu, Feng and Xia, Tian and Yu, Wenlong and Zhang, Fang and Zhang, Gengyan and Zhang, Xiaohang and Zhou, Jingwei and Zhu, Xing and Shi, Yaoyun and Chen, Jianxin and Zhao, Hui-Hai and Deng, Chunqing},
  journal = {Phys. Rev. Lett.},
  volume = {129},
  issue = {1},
  pages = {010502},
  numpages = {6},
  year = {2022},
  month = {Jun},
  publisher = {American Physical Society},
  url = {https://link.aps.org/doi/10.1103/PhysRevLett.129.010502}
}

@article{Lin.25.2,
  title = {24 Days-Stable CNOT Gate on Fluxonium Qubits with Over 99.9\% Fidelity},
  author = {Lin, Wei-Ju and Cho, Hyunheung and Chen, Yinqi and Vavilov, Maxim G. and Wang, Chen and Manucharyan, Vladimir E.},
  journal = {PRX Quantum},
  volume = {6},
  issue = {1},
  pages = {010349},
  numpages = {20},
  year = {2025},
  month = {Mar},
  publisher = {American Physical Society},
  url = {https://link.aps.org/doi/10.1103/PRXQuantum.6.010349}
}

@article{Ma.24,
  title = {Native Approach to Controlled-$Z$ Gates in Inductively Coupled Fluxonium Qubits},
  author = {Ma, Xizheng and Zhang, Gengyan and Wu, Feng and Bao, Feng and Chang, Xu and Chen, Jianjun and Deng, Hao and Gao, Ran and Gao, Xun and Hu, Lijuan and Ji, Honghong and Ku, Hsiang-Sheng and Lu, Kannan and Ma, Lu and Mao, Liyong and Song, Zhijun and Sun, Hantao and Tang, Chengchun and Wang, Fei and Wang, Hongcheng and Wang, Tenghui and Xia, Tian and Ying, Make and Zhan, Huijuan and Zhou, Tao and Zhu, Mengyu and Zhu, Qingbin and Shi, Yaoyun and Zhao, Hui-Hai and Deng, Chunqing},
  journal = {Phys. Rev. Lett.},
  volume = {132},
  issue = {6},
  pages = {060602},
  numpages = {6},
  year = {2024},
  month = {Feb},
  publisher = {American Physical Society},
  url = {https://link.aps.org/doi/10.1103/PhysRevLett.132.060602}
}

@article{Weiss.22,
  title = {Fast High-Fidelity Gates for Galvanically-Coupled Fluxonium Qubits Using Strong Flux Modulation},
  author = {Weiss, D.K. and Zhang, Helin and Ding, Chunyang and Ma, Yuwei and Schuster, David I. and Koch, Jens},
  journal = {PRX Quantum},
  volume = {3},
  issue = {4},
  pages = {040336},
  numpages = {20},
  year = {2022},
  month = {Dec},
  publisher = {American Physical Society},
  url = {https://link.aps.org/doi/10.1103/PRXQuantum.3.040336}
}

@article{Singh.26,
  title = {Fast microwave-driven two-qubit gates between fluxonium qubits with a transmon coupler},
  author = {Singh, Siddharth and Huang, Eugene Y. and Hu, Jinlun and Yilmaz, Figen and Zwanenburg, Martijn F. S. and Kumaravadivel, Piranavan and Wang, Siyu and Stefanski, Taryn V. and Andersen, Christian Kraglund},
  journal = {Phys. Rev. Appl.},
  volume = {25},
  issue = {2},
  pages = {024020},
  numpages = {16},
  year = {2026},
  month = {Feb},
  publisher = {American Physical Society},
  url = {https://link.aps.org/doi/10.1103/yxf3-jtx5}
}

@article{Lange.25,
   title={Cross-talk in superconducting qubit lattices with tunable couplers—comparing transmon and fluxonium architectures},
   volume={11},
   ISSN={2058-9565},
   url={http://dx.doi.org/10.1088/2058-9565/ae2358},
   number={1},
   journal={Quantum Sci. Technol.},
   publisher={IOP Publishing},
   author={Lange, Florian and Heunisch, Lukas and Fehske, Holger and DiVincenzo, David P and Hartmann, Michael J},
   year={2025},
   month=dec, pages={015020} }

@article{Rosenfeld.24,
  title = {High-Fidelity Two-Qubit Gates between Fluxonium Qubits with a Resonator Coupler},
  author = {Rosenfeld, Emma L. and Hann, Connor T. and Schuster, David I. and Matheny, Matthew H. and Clerk, Aashish A.},
  journal = {PRX Quantum},
  volume = {5},
  issue = {4},
  pages = {040317},
  numpages = {35},
  year = {2024},
  month = {Nov},
  publisher = {American Physical Society},
  url = {https://link.aps.org/doi/10.1103/PRXQuantum.5.040317}
}

@misc{Kugut.25,
      title={Interaction-Resilient Scalable Fluxonium Architecture with All-Microwave Gates}, 
      author={Andrei A. Kugut and Grigoriy S. Mazhorin and Ilya A. Simakov},
      eprint={2512.21189},
      archivePrefix={ArXiv},
}

@misc{Heunisch.25,
      title={Scalable Fluxonium-Transmon Architecture for Error Corrected Quantum Processors}, 
      author={Lukas Heunisch and Longxiang Huang and Stephan Tasler and Johannes Schirk and Florian Wallner and Verena Feulner and Bijita Sarma and Klaus Liegener and Christian M. F. Schneider and Stefan Filipp and Michael J. Hartmann},
      eprint={2508.09267},
      archivePrefix={ArXiv},
}

@article{Zhang.24,
  title = {Tunable Inductive Coupler for High-Fidelity Gates Between Fluxonium Qubits},
  author = {Zhang, Helin and Ding, Chunyang and Weiss, D.K. and Huang, Ziwen and Ma, Yuwei and Guinn, Charles and Sussman, Sara and Chitta, Sai Pavan and Chen, Danyang and Houck, Andrew A. and Koch, Jens and Schuster, David I.},
  journal = {PRX Quantum},
  volume = {5},
  issue = {2},
  pages = {020326},
  numpages = {18},
  year = {2024},
  month = {May},
  publisher = {American Physical Society},
  url = {https://link.aps.org/doi/10.1103/PRXQuantum.5.020326}
}

@article{Moskalenko.21,
    author = {Moskalenko, I. N. and Besedin, I. S. and Simakov, I. A. and Ustinov, A. V.},
    title = {Tunable coupling scheme for implementing two-qubit gates on fluxonium qubits},
    journal = {Applied Physics Letters},
    volume = {119},
    number = {19},
    pages = {194001},
    year = {2021},
    month = {11},
    issn = {0003-6951},
    url = {https://doi.org/10.1063/5.0064800},
}

@article{Sete.21,
  title = {Floating Tunable Coupler for Scalable Quantum Computing Architectures},
  author = {Sete, Eyob A. and Chen, Angela Q. and Manenti, Riccardo and Kulshreshtha, Shobhan and Poletto, Stefano},
  journal = {Phys. Rev. Appl.},
  volume = {15},
  issue = {6},
  pages = {064063},
  numpages = {12},
  year = {2021},
  month = {Jun},
  publisher = {American Physical Society},
  url = {https://link.aps.org/doi/10.1103/PhysRevApplied.15.064063}
}

@misc{Xiong.25,
      title={Scalable Low-overhead Superconducting Non-local Coupler with Exponentially Enhanced Connectivity}, 
      author={Haonan Xiong and Jiahui Wang and Juan Song and Jize Yang and Zenghui Bao and Yan Li and Zhen-Yu Mi and Hongyi Zhang and Hai-Feng Yu and Yipu Song and Luming Duan},
      eprint={2502.18902},
      archivePrefix={ArXiv},
}

@article{Dogan.23,
  title = {Two-Fluxonium Cross-Resonance Gate},
  author = {Dogan, Ebru and Rosenstock, Dario and Le Guevel, Lo\"{\i}ck and Xiong, Haonan and Mencia, Raymond A. and Somoroff, Aaron and Nesterov, Konstantin N. and Vavilov, Maxim G. and Manucharyan, Vladimir E. and Wang, Chen},
  journal = {Phys. Rev. Appl.},
  volume = {20},
  issue = {2},
  pages = {024011},
  numpages = {19},
  year = {2023},
  month = {Aug},
  publisher = {American Physical Society},
  url = {https://link.aps.org/doi/10.1103/PhysRevApplied.20.024011}
}

@misc{Huang.26,
      title={Exploration of Fluxonium Parameters for Capacitive Cross-Resonance Gates}, 
      author={Eugene Y. Huang and Christian Kraglund Andersen},
      eprint={2603.17936},
      archivePrefix={ArXiv},
}

@article{Goto.22,
  title = {Double-Transmon Coupler: Fast Two-Qubit Gate with No Residual Coupling for Highly Detuned Superconducting Qubits},
  author = {Goto, Hayato},
  journal = {Phys. Rev. Appl.},
  volume = {18},
  issue = {3},
  pages = {034038},
  numpages = {10},
  year = {2022},
  month = {Sep},
  publisher = {American Physical Society},
  url = {https://link.aps.org/doi/10.1103/PhysRevApplied.18.034038}
}

@article{Kubo.24,
  title = {High-performance multiqubit system with double-transmon couplers: Toward scalable superconducting quantum computers},
  author = {Kubo, Kentaro and Ho, Yinghao and Goto, Hayato},
  journal = {Phys. Rev. Appl.},
  volume = {22},
  issue = {2},
  pages = {024057},
  numpages = {21},
  year = {2024},
  month = {Aug},
  publisher = {American Physical Society},
  url = {https://link.aps.org/doi/10.1103/PhysRevApplied.22.024057}
}

@article{Chen.14,
  title = {Qubit Architecture with High Coherence and Fast Tunable Coupling},
  author = {Chen, Yu and Neill, C. and Roushan, P. and Leung, N. and Fang, M. and Barends, R. and Kelly, J. and Campbell, B. and Chen, Z. and Chiaro, B. and Dunsworth, A. and Jeffrey, E. and Megrant, A. and Mutus, J. Y. and O'Malley, P. J. J. and Quintana, C. M. and Sank, D. and Vainsencher, A. and Wenner, J. and White, T. C. and Geller, Michael R. and Cleland, A. N. and Martinis, John M.},
  journal = {Phys. Rev. Lett.},
  volume = {113},
  issue = {22},
  pages = {220502},
  numpages = {5},
  year = {2014},
  month = {Nov},
  publisher = {American Physical Society},
  url = {https://link.aps.org/doi/10.1103/PhysRevLett.113.220502}
}

@article{Arute.19,
	author = {Arute, Frank and Arya, Kunal and Babbush, Ryan and Bacon, Dave and Bardin, Joseph C. and Barends, Rami and Biswas, Rupak and Boixo, Sergio and Brandao, Fernando G. S. L. and Buell, David A. and Burkett, Brian and Chen, Yu and Chen, Zijun and Chiaro, Ben and Collins, Roberto and Courtney, William and Dunsworth, Andrew and Farhi, Edward and Foxen, Brooks and Fowler, Austin and Gidney, Craig and Giustina, Marissa and Graff, Rob and Guerin, Keith and Habegger, Steve and Harrigan, Matthew P. and Hartmann, Michael J. and Ho, Alan and Hoffmann, Markus and Huang, Trent and Humble, Travis S. and Isakov, Sergei V. and Jeffrey, Evan and Jiang, Zhang and Kafri, Dvir and Kechedzhi, Kostyantyn and Kelly, Julian and Klimov, Paul V. and Knysh, Sergey and Korotkov, Alexander and Kostritsa, Fedor and Landhuis, David and Lindmark, Mike and Lucero, Erik and Lyakh, Dmitry and Mandr{\`a}, Salvatore and McClean, Jarrod R. and McEwen, Matthew and Megrant, Anthony and Mi, Xiao and Michielsen, Kristel and Mohseni, Masoud and Mutus, Josh and Naaman, Ofer and Neeley, Matthew and Neill, Charles and Niu, Murphy Yuezhen and Ostby, Eric and Petukhov, Andre and Platt, John C. and Quintana, Chris and Rieffel, Eleanor G. and Roushan, Pedram and Rubin, Nicholas C. and Sank, Daniel and Satzinger, Kevin J. and Smelyanskiy, Vadim and Sung, Kevin J. and Trevithick, Matthew D. and Vainsencher, Amit and Villalonga, Benjamin and White, Theodore and Yao, Z. Jamie and Yeh, Ping and Zalcman, Adam and Neven, Hartmut and Martinis, John M.},
	id = {Arute2019},
	isbn = {1476-4687},
	journal = {Nature},
	number = {7779},
	pages = {505--510},
	title = {Quantum supremacy using a programmable superconducting processor},
	url = {https://doi.org/10.1038/s41586-019-1666-5},
	volume = {574},
	year = {2019}}

@article{Wu.21,
  title = {Strong Quantum Computational Advantage Using a Superconducting Quantum Processor},
  author = {Wu, Yulin and Bao, Wan-Su and Cao, Sirui and Chen, Fusheng and Chen, Ming-Cheng and Chen, Xiawei and Chung, Tung-Hsun and Deng, Hui and Du, Yajie and Fan, Daojin and Gong, Ming and Guo, Cheng and Guo, Chu and Guo, Shaojun and Han, Lianchen and Hong, Linyin and Huang, He-Liang and Huo, Yong-Heng and Li, Liping and Li, Na and Li, Shaowei and Li, Yuan and Liang, Futian and Lin, Chun and Lin, Jin and Qian, Haoran and Qiao, Dan and Rong, Hao and Su, Hong and Sun, Lihua and Wang, Liangyuan and Wang, Shiyu and Wu, Dachao and Xu, Yu and Yan, Kai and Yang, Weifeng and Yang, Yang and Ye, Yangsen and Yin, Jianghan and Ying, Chong and Yu, Jiale and Zha, Chen and Zhang, Cha and Zhang, Haibin and Zhang, Kaili and Zhang, Yiming and Zhao, Han and Zhao, Youwei and Zhou, Liang and Zhu, Qingling and Lu, Chao-Yang and Peng, Cheng-Zhi and Zhu, Xiaobo and Pan, Jian-Wei},
  journal = {Phys. Rev. Lett.},
  volume = {127},
  issue = {18},
  pages = {180501},
  numpages = {7},
  year = {2021},
  month = {Oct},
  publisher = {American Physical Society},
  url = {https://link.aps.org/doi/10.1103/PhysRevLett.127.180501}
}

@article{Walter.17,
  title = {Rapid High-Fidelity Single-Shot Dispersive Readout of Superconducting Qubits},
  author = {Walter, T. and Kurpiers, P. and Gasparinetti, S. and Magnard, P. and Poto{\v{c}}nik, A. and Salath{\'e}, Y. and Pechal, M. and Mondal, M. and Oppliger, M. and Eichler, C. and Wallraff, A.},
  journal = {Phys. Rev. Appl.},
  volume = {7},
  issue = {5},
  pages = {054020},
  year = {2017},
  month = {May},
  publisher = {American Physical Society},
  url = {https://link.aps.org/doi/10.1103/PhysRevApplied.7.054020}
}

@article{Gambetta.08,
  title = {Quantum trajectory approach to circuit QED: Quantum jumps and the Zeno effect},
  author = {Gambetta, Jay and Blais, Alexandre and Boissonneault, M. and Houck, A. A. and Schuster, D. I. and Girvin, S. M.},
  journal = {Phys. Rev. A},
  volume = {77},
  issue = {1},
  pages = {012112},
  year = {2008},
  month = {Jan},
  publisher = {American Physical Society},
  url = {https://link.aps.org/doi/10.1103/PhysRevA.77.012112}
}

@article{Eric.22,
author = {Eric J. Zhang  and Srikanth Srinivasan  and Neereja Sundaresan  and Daniela F. Bogorin  and Yves Martin  and Jared B. Hertzberg  and John Timmerwilke  and Emily J. Pritchett  and Jeng-Bang Yau  and Cindy Wang  and William Landers  and Eric P. Lewandowski  and Adinath Narasgond  and Sami Rosenblatt  and George A. Keefe  and Isaac Lauer  and Mary Beth Rothwell  and Douglas T. McClure  and Oliver E. Dial  and Jason S. Orcutt  and Markus Brink  and Jerry M. Chow },
title = {High-performance superconducting quantum processors via laser annealing of transmon qubits},
journal = {Sci. Adv.},
volume = {8},
number = {19},
pages = {eabi6690},
year = {2022},
URL = {https://www.science.org/doi/abs/10.1126/sciadv.abi6690}}

@misc{Gaurav.25,
      title={High-fidelity QND readout and measurement back-action in a Tantalum-based high-coherence fluxonium qubit}, 
      author={Gaurav Bothara and Srijita Das and Kishor V Salunkhe and Madhavi Chand and Jay Deshmukh and Meghan P Patankar and R Vijay},
      eprint={2501.16691},
      archivePrefix={ArXiv},
}

@article{Lin.18,
  title = {Demonstration of Protection of a Superconducting Qubit from Energy Decay},
  author = {Lin, Yen-Hsiang and Nguyen, Long B. and Grabon, Nicholas and San Miguel, Jonathan and Pankratova, Natalia and Manucharyan, Vladimir E.},
  journal = {Phys. Rev. Lett.},
  volume = {120},
  issue = {15},
  pages = {150503},
  numpages = {5},
  year = {2018},
  month = {Apr},
  publisher = {American Physical Society},
  doi = {10.1103/PhysRevLett.120.150503},
  url = {https://link.aps.org/doi/10.1103/PhysRevLett.120.150503}
}

@article{Earnest.18,
  title = {Realization of a $\mathrm{\ensuremath{\Lambda}}$ System with Metastable States of a Capacitively Shunted Fluxonium},
  author = {Earnest, N. and Chakram, S. and Lu, Y. and Irons, N. and Naik, R. K. and Leung, N. and Ocola, L. and Czaplewski, D. A. and Baker, B. and Lawrence, Jay and Koch, Jens and Schuster, D. I.},
  journal = {Phys. Rev. Lett.},
  volume = {120},
  issue = {15},
  pages = {150504},
  numpages = {6},
  year = {2018},
  month = {Apr},
  publisher = {American Physical Society},
  doi = {10.1103/PhysRevLett.120.150504},
  url = {https://link.aps.org/doi/10.1103/PhysRevLett.120.150504}
}

@article{Ficheux.21,
  title = {Fast Logic with Slow Qubits: Microwave-Activated Controlled-Z Gate on Low-Frequency Fluxoniums},
  author = {Ficheux, Quentin and Nguyen, Long B. and Somoroff, Aaron and Xiong, Haonan and Nesterov, Konstantin N. and Vavilov, Maxim G. and Manucharyan, Vladimir E.},
  journal = {Phys. Rev. X},
  volume = {11},
  issue = {2},
  pages = {021026},
  numpages = {16},
  year = {2021},
  month = {May},
  publisher = {American Physical Society},
  doi = {10.1103/PhysRevX.11.021026},
  url = {https://link.aps.org/doi/10.1103/PhysRevX.11.021026}
}

@Article{Acharya.25,
author={{Google Quantum AI and Collaborators}},
title={Quantum error correction below the surface code threshold},
journal={Nature},
year={2025},
month={Feb},
day={01},
volume={638},
number={8052},
pages={920-926},
issn={1476-4687},
url={https://doi.org/10.1038/s41586-024-08449-y}
}

@article{He.25,
title = {Experimental Quantum Error Correction below the Surface Code Threshold via All-Microwave Leakage Suppression},
  author = {He, Tan and Lin, Weiping and Wang, Rui and Li, Yuan and Bei, Jiahao and Cai, Jianbin and Cao, Sirui and Chen, Danning and Chen, Kefu and Chen, Xiawei and Chen, Zhe and Chen, Zhiyuan and Chen, Zihua and Chu, Wenhao and Deng, Hui and Ding, Xun and Ding, Zhuzhengqi and Fan, Bo and Fan, Daojin and Fu, Yuanhao and Gao, Dongxin and Gong, Ming and Gui, Jiacheng and Guo, Cheng and Guo, Shaojun and Han, Lianchen and Hong, Linyin and Hu, Yisen and Huang, He-Liang and Huo, Yong-Heng and Jiang, Chenyan and Jiang, Lei and Jiang, Tao and Jiang, Zuokai and Jin, Honghong and Li, Dayu and Li, Dongdong and Li, Jiaqi and Li, Jinjin and Li, Junyan and Li, Junyun and Li, Na and Li, Shaowei and Li, Yuhuai and Liang, Futian and Liao, Nanxing and Lin, Jin and Liu, Ke and Liu, Maliang and Liu, Yancheng and Lou, Haoxin and Ma, Yuwei and Nan, Kailiang and Nie, Meijuan and Niu, Le and Peng, Wenyi and Qian, Haoran and Rong, Hao and Rong, Tao and Shen, Huiyan and Shen, Qiong and Su, Hong and Su, Feifan and Sun, Chenyin and Sun, Liangchao and Sun, Tianzuo and Sun, Yingxiu and Tan, Yimeng and Tan, Jun and Tu, Wenbing and Wang, Jiafei and Wang, Biao and Wang, Chang and Wang, Chen and Wang, Chu and Wang, Jian and Wang, Shengtao and Wang, Xinzhe and Wei, Zuolin and Wu, Dachao and Wu, Gang and Wu, Yulin and Xu, Yu and Xue, Chun and Yan, Kai and Yan, Xin and Yang, Weifeng and Yang, Xinpeng and Yang, Yang and Ye, Yangsen and Ye, Zhenping and Yi, Zhengzhong and Ying, Chong and Yu, Jiale and Yu, Qinjing and Zeng, Xiangdong and Zha, Chen and Zhan, Shaoyu and Zhang, Haibin and Zhang, He and Zhang, Kaili and Zhang, Wen and Zhang, Yiming and Zhang, Yongzhuo and Zhang, Ziying and Zhao, Guming and Zhao, Xintao and Zhao, Youwei and Zhao, Zhong and Zheng, Luyuan and Zhou, Fei and Zhou, Liang and Zhou, Na and Zhou, Naibin and Zhu, Chengjun and Zhu, Qingling and Zou, Guihong and Zou, Haonan and Zhang, Qiang and Lu, Chao-Yang and Peng, Cheng-Zhi and Chen, Fusheng and Zhu, XiaoBo and Pan, Jian-Wei},
  journal = {Phys. Rev. Lett.},
  volume = {135},
  issue = {26},
  pages = {260601},
  numpages = {7},
  year = {2025},
  month = {Dec},
  publisher = {American Physical Society},
  url = {https://link.aps.org/doi/10.1103/rqkg-dw31}
}

@article{Besedin.26,
author={Besedin, Ilya
and Kerschbaum, Michael
and Knoll, Jonathan
and Hesner, Ian
and B{\"o}deker, Lukas
and Colmenarez, Luis
and Hofele, Luca
and Lacroix, Nathan
and Hellings, Christoph
and Swiadek, Fran{\c{c}}ois
and Flasby, Alexander
and Bahrami Panah, Mohsen
and Colao Zanuz, Dante
and M{\"u}ller, Markus
and Wallraff, Andreas},
title={Lattice surgery realized on two distance-three repetition codes with superconducting qubits},
journal={Nat. Phys.},
year={2026},
month={Feb},
day={01},
volume={22},
number={2},
pages={189-194},
issn={1745-2481},
url={https://doi.org/10.1038/s41567-025-03090-6}
}

@Article{Wang.26,
author={Wang, Ke
and Lu, Zhide
and Zhang, Chuanyu
and Liu, Gongyu
and Chen, Jiachen
and Wang, Yanzhe
and Wu, Yaozu
and Xu, Shibo
and Zhu, Xuhao
and Jin, Feitong
and Gao, Yu
and Tan, Ziqi
and Cui, Zhengyi
and Wang, Ning
and Zou, Yiren
and Zhang, Aosai
and Li, Tingting
and Shen, Fanhao
and Zhong, Jiarun
and Bao, Zehang
and Zhu, Zitian
and Han, Yihang
and He, Yiyang
and Shen, Jiayuan
and Wang, Han
and Yang, Jia-Nan
and Song, Zixuan
and Deng, Jinfeng
and Dong, Hang
and Sun, Zheng-Zhi
and Li, Weikang
and Ye, Qi
and Jiang, Si
and Ma, Yixuan
and Shen, Pei-Xin
and Zhang, Pengfei
and Li, Hekang
and Guo, Qiujiang
and Wang, Zhen
and Song, Chao
and Wang, H.
and Deng, Dong-Ling},
title={Demonstration of low-overhead quantum error correction codes},
journal={Nat. Phys.},
year={2026},
month={Feb},
day={01},
volume={22},
number={2},
pages={308-314},
issn={1745-2481},
url={https://doi.org/10.1038/s41567-025-03157-4}
}

@Article{Klimov.24,
author={Klimov, Paul V.
and Bengtsson, Andreas
and Quintana, Chris
and Bourassa, Alexandre
and Hong, Sabrina
and Dunsworth, Andrew
and Satzinger, Kevin J.
and Livingston, William P.
and Sivak, Volodymyr
and Niu, Murphy Yuezhen
and Andersen, Trond I.
and Zhang, Yaxing
and Chik, Desmond
and Chen, Zijun
and Neill, Charles
and Erickson, Catherine
and Grajales Dau, Alejandro
and Megrant, Anthony
and Roushan, Pedram
and Korotkov, Alexander N.
and Kelly, Julian
and Smelyanskiy, Vadim
and Chen, Yu
and Neven, Hartmut},
title={Optimizing quantum gates towards the scale of logical qubits},
journal={Nat. Commun.},
year={2024},
month={Mar},
day={18},
volume={15},
number={1},
pages={2442},
issn={2041-1723},
url={https://doi.org/10.1038/s41467-024-46623-y}
}

@misc{Valles.25,
      title={Optimizing the frequency positioning of tunable couplers in a circuit QED processor to mitigate spectator effects on quantum operations}, 
      author={S. Vallés-Sanclemente and T. H. F. Vroomans and T. R. van Abswoude and F. Brulleman and T. Stavenga and S. L. M. van der Meer and Y. Xin and A. Lawrence and V. Singh and M. A. Rol and L. DiCarlo},
      eprint={2503.13225},
      archivePrefix={ArXiv}
}

@article{Guo.18,
  title = {Dephasing-Insensitive Quantum Information Storage and Processing with Superconducting Qubits},
  author = {Guo, Qiujiang and Zheng, Shi-Biao and Wang, Jianwen and Song, Chao and Zhang, Pengfei and Li, Kemin and Liu, Wuxin and Deng, Hui and Huang, Keqiang and Zheng, Dongning and Zhu, Xiaobo and Wang, H. and Lu, C.-Y. and Pan, Jian-Wei},
  journal = {Phys. Rev. Lett.},
  volume = {121},
  issue = {13},
  pages = {130501},
  numpages = {6},
  year = {2018},
  month = {Sep},
  publisher = {American Physical Society},
  doi = {10.1103/PhysRevLett.121.130501},
  url = {https://link.aps.org/doi/10.1103/PhysRevLett.121.130501}
}

@article{Motzoi.13,
  title = {Improving frequency selection of driven pulses using derivative-based transition suppression},
  author = {Motzoi, F. and Wilhelm, F. K.},
  journal = {Phys. Rev. A},
  volume = {88},
  issue = {6},
  pages = {062318},
  numpages = {15},
  year = {2013},
  month = {Dec},
  publisher = {American Physical Society},
  doi = {10.1103/PhysRevA.88.062318},
  url = {https://link.aps.org/doi/10.1103/PhysRevA.88.062318}
}

@article{WangJH.25,
  title = {Transmon-assisted high-fidelity controlled-$Z$ gates for integer fluxonium qubits},
  author = {Wang, J.-H. and Xiong, H. and Yang, J.-Z. and Zhang, H.-Y. and Song, Y.-P. and Duan, L.-M.},
  journal = {Phys. Rev. Appl.},
  volume = {24},
  issue = {3},
  pages = {034044},
  numpages = {19},
  year = {2025},
  month = {Sep},
  publisher = {American Physical Society},
  doi = {10.1103/qmds-z7gb},
  url = {https://link.aps.org/doi/10.1103/qmds-z7gb}
}
\end{document}